%% file: _main.tex
\documentclass[
 reprint,                
 amsmath,                
 amssymb,                
 aps,                    
 physrev,                
 floatfix,              
 showkeys,               
]{revtex4-2}
\usepackage{natbib}            
\usepackage{hyperref}          
\usepackage{cleveref}          
\usepackage[utf8]{inputenc}    
\usepackage[T1]{fontenc}       
\usepackage{mathptmx}          
\usepackage{bm}                
\usepackage{siunitx}           
\usepackage{CJKutf8}           
\usepackage{booktabs}          
\usepackage{tabularx}          
\usepackage{longtable}         
\usepackage{array}             
\usepackage{makecell}          
\usepackage{multirow}          
\usepackage{threeparttable}    
\usepackage{dcolumn}           
\usepackage{graphicx}          
\usepackage[export]{adjustbox} 
\usepackage{float}             
\usepackage{subfigure}         
\usepackage{tikz}              
\usepackage{pgfplots}          
\usetikzlibrary{arrows.meta, positioning} 
\usepackage{color}             
\usepackage[dvipsnames, table]{xcolor} 
\usepackage{soul}              
\usepackage[most]{tcolorbox}   
\usepackage{enumitem}          
\usepackage{everypage}         
\usepackage{etoolbox}          
\usepackage{comment}           
\usepackage{blindtext}         
\usepackage{todonotes}         

\include{_abbrev}              

\makeatother
\begin{document}
%
\title[]{Consistent kinetic modeling of compressible flows with variable Prandtl numbers: Double-distribution quasi-equilibrium approach}
\author{R.~M.~Strässle}
\email[Email: ]{rubenst@ethz.ch}
\author{S.~A.~Hosseini}
\email[Email: ]{shosseini@ethz.ch}
\author{I.~V.~Karlin}
\email[Corresponding author email: ]{ikarlin@ethz.ch}
\affiliation{Computational Kinetics Group, Department of Mechanical and Process Engineering, ETH Zürich, 8092 Zürich, Switzerland}%
\altaffiliation[Group website: ]{\href{https://ckg.ethz.ch/}{https://ckg.ethz.ch/}}
%
%
\date{\today}

\begin{abstract}
A consistent kinetic modeling and discretization strategy for compressible flows across all Prandtl numbers and specific heat ratios is developed using the quasi-equilibrium approach within two of the most widely used double-distribution frameworks. 
The methodology ensures accurate recovery of the Navier–-Stokes–-Fourier equations, including all macroscopic moments and dissipation rates, through detailed hydrodynamic limit analysis and careful construction of equilibrium and quasi-equilibrium attractors. 
Discretization is performed using high-order velocity lattices with a static reference frame in a discrete velocity Boltzmann context to isolate key modeling aspects such as the necessary requirements on expansion and quadrature orders.
The proposed models demonstrate high accuracy, numerical stability and Galilean invariance across a wide range of Mach numbers and temperature ratios.
Separate tests for strict conservation and measurements of all dissipation rates confirm these insights for all Prandtl numbers and specific heat ratios.
Simulations {of a thermal Couette flow and} a sensitive two-dimensional shock-vortex interaction excellently reproduce viscous Navier--Stokes--Fourier-level physics.
The proposed models establish an accurate, efficient and scalable framework for kinetic simulations of compressible flows with moderate supersonic speeds and discontinuities at arbitrary Prandtl numbers and specific heat ratios, offering a valuable tool for studying complex problems in fluid dynamics 
and paving the way for future extensions to the lattice Boltzmann context, by application of correction terms, as well as high-Mach and hypersonic regimes, employing target-designed reference frames.
%
%
\end{abstract}


\maketitle

\section{Introduction\label{sec:Introduction}}

Understanding compressible and high-speed fluid flows is of paramount importance for science and engineering. 
The development  of reliable, accurate and efficient numerical methods for the simulation of compressible flows, especially on large-scale clusters, has been a topic of intense research over the past few decades~\cite{pirozzoli2011numerical, yee2018recent}. 
While different discrete approximations such as \acp{FV} and \acp{FD} to the \ac{NSF} equations were the main drivers of research in that area, the advent of the \ac{LBM} in the late 80's opened the door for a new class of numerical methods rooted in the kinetic theory of gases.

\Acp{DVBM}, such as \ac{LBM}~\cite{LBMBookKrueger, succiBook}, solve a discrete velocity version of the Boltzmann equation~\cite{Boltzmann1872weitere} obtained via a finite-order expansion of the distribution function in terms of a set of orthonormal functions such as Hermite polynomials~\cite{hosseini2023lattice}.
The dynamics of observable properties of a fluid described by the Euler or \ac{NSF} equations are recovered in the hydrodynamic limit~\cite{Chapman}.
In addition to allowing for the possibility to include physics beyond \ac{NSF} via higher-order expansions, its combination with collision models such as the \ac{BGK} approximation~\cite{bhatnagar1954model} has been shown to provide an efficient alternative to classical solvers~\cite{QianLBM, ChenLBM}. 
Simulation of compressible and high-speed flows with DVBMs has witnessed considerable advances in the past decades, illustrated most notably by formulations such as the family of \ac{UGKS} and \ac{DUGKS}~\cite{guo2021progress}, \ac{FD} and \ac{FV}-based models such as those discussed in~\cite{xu2018discrete,mieussens2000discrete} and the LBM~\cite{hosseini2024lattice}.

In the specific context of the LBM, while research has focused {primarily} on the incompressible regime and stabilization of the isothermal solver through advanced collision models and entropy constrained equilibria construction \cite{Humières_MRT, Geier_Cascaded, Geier_Cumulant, malaspinas_recursivereg, Karlin_entropicEQ, Karlin_Gibbs, Ali_reviewEntropic}, considerable advances are being reported in recent years for compressible flow simulations. These advances have been, in part, motivated by insights from kinetic theory, allowing for, e.g., proper thermal diffusivity in non-unity Prandtl number flows via models such as Shakhov \cite{Shakhov}, Holway \cite{HolwayES} or the \ac{QE} approach \cite{gorban1994, gorban2005invariant, QE_ansumali-2007}. The introduction of \ac{DDF} models~\cite{He_LBMcharacteristics}, where a second distribution carries the total or internal non-translational energy~\cite{rykov_model_1976, KarlinTwoPop, ProbingDoubleDist2024}, had considerable impact on the development of efficient compressible LBM-based solvers capturing variable specific heat ratios. The class of \ac{DDF} LBM solvers relying on standard lattices~\cite{Prasianakis2007, Saadat2019, hosseini2020compressibility, FengCorrection,li2012coupling} and hybrid solvers, modeling the energy balance equation via a FD or FV solver~\cite{FENG_FV_DBM} are practical illustrations of this impact. Models relying on higher-order lattices  properly recovering the energy balance equation~\cite{ChikatamarlaMultispeed, FrapolliMultispeed, strässle2025a-fully-conservative} without a need for correction terms were also developed in that context. While all these approaches relied on static quadratures and as a result were limited in terms of maximum achievable Mach number, the introduction of dynamic quadratures, starting with shifted lattices~\cite{Frapolli2016b,Ali_shiftedStencils, Coreixas2020adaptivevelocity} and culminating with the particles on demand (PonD) method~\cite{PonD18}, opened the door for hypersonic flow simulations~\cite{sawant2022detonation, Kallikounis22, Kallikounis23, Bhaduria23, Ji24}.

While the \ac{DDF} approach is a necessity to model variable specific heat ratios, the non-unity Prandtl number can be fixed in various ways, with the \ac{QE} approach offering the most flexible, generally applicable and stable {realization}. 
This has been shown in various publications, e.g.~\cite{Prasianakis2007, QE_ansumali-2007, KarlinTwoPop, FRAPOLLI2018ELBforthermal, saadat2021extended},
focusing on the construction of a single \ac{DDF} approach and/or single Prandtl number range, mostly notably $\mathrm{Pr}{\leq}1$.
However, while there is ample literature covering theoretical and consistency aspects of other approaches, i.e. Holway and Shakhov, a consistent and general framework 
-- w.r.t kinetic theory and the targeted macroscopic balance equations, i.e. \ac{NSF} -- combining \ac{QE} and \ac{DDF}, valid over all possible energy splitting approaches and Prandtl numbers, $\mathrm{Pr} \in [0,\infty )$, is lacking in the literature. 
The aim of this contribution is to {fill} this major gap.
Therefore, this paper presents a consistent kinetic modeling and discretization approach  for compressible flows covering all Prandtl numbers, 
using the 
\ac{QE} approach for various \ac{DDF} models, extending our previous publication for  $\mathrm{Pr}=1$~\cite{ProbingDoubleDist2024}.
Throughout the manuscript, emphasis is placed on proper model construction, in order to correctly recover all macroscopic moments and dissipation rates for \ac{NSF} level dynamics, as well as retaining strict conservation of mass, momentum and total energy.

The outline of this article is as follows:
The methodology is presented in Section~\ref{sec:methodology}, where 
{brief background and theory is given in~\ref{sec:model-background},} 
the kinetic models are firstly described in Section~\ref{sec:model-description}, the results of a detailed hydrodynamic limit analysis are provided in Section~\ref{sec:CE} and the models are discretized in Section~\ref{sec:model-discretization}.
The physical capabilities of the constructed models are validated in terms of strict conservation, proper recovery of the \ac{NSF} equations, as well as Galilean invariance, and results for {two} sensitive  benchmarks of compressible flows with non-unity Prandtl {are} presented in Section~\ref{sec:results}.
Finally, a summary and {the} conclusions are provided in Section~\ref{sec:summary}.

\section{Methodology\label{sec:methodology}}

\subsection{{Background\label{sec:model-background}}}

\subsubsection{Boltzmann--Bhatnagar-Gross-Krook {model}}

The \ac{BE} is denoted as
\begin{equation}
    \partial_t f \left(\bm{r},\bm{v},t\right) + \bm{v} \cdot \bm{\nabla} f \left(\bm{r},\bm{v},t\right)
    = \Omega_f
    , 
    \label{Boltz_cont_f}
\end{equation}
where the full Boltzmann collision integral~\cite{Boltzmann1872weitere} is modeled with the \ac{BGK} approximation~\cite{bhatnagar1954model},
\begin{equation}
    \Omega_f = -\frac{1}{\tau} \left [f \left(\bm{r},\bm{v},t\right) -f^{\rm eq} \left(\bm{r},\bm{v},t\right)\right].
    \label{BGK_f}
\end{equation}
Particle velocity is designated by $\bm{v}$ while $\bm{r}$ marks the position in space and $t$ the time. 
The probability distribution function and the local equilibrium distribution function are represented by $f \left(\bm{r},\bm{v},t\right)$ and $f^{\rm eq} \left(\bm{r},\bm{v},t\right)$, respectively.
The parameter $\tau$ is the relaxation time that controls the relaxation rate of the distribution function towards the equilibrium given by the local \ac{MB} distribution function for monatomic particles,
\begin{equation}
    f^{\rm eq}\left (\bm{r} , \bm{v},t\right ) 
    = \frac{n }{(2\pi R T)^{D/2}} \exp{\left[-\frac{{{|}\bm{v}-\bm{u}{|}}^2}{2RT}\right]}
    ,\label{eq:MB-equilibrium}
\end{equation}
where $n=\rho / m$ is the particle number density with the particle mass $m$ and mass density $\rho$, $D$ is the dimension of the physical space and $R$ designates the specific gas constant. 
The local equilibrium distribution function is parametrized by mass density $\rho$, velocity $\bm{u}$ and temperature $T$, which are found from, {using the notation $v^2 = \bm{v}\cdot\bm{v}$,}
\begin{gather}
    \rho = m \int f d\bm{v} = m \int f^{\rm eq} d\bm{v}
    ,
    \label{density_1}
\\
    \rho \bm{u} = m \int \bm{v} f d\bm{v} = m \int \bm{v} f^{\rm eq} d\bm{v}
    ,
    \label{momentum_1}
    \\
    E = m \int \frac{v^2}{2} f d\bm{v} = m \int \frac{v^2}{2} f^{\rm eq} d\bm{v}
    ,
    \label{energy_1}
\end{gather}
where $\rho \bm{u}$ {represents} the momentum vector and $E$ is the total energy.
{
Thus, the temperature is obtained with the definition of the total energy for a monatomic perfect gas from the sum of the internal energy, $U^{\rm tr}=\rho C_v^{\rm tr} T$, and the kinetic energy, $K=\rho u^2/2$, i.e.,
\begin{equation}
    E = U^{\rm tr} + K = \rho C_v^{\rm tr} T + \rho \frac{u^2}{2}
    ,\label{eq:Total_energy_MB_tr}
\end{equation}
where, due to the monatomic gas which is considered, the internal energy only contains translational modes, leading to the restriction 
\begin{equation}
    C_v^{\rm tr} = DR/2
    .\label{C_v_tr}
\end{equation}
}
For simplicity, equations are written with $m=1$ for the remainder of this manuscript.

The \ac{MB} \ac{EDF} annuls the Boltzmann collision integral, and is also the minimizer of the $H$-function,
\begin{equation}
    H(f) = \int f \ln \left( f \right) d \bm{v}
    ,\label{eq:H-function}
\end{equation}
under constraints of locally conserved moments, density, momentum and total energy, 
{i.e., }
\begin{equation}
    {
    f^{\rm eq} = \mathrm{argmin} \ H (f) \Big|_{\rho, \rho \bm{u}, E}
    \label{eq:minH_f_DDF}
    .
    }
\end{equation}
{
The BGK approximation fulfills Boltzmann’s $H$-theorem, which states that, for solutions of the Boltzmann equation, the production of the $H$-function due to the collision integral is non-positive definite,
\begin{equation}
    \sigma = \int (\ln f)\,\Omega_f\,\mathrm{d}\bm{v} \le 0
    ,\label{eq:Htheorem}
\end{equation}
expressing irreversible entropy production and a monotonic decay of $H$ toward its minimum.
At its minimum, which is uniquely attained at the local \ac{MB} equilibrium, the $H$-function is directly related to the thermodynamic entropy $s$ of a gas through
\begin{equation}
    s = -k_B H(f^{\rm eq}).
\end{equation}
The $H$-function may therefore be interpreted as a non-equilibrium extension of the thermodynamic entropy when the distribution function departs from the Maxwell--Boltzmann family, and the monotonic decrease of $H$ implied by the $H$-theorem corresponds to non-negative entropy production in accordance with the second law of thermodynamics.
}

\begin{figure*}
    \centering
    \hspace{1cm}
    \includegraphics[width=0.25\linewidth]{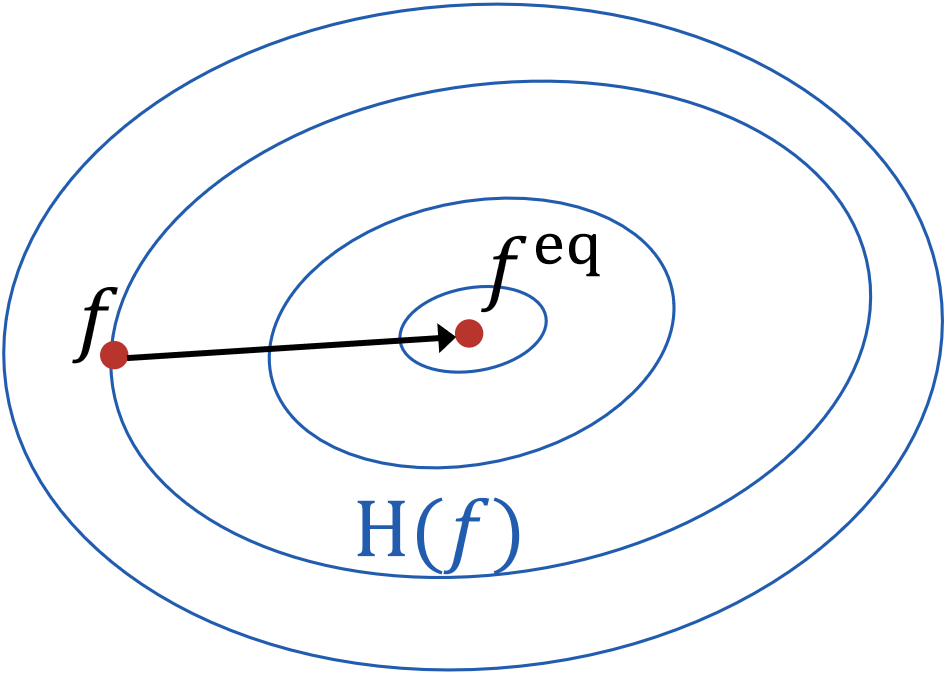}%
    \hfill
    \includegraphics[width=0.25\linewidth]{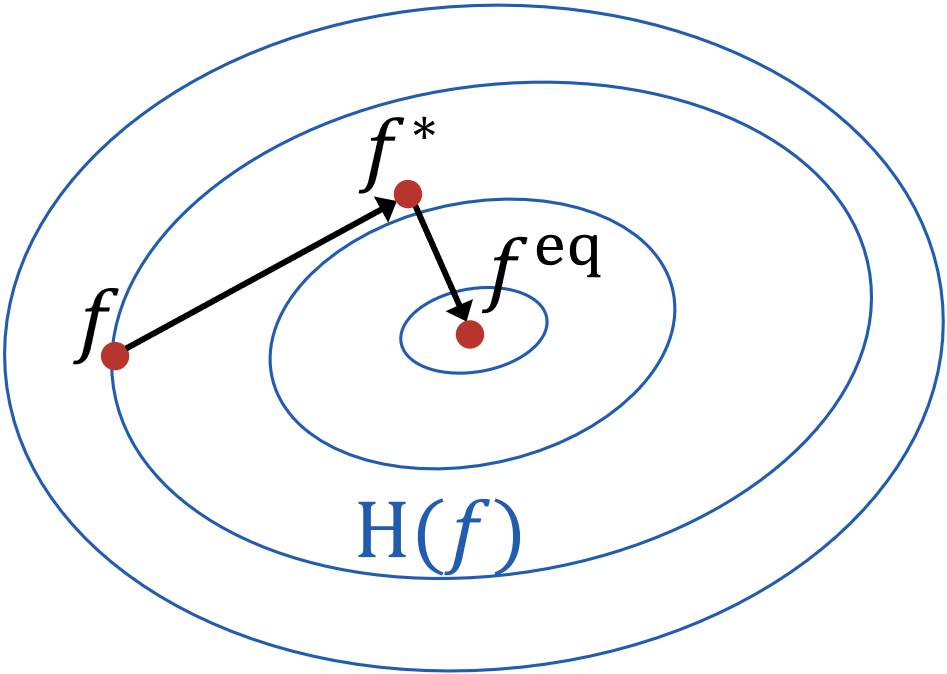}%
    \hfill 
    \includegraphics[width=0.25\linewidth]{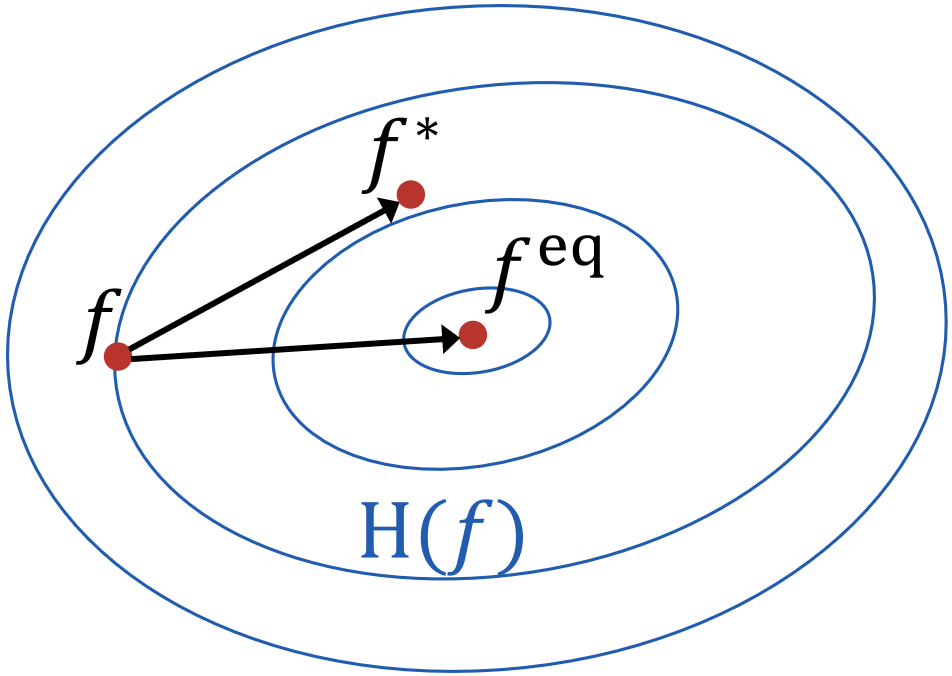}%
    \hspace{1cm}
    \caption{{
    Schematic of the relaxation process(es) in the BGK and QE-BGK kinetic models depicted with contour lines of the $H$-function. 
    From left to right: 
    BGK, Eq. \eqref{BGK_f}; 
    QE-BGK in serial form, Eq. \eqref{eq:Qstar};
    QE-BGK in parallel form, Eq. \eqref{eq:Qstar1}.
    }}
    \label{fig:BGKandQEBGKs}%
\end{figure*}

{
In the next section, the generic construction of quasi-equilibrium kinetic models shall be refreshed, before proceeding to a discussion of modeling general compressible flows, where the \ac{QE} formalism is revisited.
}

{
\subsubsection{Quasi-equilibrium kinetic models}

The generic construction of \ac{QE} kinetic models \cite{gorban1994, Levermore1996Moment} starts by denoting $M$ as the set of locally conserved moments, i.e., Eqs. \eqref{density_1}, \eqref{momentum_1} and \eqref{energy_1}, as
\begin{equation}
    M(f)=\{M_0(f),\dots, M_{D+1}(f)\},
\end{equation}
with 
\begin{equation}
    M_k(f)=\int m_{k}(\bm{v}) f d\bm{v}, \quad k=0,\dots,D+1.
\end{equation} 
The local \ac{EDF}, i.e., Eq. \eqref{eq:MB-equilibrium}, can be written as $f^{\rm eq}(M)$, parametrized by the set of conserved moments.
Furthermore, another linearly independent set of higher-order moments $N$ can be written as 
\begin{equation}
    N(f)=\{N_0(f),\dots, N_{l}(f)\},
\end{equation}
with
\begin{equation}
    N_k(f)=\int n_k(\bm{v}) fd\bm{v}, \quad k=0,\dots,l.
\end{equation}
The \ac{QEDF} is defined as a minimizer of the $H$-function, i.e., Eq. \eqref{eq:H-function}, subject to the fixed set of moments $M$ and $N$ as
\begin{align}
    f^*(M,N)={\rm{argmin}}\ H {(f)} \Big|_{M,N}
    .\label{eq:QEsingleArgmin}
\end{align}
A formal solution is represented as 
\begin{equation}
    \ln f^*=\sum_{k=0}^{D+1}\lambda_km_k+\sum_{k=0}^l\beta_kn_k
    ,\label{eq:lnfstar}
\end{equation}
where $\lambda_k$ and $\beta_k$ are Lagrange multipliers.
If the solution exists, it is unique due to the convexity of the $H$-function.
This is guaranteed if the highest-order moment $n_l\sim |\bm{v}|^l$ is even. 
By the construction, the \ac{QEDF} satisfies the following consistency conditions,
\begin{align}
    M(f^*(M,N))&=M
    ,\label{eq:Mstar=M}
    \\
    N(f^*(M,N))&=N
    ,\label{eq:Nstar=N}
    \\
    f^*(M,N(f^{\rm eq}(M)))&=f^{\rm eq}(M)
    .\label{eq:fstar(feq)=feq}
\end{align}
The latter condition, Eq. \eqref{eq:fstar(feq)=feq}, characterizes the \ac{EDF} as a special case of the \ac{QEDF}.

Using \ac{BGK} approximations \cite{bhatnagar1954model}, the \ac{QE} kinetic model can be written as a \ac{QE-BGK}, i.e., a two-step relaxation to the equilibrium through the intermediate quasi-equilibrium state.
In expanded notation this reads,
\begin{multline}
    \Omega_f=-\frac{1}{\tau_1}(f-f^*(M(f),N(f))) \\ -\frac{1}{\tau_2}(f^*(M(f),N(f))-f^{\rm eq}(M(f)))
    .\label{eq:Qstar}
\end{multline}
A schematic of the \ac{BGK} and the \ac{QE-BGK} collision can be found in Fig. \ref{fig:BGKandQEBGKs}. 
Here the first stage is a relaxation from the current state $f$ to the intermediate quasi-equilibrium $f^*$, with a relaxation time $\tau_1$, while the second stage is the relaxation from the quasi-equilibrium $f^*$ to the equilibrium $f^{\rm eq}$, with the relaxation time $\tau_2$. 
When both relaxation rates are equal, $\tau_1=\tau_2$, the intermediate state $f^*$ is not visited and the \ac{QE-BGK} relaxation, Eq. \eqref{eq:Qstar}, reduces to the \ac{BGK} relaxation, Eq. \eqref{BGK_f}. 
The representation of the \ac{QE} kinetic model in the form of Eq.  \eqref{eq:Qstar} can be termed sequential.
Another telling form, which can be termed a parallel (see Fig. \ref{fig:BGKandQEBGKs} for a schematic), is obtained upon rewriting Eq. \eqref{eq:Qstar} as
\begin{align}
    \Omega_f&=-\frac{1}{\tau_1}(f-f^*)-\frac{1}{\tau_2}(f^*-f^{\rm eq})-\frac{1}{\tau_2}f+\frac{1}{\tau_2}f\nonumber\\
       &=-\left(\frac{1}{\tau_1}-\frac{1}{\tau_2}\right)(f-f^*)-\frac{1}{\tau_2}(f-f^{\rm eq})
       .\label{eq:Qstar1}
\end{align}
Here, the current state $f$ relaxes in parallel to both the \ac{EDF}, $f^{\rm eq}$, (with the time $\tau_2$) and to the intermediate \ac{QEDF}, $f^*$. 
The latter relaxation is characterized with the effective relaxation time $\theta$,
\begin{equation}
    \theta=\frac{\tau_1\tau_2}{\tau_2-\tau_1}
    .\label{eq:taueff}
\end{equation}
Positivity of the effective relaxation time \eqref{eq:taueff} requires the inequality of the relaxation times,
\begin{equation}
    \tau_1\le \tau_2
    .\label{eq:slowfast}
\end{equation}
The latter hierarchy of relaxation times allows to interpret the first stage of relaxation (from $f$ to $f^*$) in Eq. \eqref{eq:Qstar} as a "fast" process, as compared to the "slow" second stage (from $f^*$ to $f^{\rm eq}$). 
Consequently, the higher-order moments $N$ are termed as slow moments, as compared to all other higher-order moments that are linearly independent from $N$ and $M$. 
Moreover, the fundamental hierarchy of relaxation times, i.e., Eq. \eqref{eq:slowfast}, guarantees the $H$-theorem for the \ac{QE} kinetic models. 
This can be found when computing the entropy production, cf. Eq. \eqref{eq:Htheorem}, with the relaxation term in the form of Eq. \eqref{eq:Qstar1}, and taking into account the inequality in Eq. \eqref{eq:slowfast}, as
\begin{align}
    \sigma&=\int (\ln f) \Omega_fd\bm{v}\nonumber\\
    &=-\left(\frac{1}{\tau_1}-\frac{1}{\tau_2}\right)\int (\ln f)(f-f^*)d\bm{v} 
    \nonumber\\&\quad\quad-\frac{1}{\tau_2}\int(\ln f)(f-f^{\rm eq})d\bm{v}\nonumber\\
    &=-\left(\frac{1}{\tau_1}-\frac{1}{\tau_2}\right)\int \ln \left(\frac{f}{f^*}\right)(f-f^*)d\bm{v}
    \nonumber\\&\quad\quad-\frac{1}{\tau_2}\int\ln \left(\frac{f}{f^{\rm eq}}\right)(f-f^{\rm eq})d\bm{v}\nonumber\\
    &\leq 0, \label{eq:QEmodel_f_Htheorem}
\end{align}
where Eq. \eqref{eq:lnfstar} was used.

In practical applications, only approximate expressions for the quasi-equilibrium distribution functions are usually available, such as Grad's moment approximation \cite{grad1949kinetic} or the so-called triangle entropy method \cite{gorban1991quasi}. 
However, the fundamental relaxation times hierarchy, i.e., Eq. \eqref{eq:slowfast}, has to be maintained regardless of whether the exact quasi-equilibrium or an approximation thereof is used. 
}

\subsection{{Kinetic m}odel description\label{sec:model-description}}

\subsubsection{{Extension to} compressible flows}

The \ac{BGK-BE}, as discussed in Section \ref{sec:model-background}, has few well-known shortcomings that need to be addressed before being used to model general compressible flow hydrodynamics: 
\begin{itemize}
    \item Variable specific heat capacity valid for polyatomic molecules: The distribution function has to be extended in order to account for the internal roto-vibrational degrees of freedom of the gas molecules in a polyatomic gas.
    \item Variable Prandtl number: The \ac{BGK} collision operator results in the restriction of a Prandtl number of unity.
\end{itemize}

{
To account for the first issue (variable specific heat capacity valid for polyatomic molecules), a widely used approach was originally introduced by Rykov~\cite{rykov_model_1976}, who extended the BGK framework to polyatomic gases by introducing additional relaxation mechanisms to account for the additional degrees of freedom allowing to model flows with variable adiabatic exponents $\gamma$.
For computational convenience, this idea was later reformulated, mainly based on the lattice Boltzmann picture, by introducing a second distribution function $g$, evolving according to another Boltzmann transport equation, which explicitly represents the evolution of internal roto-vibrational energy modes of the gas. 
This description further allowed to extend the second distribution to carry some, or the full, part of the total energy, as the $g$- and $f$-distributions can be linked to constitute the necessary information~\cite{ProbingDoubleDist2024}. 
It shall be noted that, in the dilute-gas limit with monoatomic particles the double-distribution approach does not strictly reduce to the classical single-distribution BGK equation, as the conservation of energy is represented differently; this distinction does, however, not affect the recovery of the intended macroscopic hydrodynamic equations.

An option to lift the second limitation (variable Prandtl number) is given} 
by the application of the generalized-BGK collision operator, guided by the \ac{QE} approximation approach~\cite{gorban1994, gorban1991quasi, Gorban2006Quasi} 
{(see previous section)}. 
The latter component decomposes the dynamics of the system into fast and slow modes, which is very practical for systems possessing different time scales of physical processes. Of interest here is a Prandtl number of non-unity,
\begin{equation}
  {\rm Pr} = \frac{C_p \mu}{\kappa} = \frac{\nu}{\alpha}
  ,
\end{equation}
which expresses the ratio of momentum and thermal diffusion. In the case where the heat flux is regarded as the "slow" variable, the thermal conductivity $\kappa$ (or thermal diffusivity $\alpha$) can be related to the "slow" relaxation time, while the dynamic (shear) viscosity $\mu$ (or kinematic viscosity $\nu$) is proportional to the "fast" relaxation time, and vice versa.

\subsubsection{{Double-distribution quasi-equilibrium kinetic model}}

{Combining the double-distribution and the \ac{QE} approach, variable adiabatic exponents can be introduced and the Prandtl number can be controlled independently.} 
The transport equations for $f$ and $g$ become
{
\begin{multline}
    \partial_t f\left(\bm{r},\bm{v},t\right) 
    + \bm{v} \cdot \bm{\nabla} f\left(\bm{r},\bm{v},t\right) 
    = 
    - \frac{1}{\tau_1} \left[  f\left(\bm{r},\bm{v},t\right) - f^{*} \left(\bm{r},\bm{v},t\right) \right] 
\\
    - \frac{1}{\tau_2} \left[ f^{*} \left(\bm{r},\bm{v},t\right) - f^{\rm eq}\left(\bm{r},\bm{v},t\right) \right]
,\label{new_collision_f_serial}
\end{multline}
and
\begin{multline}
    \partial_t g\left(\bm{r},\bm{v},t\right) 
    + \bm{v} \cdot \bm{\nabla} g\left(\bm{r},\bm{v},t\right) 
    = 
    - \frac{1}{\tau_1} \left[  g\left(\bm{r},\bm{v},t\right) - g^{*} \left(\bm{r},\bm{v},t\right) \right] 
\\
    - \frac{1}{\tau_2} \left[ g^{*} \left(\bm{r},\bm{v},t\right) - g^{\rm eq}\left(\bm{r},\bm{v},t\right) \right]
    .\label{new_collision_g_serial}
\end{multline}
The same consideration as in the single distribution \ac{QE} model with 
the hierarchy of relaxation times, 
cf. Eq. \eqref{eq:slowfast}, carries over to the double-distribution setting, i.e. 
\begin{equation}
    \tau_1 \leq \tau_2.\label{eq:slowfast_doubledist}
\end{equation}
Finally, it shall be noted that the necessity of introducing an intermediate \ac{QE} state when employing multiple relaxation times in the double-distribution framework has been discussed in detail in \cite{KarlinTwoPop}.
}

In the following sections,
{the construction of all \acp{EDF} and \acp{QEDF} are specified in detail.}

\subsubsection{{Variable adiabatic exponent: Energy splits and second distribution}}

{
In the double distribution setting, the definition of the first reduced
distribution function follows that of Rykov \cite{rykov_model_1976}.
The density and momentum are computed from the first reduced distribution function, $f$, as
\begin{gather}
    \rho = \int f d\bm{v}
        = \int f^{\rm eq} d\bm{v} 
    ,\label{eq_con_f1}
\\
    \rho \bm{u} = \int \bm{v} f d\bm{v}
        = \int \bm{v} f^{\rm eq} d\bm{v}
    .\label{eq_con_f2}
\end{gather}
The equilibrium, $f^{\rm eq}$, is defined as the \ac{MB} distribution, i.e. Eq. \eqref{eq:MB-equilibrium}, which corresponds to the minimizer of the $H$-function under conserved density, momentum and total energy.
Therefore, the expressions \eqref{eq_con_f1} and \eqref{eq_con_f2} are the same as in Eqs. \eqref{density_1} and \eqref{momentum_1} used in the single-distribution case.
However, the total energy, which is used in the minimization, differs from Eq. \eqref{energy_1} and \eqref{eq:Total_energy_MB_tr} in the double distribution setting, as also non-translational degrees of freedom are to be considered, which is where the second distribution comes into play.
With the kinetic energy $K$ and the full internal energy, $U=\rho C_vT$, where $C_v$ is not anymore restricted to a monoatomic gas and thus Eq. \eqref{C_v_tr}, 
the definition of the total energy is written as}
\begin{equation}
    E = U + K = \rho C_v T + \rho \frac{u^2}{2}
    .\label{eq:Total_energy_MB}
\end{equation}
{In other words, while replacing the temperature $T$ in the Maxwell-Boltzmann equilibrium \eqref{eq:MB-equilibrium} through total energy as in Eq. \eqref{eq:Total_energy_MB} instead of Eq. \eqref{eq:Total_energy_MB_tr}, the reduced $f^{\rm eq}$ becomes dependent on the second distribution function $g$, which carries parts of the additional energy due to the presence of non-translational degrees of freedom.
}

{Different parts of energy can be carried by the second distribution, $g$.
In this work, the forms where either the total energy or the internal non-translational energy are put on $g$ were considered, respectively. 
Note that other energy splits between $f$ and $g$ are also possible, as can for example be read in a comparative study in~\cite{ProbingDoubleDist2024}. }
However subsequent modification to Eqs.~\eqref{new_collision_f_serial} and~\eqref{new_collision_g_serial} leading to appearance of non-conservative and non-local source terms makes these splits less computationally attractive. 
{In addition to} that, other advantages and disadvantages were {found,} which will also become visible in the remainder of this manuscript.
Hereafter, equations and remarks concerning {uniquely} the total {or} internal non-translational energy split are labeled by the roman numbers ($\mathrm{I}$) {or} ($\mathrm{II}$), respectively, for clearer visibility.

{Using the internal energy associated with the non-translational degrees of freedom for convenience, here denoted as}
\begin{equation}
    {U^{\rm ntr} = U - U^{\rm tr} }= U - \rho \frac{DR}{2}T
    ,
\end{equation} 
{the second distribution can now be defined with respect to the total energy as}
\begin{align}
    \begin{aligned}[b]
        &\textcolor{black}{(\mathrm{I})}\phantom{\mathrm{I}} \quad 
        E = \int g d\bm{v} = \int g^{\rm eq} d\bm{v}
        ,
    \\
        &\textcolor{black}{(\mathrm{II})} \quad
        E = \int g + \frac{v^2}{2} f d\bm{v} = \int g^{\rm eq} + \frac{v^2}{2} f^{\rm eq} d\bm{v}
        ,
    \end{aligned}
\end{align}
for the total and internal non-translational energy split, respectively.
{
Consequently, the equilibrium of the second distribution is defined as}
\begin{align}
    \begin{aligned}[b]
        &\textcolor{black}{(\mathrm{I})}\phantom{\mathrm{I}} \quad
        g^{\rm eq} = \left( {\frac{U^{\rm ntr}}{\rho}} + \frac{v^2}{2} \right) f^{\rm eq}
        = \left (C_vT-\frac{RDT}{2} + \frac{v^2}{2} \right) f^{\rm eq}
        ,
    \\
        &\textcolor{black}{(\mathrm{II})} \quad
        g^{\rm eq} = {\frac{U^{\rm ntr}}{\rho}} f^{\rm eq}
        = \left (C_v T - \frac{RDT}{2} \right) f^{\rm eq}
        ,
    \end{aligned}
    \label{energy_eq_1}
\end{align}
respectively.
{Note that $g^{\rm eq}$ is therefore a reparameterization of $f^{\rm eq}$.}

{
All explicit expressions for the relevant equilibrium and conserved moments of $f$ and $g$ are listed in Appendix \ref{appendix:EqMoments}. 
}

\subsubsection{{Variable Prandtl number: Constraints for the quasi-equilibrium states}}

{The \acp{QEDF} in the double-distribution setting are defined as the minimizer of the $H$-function subject to the locally conserved density, momentum and total energy and additional quasi-conserved slow fields, analogous to the single-distribution setting, cf. Eq. \eqref{eq:QEsingleArgmin}.}
In case of applying the \ac{QE} notion to variable Prandtl numbers, the pressure tensor and the heat flux vector mark the quasi-conserved fields, with altering conditions depending on the Prandtl number {$\{ \mathrm{Pr}\leq1$, $\mathrm{Pr}\geq1 \}$}. 
Hence, explicitly, in order to recover the NSF equations with variable Prandtl numbers, the \acp{QEDF} are required to satisfy the conservation of mass and momentum as
\begin{gather}
    \rho = \int f^{*} d\bm{v} 
        = \int f d\bm{v}
        = \int f^{\rm eq} d\bm{v} 
    ,\label{eq_con_fstar1}
\\
    \rho \bm{u} = \int \bm{v}  f^{*} d\bm{v} 
        = \int \bm{v} f d\bm{v}
        = \int \bm{v} f^{\rm eq} d\bm{v}
    ,\label{eq_con_fstar2}
\end{gather}
and total energy as
\begin{align}
    \begin{aligned}[b]
        &\textcolor{black}{(\mathrm{I})}\phantom{\mathrm{I}} \quad
        E = 
        \int g^{*} d\bm{v} 
        = \int  g d\bm{v} 
        = \int g^{\rm eq} d\bm{v} 
        ,
    \\
        &\begin{aligned}[b]
        \textcolor{black}{(\mathrm{II})} \quad
        E = 
        \int g^{*} + \frac{v^2}{2} f^{*}  d\bm{v} 
        =& \int g + \frac{v^2}{2} f d\bm{v}
        \\
        &= \int g^{\rm eq} + \frac{v^2}{2} f^{\rm eq} d\bm{v} 
        .
        \end{aligned}
        \label{eq_con_gstar_total}
    \end{aligned}
\end{align}
For $\mathrm{Pr} {\leq} 1$, in addition to Eqs. \eqref{eq_con_fstar1} to \eqref{eq_con_gstar_total}, 
the \acp{QEDF} have to satisfy for the pressure tensor, 
\begin{equation}
    \int\bm{v}\otimes\bm{v} f^{*} d\bm{v} 
    = \int\bm{v}\otimes\bm{v} f^{\rm eq} d\bm{v}
    ,\label{eq:QE_Pr<1_conserved}
\end{equation}
and for the heat flux vector, 
\begin{align}
    \begin{aligned}[b]
        &\begin{aligned}
            \textcolor{black}{(\mathrm{I})}\phantom{\mathrm{I}} \quad
            \int \bm{v} g^{*} d\bm{v} 
            - \bm{u}\cdot &
        \int \bm{v}\otimes\bm{v}f^{*} d\bm{v}
        \\ 
            &= \int \bm{v} g d\bm{v}  
            - \bm{u}\cdot\int \bm{v}\otimes\bm{v}f d\bm{v}
            ,
        \end{aligned}
        \\  
        &\begin{aligned}[b]
            \textcolor{black}{(\mathrm{II})} \quad
            \int \bm{v} g^{*}& + \bm{v}\frac{v^2}{2} f^{*} d\bm{v}  
            - \bm{u}\cdot 
            \int \bm{v}\otimes\bm{v}f^{*} d\bm{v}
        \\ 
            &= \int \bm{v} g + \bm{v} \frac{v^2}{2} f d\bm{v} 
            - \bm{u}\cdot\int \bm{v}\otimes\bm{v}f d\bm{v}
            .
        \end{aligned}
    \end{aligned}
    \label{eq:QE_Pr<1_quasiconserved}
\end{align}
For $\mathrm{Pr} {\geq} 1$, the \acp{QEDF} are required to satisfy for
the pressure tensor,
\begin{equation}
    \int\bm{v}\otimes\bm{v} f^{*} d\bm{v} 
    = \int\bm{v}\otimes\bm{v} f d\bm{v}
    ,\label{eq:QE_Pr>1_quasiconserved}
\end{equation}
and for the heat flux vector,
\begin{align}
    \begin{aligned}[b]
        &\begin{aligned}
            \textcolor{black}{(\mathrm{I})}\phantom{\mathrm{I}} \quad
            \int \bm{v} g^{*} d\bm{v} 
            - \bm{u}\cdot &
            \int \bm{v}\otimes\bm{v}f^* d\bm{v}
        \\ 
            &= \int \bm{v} g^{\rm eq} d\bm{v}  
            - \bm{u}\cdot\int \bm{v}\otimes\bm{v}f^{\rm eq} d\bm{v}
            ,
        \end{aligned}
        \\  
        &\begin{aligned}
            \textcolor{black}{(\mathrm{II})} \quad
            \int \bm{v} &g^{*} + \bm{v}\frac{v^2}{2} f^{*} d\bm{v}  
            - \bm{u}\cdot
            \int \bm{v}\otimes\bm{v}f^* d\bm{v}
        \\ 
            &= \int \bm{v} g^{\rm eq} + \bm{v} \frac{v^2}{2} f^{\rm eq} d\bm{v} 
            - \bm{u}\cdot\int \bm{v}\otimes\bm{v}f^{\rm eq} d\bm{v}
            .
        \end{aligned}
    \end{aligned}
    \label{eq:QE_Pr>1_conserved}
\end{align}
In the limit of $\mathrm{Pr}=1$, no quasi-conserved fields are present; 
{only the constraints on the conserved fields hold, and thus the \acp{QEDF} are equivalent to the \acp{EDF}, i.e. $\{f^*,g^*\} = \{f^{\rm eq},g^{\rm eq}\}$. 

Note that the key difference in the above constraints on the quasi-conserved fields between the cases of $\mathrm{Pr}{\leq}1$ and $\mathrm{Pr}{\geq}1$, 
i.e. Eqs. \eqref{eq:QE_Pr<1_conserved} and \eqref{eq:QE_Pr<1_quasiconserved} vs. Eqs. \eqref{eq:QE_Pr>1_quasiconserved} and \eqref{eq:QE_Pr>1_conserved},
lies in the \ac{QE} pressure tensor and the \ac{QE} heat flux vector:
In one case ($\mathrm{Pr}{\leq}1$) the \ac{QE} pressure tensor is set to equilibrium while the \ac{QE} heat flux is set to the full off-equilibrium distribution, whereas in the other case ($\mathrm{Pr}{\geq}1$) the \ac{QE} pressure tensor is off-equilibrium and the \ac{QE} heat flux is set to equilibrium.
Further, note that herein, the difference in the expressions between the splits lies in the fact that the total energy $E$ is composed of contributions from $g$ only in the case of the total energy split, e.g. $\int g^{*} d\bm{v}$, whereas the internal non-translational energy split also contains contributions from $f$, e.g., $\int g^{*} + \frac{v^2}{2} f^{*} d\bm{v}$, cf. Eq. \eqref{eq_con_gstar_total}.
This difference also carries over to the higher-order expressions, i.e., to the \ac{QE} heat flux.
}

Together with the expressions for the relevant equilibrium and conserved moments in Appendix~\ref{appendix:EqMoments}, the specifications stated in this section (Section \ref{sec:model-description}) complete all necessary information for the constructed kinetic models.
Next, these models are assessed to prove convergence to a hydrodynamic limit.

\subsection{Hydrodynamic limit\label{sec:CE}}

The conclusions of the multiscale analysis in the form of the Chapman-Enskog expansion~\cite{Chapman} shall be highlighted at this point. The detailed derivations are attached in Appendix~\ref{appendix:CE} for the interested reader.
The specified system recovers the \ac{NSF} equations in the hydrodynamic limit, i.e.,
\begin{gather}
    \partial_t\rho + \bm{\nabla} \cdot \rho \bm{u} 
    = 0
    ,
    \\
    \partial_t (\rho \bm{u}) + \bm{\nabla} \cdot \left( \rho \bm{u} \otimes \bm{u} + p\bm{I} {+}\bm{\tau}_{\rm NS} \right) 
    = 0
    ,
    \\
    \partial_t E + \bm{\nabla} \cdot \left[ (E + p)\bm{u} + \bm{q}_{\rm F} {+}
    \bm{q}_{\rm H}
    \right] 
    = 0
    ,
\end{gather}
where the dissipative mechanisms are in the form of the Navier--Stokes stress tensor
\begin{multline}
    \bm{\tau}_{\rm NS} = 
    {-}\mu\left[ \bm{\nabla}\bm{u} + \bm{\nabla}\bm{u}^{\dagger}- \frac{R}{C_v}(\bm{\nabla}\cdot\bm{u})\bm{I} \right]
    \\ 
    = {-}\mu \left[\bm{\nabla}\bm{u} + \bm{\nabla}\bm{u}^\dagger - \frac{2}{D}(\bm{\nabla}\cdot\bm{u})\bm{I} \right] 
    {-} \eta (\bm{\nabla}\cdot\bm{u})\bm{I}
    ,\label{eq:Tns}
\end{multline}
the Fourier heat flux
\begin{equation}
    \bm{q}_{\rm F} = - \kappa \bm{\nabla}T
    \label{eq:Fourier}
    ,
\end{equation}
and the viscous heating vector
\begin{multline}
    \bm{q}_{\rm H} = \bm{u} \cdot \bm{\tau}_{\rm NS} =
   {-} \mu \Bigl[ \bm{u}\cdot\bm{\nabla}\bm{u} + \bm{u}\cdot\bm{\nabla}\bm{u}^{\dagger} - \frac{R}{C_v}{\bm{u}(\bm{\nabla}\cdot\bm{u})}\Bigr]
    \\ 
    ={-}\mu \left[\bm{u}\cdot\bm{\nabla}\bm{u} + \bm{u}\cdot\bm{\nabla}\bm{u}^\dagger - \frac{2}{D}{\bm{u}(\bm{\nabla}\cdot\bm{u})} \right] 
    {-\eta \bm{u}(\bm{\nabla}\cdot\bm{u})}
    .\label{eq:viscHeating}
\end{multline}
{From the multiscale analysis is is further found that}
the relaxation parameters $\tau_1$ and $\tau_2$ {in Eqs. \eqref{new_collision_f_serial} and \eqref{new_collision_g_serial}} are related to the shear viscosity $\mu$, bulk viscosity $\eta$ and thermal conductivity $\kappa$ as, for ${\rm Pr\leq1}$,
\begin{gather}
    \label{visc_shear_1}
    \nu = \frac{\mu}{\rho} =\tau_1    R T
    , 
\\
    \label{visc_bluk_1}
    \zeta = \frac{\eta}{\rho} = \left(\frac{2}{D}-\frac{R}{C_v}\right) \tau_1 R T
    ,
\\
    \label{visc_thermal_1}
    \alpha = \frac{\kappa}{C_p\rho} = \tau_2 R T
    ,
\end{gather}
while for ${\rm Pr{\geq}1}$
\begin{gather}
    \label{visc_shear_2}
    \nu = \frac{\mu}{\rho} = \tau_2    R T
    , 
\\
    \label{visc_bluk_2}
    \zeta = \frac{\eta}{\rho} = \left(\frac{2}{D}-\frac{R}{C_v}\right) \tau_2 R T
    ,
\\
    \label{visc_thermal_2}
    \alpha = \frac{\kappa}{C_p\rho} = \tau_1 R T
    ,
\end{gather} 
and can therefore be found from the imposed dynamic shear viscosity and Prandtl number. 
{In order to respect the relaxation times hierarchy, cf. Eq. \eqref{eq:slowfast_doubledist}, it becomes evident that the Prandtl number and the relaxation parameters are related through 
\begin{equation}
    \mathrm{Pr} =
    \begin{cases}
        \frac{\tau_1}{\tau_2}, & \mathrm{Pr} \leq 1,
        \\
        \frac{\tau_2}{\tau_1}, & \mathrm{Pr} {\geq} 1.
    \end{cases} 
    \label{eq:Pr_times_cases}
\end{equation} 
}

Next, the discretization of the model in phase space is discussed.

\subsection{Phase-space discretization\label{sec:model-discretization}}


\subsubsection{Discrete velocity system}
The phase-space is discretized with a set of $Q$ discrete velocities $\bm{v}_i$, where $i \in \{0, Q-1\}$. 
Discretization is operated via expansion using the Hermite orthonormal polynomials and application of the Gauss-Hermite quadrature. 
As this is standard practice, further description is omitted here for the sake of readability, however, the interested reader can find more details in Appendix \ref{appendix:Gauss-Hermite} {or refer to} 
\cite{hosseini2023lattice, LBMBookKrueger}. 
The phase-space-discrete system of hyperbolic \acp{PDE} then reads,
{
\begin{multline}
        \partial_t f_i \left(\bm{r},t\right) 
        + \bm{v}_i \cdot \bm{\nabla} f_i \left(\bm{r},t\right) 
        = 
        -\frac{1}{\tau_1} \left[ f_i  \left(\bm{r},t\right) - f_i^{*} \left(\bm{r},t\right) \right] 
    \\
        - \frac{1}{\tau_2} \left[ f_i^{*} \left(\bm{r},t\right)  
        -  f_i^{\rm eq} \left(\bm{r},t\right)  \right]
        .\label{new_collision_fi}
\end{multline}
and,
\begin{multline}
        \partial_t g_i \left(\bm{r},t\right) 
        + \bm{v}_i \cdot \bm{\nabla} g_i \left(\bm{r},t\right) 
        = 
        -\frac{1}{\tau_1} \left[ g_i  \left(\bm{r},t\right) - g_i^{*} \left(\bm{r},t\right) \right] 
    \\
        - \frac{1}{\tau_2} \left[ g_i^{*} \left(\bm{r},t\right)  
        -  g_i^{\rm eq} \left(\bm{r},t\right)  \right]
        .\label{new_collision_gi}
\end{multline}
}
Moments of distribution functions are computed as numerical quadratures, e.g. the conserved density and momentum as
\begin{gather}
    \label{eq_quadratur1}
    \rho = \sum_{i=0}^{Q-1} f_i = \sum_{i=0}^{Q-1} f^{\rm eq}_i = \sum_{i=0}^{Q-1} f^{*}_i
    , 
\\
    \label{eq_quadratur2}
    \rho\bm{u} = \sum_{i=0}^{Q-1} \bm{v}_i f_i = \sum_{i=0}^{Q-1} \bm{v}_i f^{\rm eq}_i = \sum_{i=0}^{Q-1} \bm{v}_i f^{*}_i
    ,  
\end{gather}
and total energy as 
\begin{align}
    \begin{aligned}[b]
        &\textcolor{black}{(\mathrm{I})}\phantom{\mathrm{I}} \quad
        E = 
        \sum_{i=0}^{Q-1}  g_i = \sum_{i=0}^{Q-1}  g^{\rm eq}_i = \sum_{i=0}^{Q-1} g^{*}_i
        ,
    \\
        &\textcolor{black}{(\mathrm{II})} \quad
        E = \sum_{i=0}^{Q-1} g_i + \frac{v_i^2}{2}  f_i = \sum_{i=0}^{Q-1} g^{\rm eq}_i + \frac{v_i^2}{2} f^{\rm eq}_i = \sum_{i=0}^{Q-1} g^*_i + \frac{v_i^2}{2}  f^*_i
        . 
    \label{eq_quadratur3_intnontrans}
    \end{aligned}    
\end{align}

\subsubsection{Discrete equilibria}
The discrete $f$- and $g$-equilibria can be reconstructed as a finite-order Grad expansion~\cite{grad1949kinetic},
\begin{align}
     f^{\rm eq}_{i} 
     &= w_i \sum_{n=0}^{N} \frac{1}{n!(RT_{ref})^n} 
     \bm{a}_{n}^{\rm eq}(f) : \bm{\mathcal{H}}_{n}(\bm{v}_i)
     ,
 \\
     g^{\rm eq}_{i} 
     &= w_i \sum_{n=0}^{N} \frac{1}{n!(RT_{ref})^n} 
     \bm{a}_{n}^{\rm eq}(g) : \bm{\mathcal{H}}_{n}(\bm{v}_i)
     ,\label{eq:Grad_g}
\end{align}
where the series is truncated at the order $N$. The weights and the reference temperature of the velocity set are given by $w_i$ and $T_{ref}$, respectively. Both discrete equilibria are written as expansions parametrized by $\bm{\mathcal{H}}_{n}(\bm{v}_i)$ and $\bm{a}_{n}(\{f,g\})$, where
$\bm{\mathcal{H}}_{n}(\bm{v}_i)$ is the Hermite polynomial tensor of order $n$ of the $i$-the particle velocity and $\bm{a}_{n}$ is the corresponding coefficient tensor accounting for the required set of equilibrium moments. 
More details and the explicit expressions for the Hermite polynomials and coefficients of the Grad expansion are given in Appendix~\ref{appendix:Grad-Hermite}. The set of equilibrium moments can directly be drawn from Appendix~\ref{appendix:EqMoments}.

\subsubsection{Requirements on the phase-space discretization\label{sec:Requirements}}

When going from a continuous phase-space to a discrete velocity set, there are a minimum number of requirements that need to be satisfied. To capture the fundamental flow physical properties of the \ac{NSF} equations in the hydrodynamic limit, all moments appearing in the phase-space continuous multiscale expansion have to be matched when computed with discrete quadratures.
The detailed considerations can be found in Appendix~\ref{appendix:requirements}.
In summary, for the $f$-distribution, equilibrium moments up to order three have to be properly recovered for the total energy split. For the non-translational {energy} split, fourth-order equilibrium moments have to be properly recovered as well. For the $g$-distribution, regardless of the split, equilibrium moments up to order two need to be properly recovered.
These orders of expansion require higher-order velocity sets, cf.~\cite{FrapolliMultispeed} or standard nearest-neighbor velocity sets with correction terms to the otherwise incorrectly recovered higher-order moments~\cite{saadat2021extended, Prasianakis2007}. 

Note that the discrete $g$-equilibria, if applied with the same velocity set as $f$, can also be directly parametrized by the discrete $f$-equilibria as in the phase-space continuous kinetic model, cf. Eq.~\eqref{energy_eq_1}.
{The detailed considerations can be found in} Appendix~\ref{appendix:directParamEq}.
{In summary,} while this parametrization simplifies the computation of the $g_i^{\rm eq}$ for the internal non-translational {energy} split, 
for the total energy split it increases the 
{orders of the Grad expansion and the quadrature.}

In this work, all equilibria were constructed with the Grad--Hermite expansion
{and the solution via higher-order velocity sets was considered.
The direct parametrization of $g_i$ was not considered, in order to apply the minimal Hermite-based higher-order velocity sets {(in two dimensions)}, i.e., the D2Q16 and the D2Q25 were used for the total and internal non-translational energy splits, respectively.}
The details of the mentioned velocity sets are listed in Appendix~\ref{appendix:latticespecs}.

\subsubsection{Discrete quasi-equilibria}
{Finally}, the construction of the \ac{QE}-populations remains to be clarified.
They are also {constructed using the Grad--Hermite expansion} to the same order as discussed before.
However, a different set of moments is accounted for in the coefficient tensors  $\bm{a}_{n}^*(\{f,g\})$.
Due to the similarity in the derivation of equilibria (minimization of discrete $H$-function under constraints of the conserved moments, cf. constraints in Eqs.~\eqref{eq_con_fstar1} to~\eqref{eq_con_gstar_total}), the quasi-equilibria generally contain the equilibrium moments with the 
{exception of some moments coming from the constraints for} 
the quasi-conserved fields, i.e. Eqs.~\eqref{eq:QE_Pr<1_conserved} to~\eqref{eq:QE_Pr>1_conserved}.
This means that the quasi-equilibria can be constructed as a "new" Grad--Hermite expansion, or as a {"modification"} to the computed equilibrium populations at the affected order. 
To be concise, the full expansions of all distributions are used herein. 
{For this, the \ac{QE} moments to use in the coefficient tensors are as follows.}

For 
\textcolor{black}{
$\mathrm{Pr} < 1$,
}
{one has}
\begin{equation}
    \textcolor{black}{(\mathrm{I, \ II})} \quad 
    \sum_{i=0}^{Q-1} \bm{v}_i\otimes\bm{v}_i f_i^{*} = \sum_{i=0}^{Q-1} \bm{v}_i\otimes\bm{v}_i f_i^{\rm eq}
    ,\label{eq:disc_QE_Pr<1_conserved}
\end{equation}
for both splits. 
In addition, for the total energy split one obtains
\begin{equation}
    \textcolor{black}{(\mathrm{I})} \quad
            \sum_{i=0}^{Q-1} \bm{v}_i g_i^{*}   
            = \sum_{i=0}^{Q-1} \bm{v}_i g_i    
            - \bm{u}\cdot
        \Bigl(            
            \sum_{i=0}^{Q-1} \bm{v}_i\otimes\bm{v}_if_i  
        -\sum_{i=0}^{Q-1} \bm{v}_i\otimes\bm{v}_if_i^{\rm eq} \Bigr)
        ,\label{eq:disc_QE_Pr<1_quasiconserved}
\end{equation}
{and for the internal non-translational energy split, the explicit expressions read}
\begin{equation}
    \textcolor{black}{(\mathrm{II})} \quad
    \sum_{i=0}^{Q-1} \bm{v}_i g_i^{*}   
    = \sum_{i=0}^{Q-1} \bm{v}_i g_i   
    ,
\end{equation}
and
\begin{multline}
    \textcolor{black}{(\mathrm{II})} \quad
    \sum_{i=0}^{Q-1} \bm{v}_i\otimes\bm{v}_i\otimes\bm{v}_i f_i^{*} =  \sum_{i=0}^{Q-1} \bm{v}_i\otimes\bm{v}_i\otimes\bm{v}_i f_i    
    \\
    - 2 \bm{u} \otimes
    \Bigl(            
        \sum_{i=0}^{Q-1} \bm{v}_i\otimes\bm{v}_if_i  
    -\sum_{i=0}^{Q-1} \bm{v}_i\otimes\bm{v}_if_i^{\rm eq} \Bigr)
    ,
\end{multline}
where Eq.~\eqref{eq:disc_QE_Pr<1_conserved} was applied.
{The latter expression for the third-order moment of $f^*$ is slightly over-constrained, as in \cite{2016Entropic}.}
For 
\textcolor{black}{
$\mathrm{Pr} > 1$,
} 
{one has}
\begin{equation}
    \textcolor{black}{(\mathrm{I})} \quad
    \sum_{i=0}^{Q-1}\bm{v}_i\otimes\bm{v}_i f_i^{*}   
    = \sum_{i=0}^{Q-1}\bm{v}_i\otimes\bm{v}_i f_i  
    ,\label{eq:disc_QE_Pr>1_quasiconserved}
\end{equation}
{for the total energy split, whereas for the internal on-translational split one obtains}
\begin{equation}
        \textcolor{black}{(\mathrm{II})} \quad
    \sum_{i=0}^{Q-1}\bm{v}_i\otimes\bm{v}_i f_i^{*}   
    = \sum_{i=0}^{Q-1}\bm{v}_i\otimes\bm{v}_i f_i  
    - \frac{\bm{I}}{D} \sum_{i=0}^{Q-1}v_i^2 \left(f_i - f_i^{\rm eq}\right)
    ,\label{eq:disc_QE_Pr>1_quasiconserved_conservationFix}
\end{equation}
in order to respect Eq.~\eqref{eq_quadratur3_intnontrans} (conservation of total energy) at the same time.
{Additionally, for the total energy split one obtains}
{and}
\begin{multline}
    \textcolor{black}{(\mathrm{I})}\phantom{\mathrm{I}} \quad
        \sum_{i=0}^{Q-1} \bm{v}_i g_i^{*}   
        = \sum_{i=0}^{Q-1} \bm{v}_i g_i^{\rm eq}    
       \\ + \bm{u}\cdot
    \Bigl(
        \sum_{i=0}^{Q-1} \bm{v}_i\otimes\bm{v}_if_i 
    -  \sum_{i=0}^{Q-1} \bm{v}_i\otimes\bm{v}_if_i^{\rm eq}  
    \Bigr)
    ,\label{eq:disc_QE_Pr>1_conserved}
\end{multline}
{and for the internal on-translational split the explicit expressions are}
\begin{equation}
    \textcolor{black}{(\mathrm{II})} \quad
    \sum_{i=0}^{Q-1} \bm{v}_i g_i^{*} = \sum_{i=0}^{Q-1} \bm{v}_i g_i^{\rm eq} 
    ,
\end{equation}
and 
\begin{multline}
        \textcolor{black}{(\mathrm{II})} \quad
          \sum_{i=0}^{Q-1} \bm{v}_i\otimes\bm{v}_i\otimes\bm{v}_i f_i^{*} =  \sum_{i=0}^{Q-1} \bm{v}_i\otimes\bm{v}_i\otimes\bm{v}_i f_i^{\rm eq}
        \\  \quad + 2 \bm{u} \otimes
        \Bigl(
        \sum_{i=0}^{Q-1}\bm{v}_i\otimes\bm{v}_i \left(f_i - f_i^{\rm eq}\right)
        - \frac{\bm{I}}{D} \sum_{i=0}^{Q-1}v_i^2 \left(f_i - f_i^{\rm eq}\right) 
        \Bigr)
        . \label{eq:disc_QE_Pr>1_conserved2}  
\end{multline}
where Eq.~\eqref{eq:disc_QE_Pr>1_quasiconserved} was applied.
{
The latter expression for the third-order moment of $f^*$ is again slightly over-constrained, as in \cite{2016Entropic}.
Note that all above expressions satisfy the constraints in Eqs. \eqref{eq_con_fstar1} to \eqref{eq:QE_Pr>1_conserved}.
}

Further, note that the derived conditions can also be formulated in another Hermite basis, for example with $\mathcal{H}_{xy}$, $\mathcal{H}_{xx}-\mathcal{H}_{yy}$ and $\mathcal{H}_{xx}+\mathcal{H}_{yy}$ in the second-order contributions for the Grad expansions, in order to impose conditions on the trace and traceless parts of the second-order moments directly, which could be useful for, e.g., incorporating {a contacted moment.}

\begin{table*}[t]
\footnotesize
\centering
\begin{tabular}{|l|l|lllll|l|}
    \hline
    \multicolumn{8}{|c|}{\textcolor{black}{$(\mathrm{I})$} Total energy split} \\
    \hline
    Prandtl & Pop. & \multicolumn{5}{c|}{Set of moments} & Comment \\
    \hline
    \multirow{2}{*}{any}  & $f_i^{\rm eq}$  & $  M_{0}(f^{\rm eq}),$ 
                                            & $\bm{M}_{1}(f^{\rm eq}),$ 
                                            & $\bm{M}_{2}(f^{\rm eq}),$ 
                                            & $\bm{M}_{3}(f^{\rm eq})$, 
                                            & - & - \\
                          & $g_i^{\rm eq}$  & $  M_{0}(g^{\rm eq}),$ 
                                            & $\bm{M}_{1}(g^{\rm eq}),$ 
                                            & $\bm{M}_{2}(g^{\rm eq}),$ 
                                            & -, 
                                            & - & - \\
    \cline{1-1}
    \multirow{2}{*}{Pr$\ =1$} & $f_i^{*}$       & $  M_{0}(f^{\rm eq}),$ 
                                                & $\bm{M}_{1}(f^{\rm eq}),$ 
                                                & $\bm{M}_{2}(f^{\rm eq}),$ 
                                                & $\bm{M}_{3}(f^{\rm eq})$, 
                                                & - & $f_i^{*} = f_i^{\rm eq}$ \\
                              & $g_i^{*}$       & $  M_{0}(g^{\rm eq}),$ 
                                                & $\bm{M}_{1}(g^{\rm eq}),$ 
                                                & $\bm{M}_{2}(g^{\rm eq}),$ 
                                                & -, 
                                                & - & $g_i^{*} = g_i^{\rm eq}$ \\
    \cline{1-1}
    \multirow{2}{*}{Pr$\ <1$} & $f_i^{*}$       & $  M_{0}(f^{\rm eq}),$ 
                                                & $\bm{M}_{1}(f^{\rm eq}),$ 
                                                & $\bm{M}_{2}(f^{\rm eq}),$ 
                                                & $\bm{M}_{3}(f^{\rm eq})$, 
                                                & - & $f_i^{*} = f_i^{\rm eq}$ \\
                              & $g_i^{*}$       & $  M_{0}(g^{\rm eq}),$ 
                                                & \cellcolor{green!20} $\bm{M}_{1}(g)-\bm{u}\cdot\bigl(\bm{M}_{2}(f)-\bm{M}_{2}(f^{\rm eq})\bigr),$ 
                                                & $\bm{M}_{2}(g^{\rm eq}),$ 
                                                & -, 
                                                & - & - \\
    \cline{1-1}
    \multirow{2}{*}{Pr$\ >1$}  & $f_i^{*}$       & $  M_{0}(f^{\rm eq}),$ 
                                                & $\bm{M}_{1}(f^{\rm eq}),$ 
                                                & \cellcolor{green!20} $\bm{M}_{2}(f),$ 
                                                & $\bm{M}_{3}(f^{\rm eq})$, 
                                                & - & - \\
                               & $g_i^{*}$      & $  M_{0}(g^{\rm eq}),$ 
                                                & \cellcolor{green!20} $\bm{M}_{1}(g^{\rm eq})+\bm{u}\cdot\bigl(\bm{M}_{2}(f)-\bm{M}_{2}(f^{\rm eq})\bigr),$ 
                                                & $\bm{M}_{2}(g^{\rm eq}),$ 
                                                & -, 
                                                & - & - \\
    \hline
    \hline
    \multicolumn{8}{|c|}{\textcolor{black}{($\mathrm{II}$)} Internal non-translational energy split} \\
    \hline
    Prandtl & Pop. & \multicolumn{5}{c|}{Set of moments} & Comment \\
    \hline
    \multirow{2}{*}{any}  & $f_i^{\rm eq}$  & $ M_{0}(f^{\rm eq}),$ 
                                            & $\bm{M}_{1}(f^{\rm eq}),$ 
                                            & $\bm{M}_{2}(f^{\rm eq}),$ 
                                            & $\bm{M}_{3}(f^{\rm eq})$, 
                                            & $\bm{M}_{4}(f^{\rm eq})$ & - \\
                          & $g_i^{\rm eq}$  & $ M_{0}(g^{\rm eq}),$ 
                                            & $\bm{M}_{1}(g^{\rm eq}),$ 
                                            & $\bm{M}_{2}(g^{\rm eq}),$ 
                                            & -,
                                            & - & - \\
    \cline{1-1}
    \multirow{2}{*}{Pr$\ =1$} & $f_i^{*}$       & $ M_{0}(f^{\rm eq}),$ 
                                                & $\bm{M}_{1}(f^{\rm eq}),$ 
                                                & $\bm{M}_{2}(f^{\rm eq}),$ 
                                                & $\bm{M}_{3}(f^{\rm eq})$, 
                                                & $\bm{M}_{4}(f^{\rm eq})$ & $f_i^{*} = f_i^{\rm eq}$ \\
                              & $g_i^{*}$       & $  M_{0}(g^{\rm eq}),$ 
                                                & $\bm{M}_{1}(g^{\rm eq}),$ 
                                                & $\bm{M}_{2}(g^{\rm eq}),$ 
                                                & -, 
                                                & - & $g_i^{*} = g_i^{\rm eq}$ \\
    \cline{1-1}
    \multirow{2}{*}{Pr$\ <1$} & $f_i^{*}$       & $ M_{0}(f^{\rm eq}),$ 
                                                & $\bm{M}_{1}(f^{\rm eq}),$ 
                                                & $\bm{M}_{2}(f^{\rm eq}),$ 
                                                & \cellcolor{green!20} $\bm{M}_{3}(f) -2\bm{u}\otimes\bigl(\bm{M}_{2}(f)-\bm{M}_{2}(f^{\rm eq})\bigr)$, 
                                                & $\bm{M}_{4}(f^{\rm eq})$ & - \\
                              & $g_i^{*}$       & $ M_{0}(g^{\rm eq}),$ 
                                                & \cellcolor{green!20} $\bm{M}_{1}(g),$ 
                                                & $\bm{M}_{2}(g^{\rm eq}),$ 
                                                & -, 
                                                & - & - \\
    \cline{1-1}
    \multirow{2}{*}{Pr$\ >1$} & $f_i^{*}$       & $ M_{0}(f^{\rm eq}),$ 
                                                & $\bm{M}_{1}(f^{\rm eq}),$ 
                                                & \cellcolor{green!20} $\bm{M}_{2}(f) 
                                                - \frac{\bm{I}}{D} \mathrm{tr}\bigl(\bm{M}_{2}(f)-\bm{M}_{2}(f^{\rm eq})\bigr),$
                                                & \makecell{\cellcolor{green!20}
                                                    $\bm{M}_{3}(f^{\rm eq}) +2\bm{u}\otimes\bigl[(\bm{M}_{2}(f)-\bm{M}_{2}(f^{\rm eq}))$ 
                                                \\
                                                    $\phantom{\bm{M}_{3}(f^{\rm eq}) \ \ }-\frac{\bm{I}}{D} \mathrm{tr}(\bm{M}_{2}(f)-\bm{M}_{2}(f^{\rm eq})) \bigr]$,
                                                } 
                                                & $\bm{M}_{4}(f^{\rm eq})$ & - \\
                               & $g_i^{*}$      & $ M_{0}(g^{\rm eq}),$ 
                                                & $\bm{M}_{1}(g^{\rm eq}),$ 
                                                & $\bm{M}_{2}(g^{\rm eq}),$ 
                                                & -, 
                                                & - & $g_i^{*} = g_i^{\rm eq}$ \\
    \hline
\end{tabular}
\caption{
    Summary of moments for the construction of equilibrium and quasi-equilibrium populations using Grad--Hermite expansions.
}
\label{tab:SummaryOfPopulations}
\end{table*}

{\subsubsection{Summary of the discrete equilibria and quasi-equilibria construction}}

A summary of the construction of all discrete equilibria and quasi-equilibria can be found in Table~\ref{tab:SummaryOfPopulations}.
The expressions for the equilibrium moments can be drawn from  Appendix~\ref{appendix:EqMoments}. 
All expressions that differ from the equilibria construction are indicated with colored background. 
The following notation is used in the table,
\begin{gather}
    \bm{M}_{n}(\{f,g\}) = \sum_{i=0}^{Q-1} \overbrace{v_{i} \otimes\dots\otimes v_{i}}^{n} \{f_i,g_i\},\\ 
    \bm{M}_{n}(\{f^{\rm eq},g^{\rm eq}\}) = \sum_{i=0}^{Q-1} \overbrace{v_{i} \otimes\dots\otimes v_{i}}^{n} \{f_i^{\rm eq},g_i^{\rm eq}\}.
\end{gather}

\section{Validation, results and discussion\label{sec:results}}

\subsection{Discrete velocity Boltzmann implementation}
\subsubsection{Time-explicit finite-volume scheme}
As the Hermite-based higher-order velocity sets employed in this work do not propagate on-lattice, instead of using a semi-Lagrangian approach, e.g.~\cite{BardowMultispeedSSLBM, Kallikounis_multiscale, Wilde_SSLBM_CF, Wilde_CubatureRules}, a time-explicit finite-volume scheme in the form of a discrete velocity Boltzmann solver is used for the fully conservative discretization in space and time, similar as in~\citet{strässle2025a-fully-conservative,Ji24}.
For simplicity, a first-order Euler-forward discretization in time was applied.
Note that the resulting distribution functions and moments should be understood as volumetric averages over a cell. The resulting fully discretized system of hyperbolic \acp{PDE} are written as  
{
\begin{multline}
\label{eq:finalDiscretePDE}
    \{f_i,g_i\} (\bm{r}, t+ \delta t) = \{f_i,g_i\}(\bm{r},t) - \frac{\delta t}{\delta V} \sum_{\sigma \in \Theta}  \{\mathcal{F}_i,\mathcal{G}_i\} (\sigma, t)
\\
    - \frac{\delta t}{\tau_1} \Bigl[ \{f_i,g_i\}(\bm{r},t) - \{f_i^*,g_i^*\}(\bm{r},t) \Bigr] 
\\
    - \frac{\delta t}{\tau_2} \Bigl[ \{f_i^*,g_i^*\}(\bm{r},t) - \{f_i^{\rm eq},g_i^{\rm eq}\}(\bm{r},t) \Bigr] 
    .  
\end{multline}
}
Here $\delta t$ is the time-step size, $\delta V$ the volume of the cell and $\{\mathcal{F}_i,\mathcal{G}_i\}$ fluxes through cell boundaries $\sigma$. The estimation of the fluxes through the cell boundaries is the key ingredient in the discretization of space, both in terms of accuracy and stability.

\begin{figure*}
    \centering
    \includegraphics[width=\linewidth]{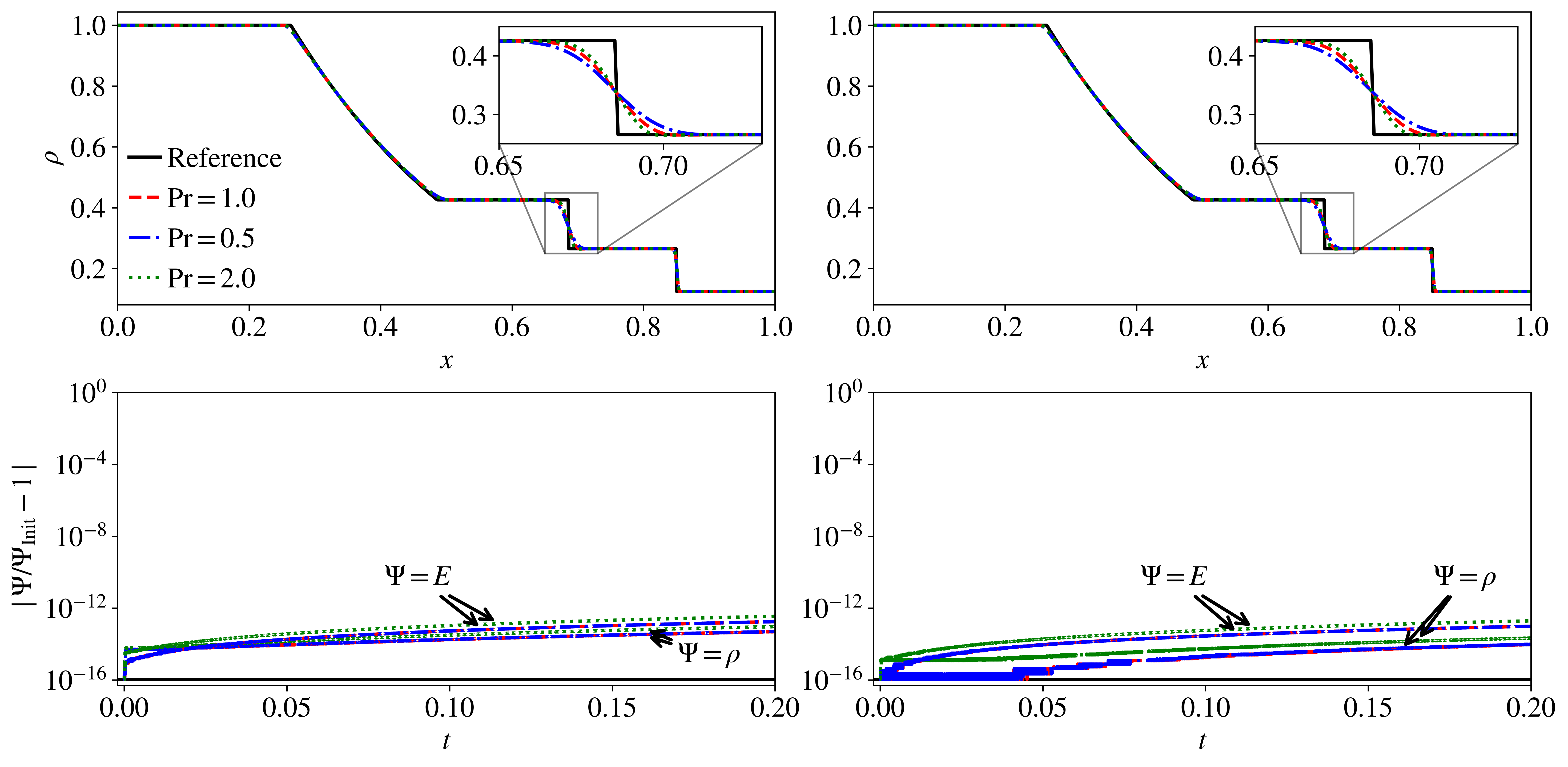}%
    \caption{
    Results of the shock tube problem with $\mathrm{Pr} = \{0.5,1,2\}$ for the total energy split in the left column and internal-non-translational in the right column. 
    The top row depicts a comparison of the density with the reference (Riemann solution for inviscid problem), whereas the bottom row shows the relative error in conservation of mass and total energy, where the machine epsilon for double precision is depicted as a reference.
    }
    \label{fig:ConservationSod1}%
\end{figure*}

\subsubsection{Flux reconstruction}

The fluxes in Eq.~\eqref{eq:finalDiscretePDE} are computed as 
\begin{equation}
        \{\mathcal{F}_i, \mathcal{G}_i \} (\sigma) = 
        \bm{v}_i \cdot \bm{n}(\sigma)  
        \{{f}_i, {g}_i \} (\sigma) 
        \delta A(\sigma)
    ,
\end{equation}
where $\bm{n}$ is the surface normal vector and $\delta A$ the infinitesimal area of the discrete surface $\sigma$ of the Cuboid's hull $\Theta$ with volume $\delta V$.
The distribution functions require interpolation to the interfaces in an accurate, yet stable, manner by introducing more numerical dissipation if necessary.
In this work, a nearest neighbor deformation (NND) interpolation scheme~\cite{NND1, NND2} was applied together with a generalized van Leer limiter~\cite{Harten1987Uniformly, Harten1987Uniformly}, which takes into account the ratio of successive slopes $a$, $b$~\cite{VanLeer_muscle, limiterRoe1986}. 
On a uniform Cartesian grid, the scheme reads
\begin{align}
    v_{i,x - \delta x/2} \geq \text{$0$:} \ \
      &\{f_i, g_i\}_{x - \delta x/2} = \{f_i, g_i\}_{x - \delta x} + \frac{b}{2} \phi \left(a, b\right); 
    \nonumber \\&\qquad\quad a=\{f_i, g_i\}_{x - \delta x} - \{f_i, g_i\}_{x - 2 \delta x}, 
    \nonumber \\&\qquad\quad b=\{f_i, g_i\}_{x} - \{f_i, g_i\}_{x - \delta x},
    \nonumber \\
    v_{i,x - \delta x/2} < \text{$0$:} \ \
      &\{f_i, g_i\}_{x - \delta x/2} = \{f_i, g_i\}_{x} + \frac{b}{2} \phi \left(a, b\right); 
    \nonumber \\&\qquad\quad a=\{f_i, g_i\}_{x} - \{f_i, g_i\}_{x + \delta x}, 
    \nonumber \\&\qquad\quad b= \{f_i, g_i\}_{x - \delta x} - \{f_i, g_i\}_{x},
    \nonumber \\
\end{align}
here denoted only for the discrete surface located at \mbox{$\sigma=x-\delta x/2$}, 
with
\begin{align}
    &\phi(a, b) = \max\left(0, \min\left[ \beta \frac{a}{b}, \frac{1}{2}\left(1 + \frac{a}{b}\right), \beta \right] \right);
    & &\beta \in [1, 2]
    ,
\end{align}
where a free parameter of $\beta = 1$ would lead to the most dissipation, reducing the limiter to a classical minmod.

\subsubsection{Simulation parameters}
In all simulations, unless otherwise stated, the parameters for the presented results were set to 
\mbox{$\mu = \num{5e-5}$} and \mbox{$\gamma=1.4$} with the gas constant set to $R=1$.
The cases were run to solve for the NSF solution, i.e. the requirements outlined in Section~\ref{sec:Requirements} concerning the Grad expansion and the quadrature order of the velocity set were respected such that the NSF equations are recovered in the hydrodynamic limit.
The free parameter in the flux limiter was set to $\beta = 1.2$.
\Acp{IC} were supplied by means of equilibrium populations.
1-D tests were run as pseudo 1-D with periodic \acp{BC} in the pseudo direction, otherwise von Neumann \acp{BC} were applied.
The time step was estimated with
{
\begin{equation}
    \delta t = \frac{\min(\delta x,\delta y)}{|{\bm{u}| + c_s}} \text{CFL}
    ,
\end{equation}
}
where the CFL number was uniformly set to $0.01$ in order to also stably simulate higher Mach numbers.
Physical quantities are reported as non-dimensional values whenever no units are mentioned.

\subsection{Conservation properties}

First, the conservation properties of the discrete kinetic models were addressed.
This was tested using a Sod shock tube~\cite{Sodtube} with the \acp{IC}
\begin{equation}
  (\rho, p, u_x) =
    \begin{cases}
      (1, 1, 0), & 0 \leq x \leq 0.5,\\
      (0.125, 0.1, 0), & 0.5 < x \leq 1,
    \end{cases}       
\end{equation}
in an extended fully periodic domain, $x \in [-0.5,1.5]$. 
A spatial resolution of $\delta x = L_x/1024$ was applied such that the relevant region in $x \in [0,1]$ contains $512$ cells.

The results for the cropped region $x \in [0, 1]$ at $t=0.2$s are depicted in Fig.~\ref{fig:ConservationSod1} (top row) together with the reference obtained from the Riemann solution of the inviscid problem.
The relative errors in conservation of mass and total energy in the total domain are shown in the bottom row.
All models (total and internal non-translational energy split for Prandtl number $\mathrm{Pr} = \{0.5,1,2\}$) capture the reference solution with good agreement.
The same conservation errors, which are in the order of the rounding machine precision ($2^{-53} \approx \num{1.11e-16}$ for double precision) on each cell interface, are achieved for all models, hence it is apparent that all constructed models are strictly conservative.
Further, it can be seen that the dissipation around the shocks slightly change for $\mathrm{Pr} = \{0.5,1,2\}$, however, importantly, the positions of all characteristic waves in this problem (shock, contact discontinuity and rarefraction wave) do not depend on the Prandtl number. 
This further confirms the models' correctness, as the position of these waves is determined by the Euler level solution, i.e. the dispersion of the hydrodynamic modes and the speed of sound, which are not affected by the dissipation of hydrodynamic modes such as the Prandtl-dependent thermal dissipation rate.
The dispersion and dissipation of hydrodynamic modes are therefore validated next for a variety of imposed parameters.

\subsection{Dispersion of hydrodynamic modes: Speed of sound\label{sec:speedSound}}

\begin{figure*}
    \centering
    \includegraphics[width=\linewidth]{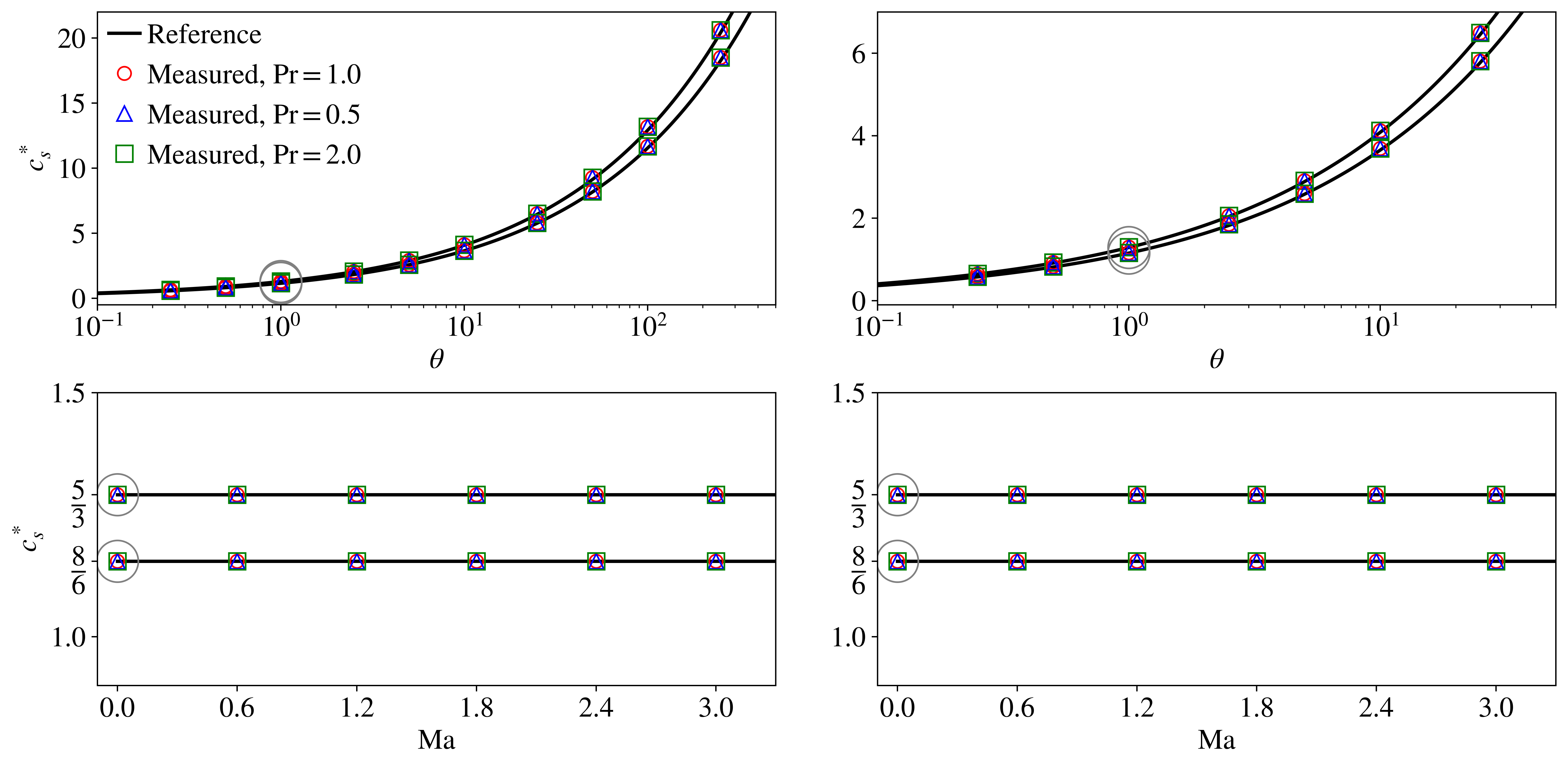}%
    \caption{
    Results of the dispersion tests with $\gamma = \{5/3, 8/6\}$ and $\mathrm{Pr} = \{0.5,1,2\}$ for the total energy split in the left column and internal-non-translational in the right column. 
    The top row depicts a comparison of the measured normalized speed of sound $c_s^* = \sqrt{\gamma R \theta}$  at various normalized temperatures $\theta = T/T_{ref}$ for $\mathrm{Ma} = 0$, i.e. $u_x = u_0 = 0$, whereas the bottom row shows $c_s^*$ at various Mach numbers of the mean flow at $\theta = 1$, i.e. $T = T_0 = T_{ref}$.
    The gray circles indicate corresponding measurement points between the two rows.
    }
    \label{fig:SpeedSound}%
\end{figure*}

The dispersion rates were probed to further assess the correct behavior concerning dispersion of hydrodynamic eigen-modes.
For this, a standard test was used as, e.g., given in~\cite{strässle2025a-fully-conservative, Ji24, Saadat2019, hosseini2020compressibility, ProbingDoubleDist2024}.
Note that all results reported here correspond to converged simulations in space and time. 
All setups were run in a fully periodic pseudo 1-D domain $x \in [0,1]$ with a resolution of $\delta x = L_x/512$.

The temperature dependence of the speed of sound was investigated by means of a freely traveling pressure front.
To that end, the domain was divided into two regions with
\begin{equation}
    p =
    \begin{cases}
        1 + A, & 0 \leq x \leq 0.5,\\
        1, & 0.5 < x \leq 1,
    \end{cases}       
\end{equation} 
with amplitude $A=\num{1e-6}$. A uniform temperature $T=T_0$ as well as velocity $u_x = u_0$ was applied in both regions.
The velocity was derived from the Mach number as $u_0 = \mathrm{Ma} \sqrt{\gamma R T_0}$ and varied in subsequent simulations.
Two different specific heat ratios, namely $\gamma = 5/3$ and $\gamma = 8/6$, were assessed for various temperatures.
The speed of sound was computed by tracking the shock front relative to the mean flow velocity $u_x = u_0$ over time and comparing it with the analytical value of $c_s = \sqrt{\gamma R T}$. 

Fig.~\ref{fig:SpeedSound} demonstrates that all constructed models for different specific heat ratios and Prandtl numbers can correctly capture the speed of sound over a wide temperature range spanning four orders of magnitude with the total energy split and three orders of magnitude with the internal non-translational split, respectively.
The same applies over a wide Mach number range up to $\mathrm{Ma} \approx 3.0$, 
for both energy splits, where deviations eventually start occurring due to spurious numerical oscillations.

The difference in the reported range of temperature is tied to the stability limits associated to the employed velocity sets. It is known that the D2Q16, which was employed with the total energy split, possesses an increased temperature range as compared to the D2Q25 velocity set, which was employed with the internal non-translational split.
These results confirm that the models were correctly constructed for what concerns the Euler level solutions and dispersion of eigen-modes.

\subsection{Dissipation of hydrodynamic modes: Shear, bulk and entropic modes}

\begin{figure*}
    \centering
    \includegraphics[width=\linewidth]{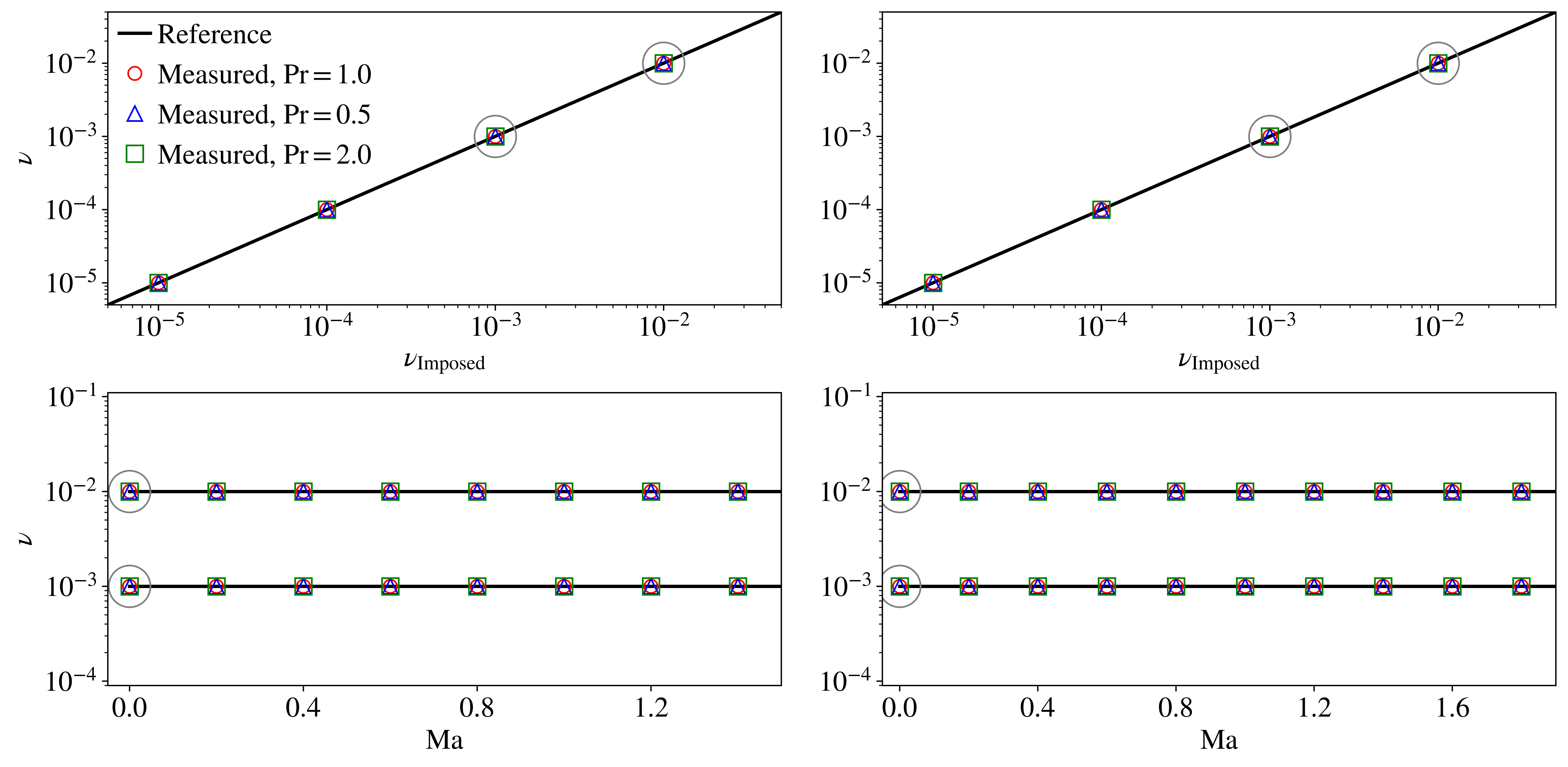}%
    \caption{
    Results of the dissipation tests for the  shear mode with $\mathrm{Pr} = \{0.5,1,2\}$ for the total energy split in the left column and internal-non-translational in the right column. The measured versus imposed values of kinematic shear viscosity 
    is depicted in the top row at $\mathrm{Ma} = 0$, whereas the bottom row shows the measured values at various Mach numbers. The gray circles indicate corresponding measurement points between the two rows.
    }
    \label{fig:Shear}%
\end{figure*}

\begin{figure*}
    \centering
    \includegraphics[width=\linewidth]{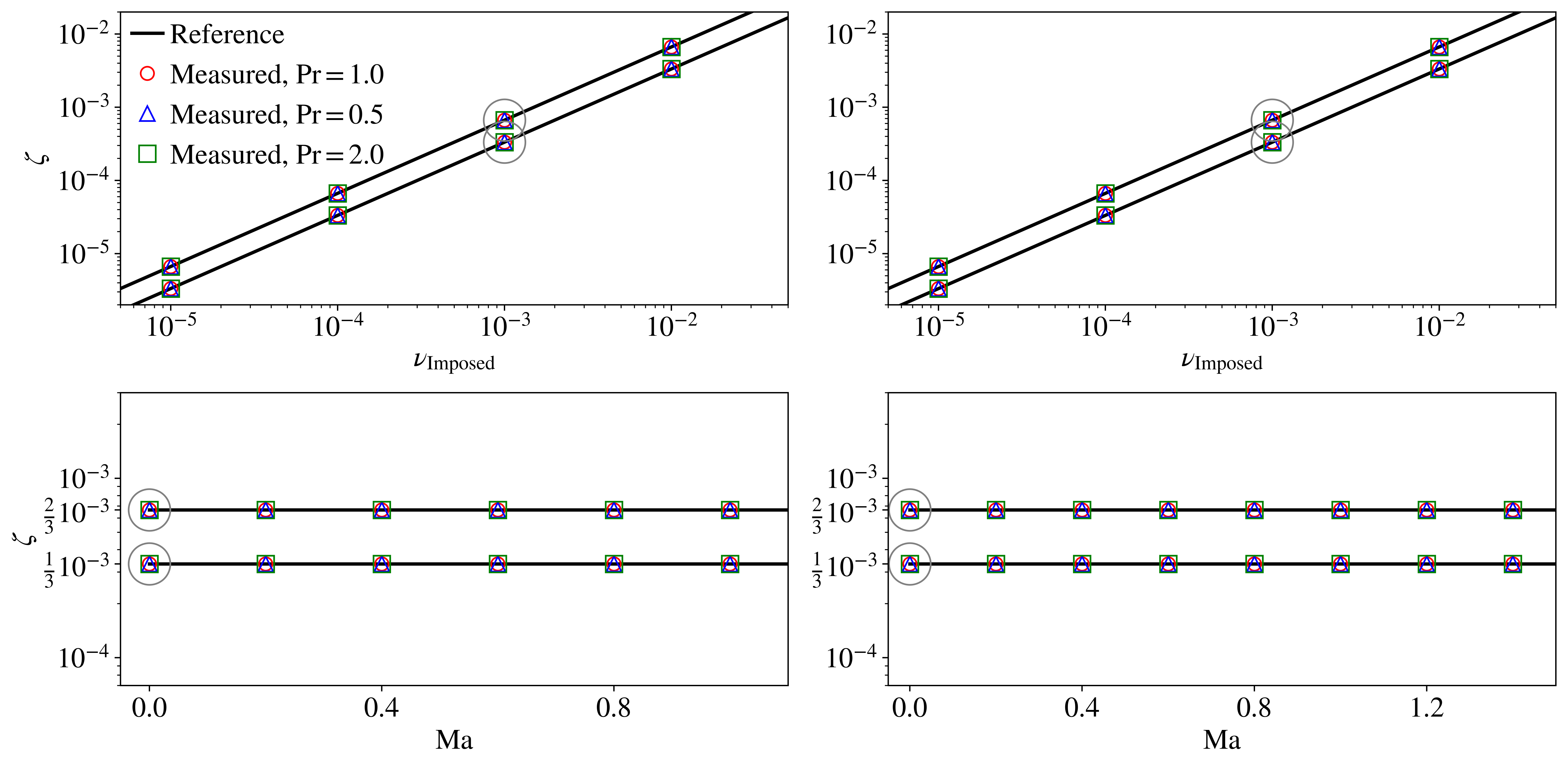}%
    \caption{
    Results of the dissipation tests for the 
bulk mode 
    with $\gamma = \{5/3, 8/6\}$ and $\mathrm{Pr} = \{0.5,1,2\}$ for the total energy split in the left column and internal-non-translational in the right column. 
    The measured versus imposed values of
kinematic bulk viscosity 
    is depicted in the top row at $\mathrm{Ma} = 0$, whereas the bottom row shows the measured values at various Mach numbers.
    The gray circles indicate corresponding measurement points between the two rows.
    }
    \label{fig:Bulk}%
\end{figure*}

To further assess the correct behavior concerning dissipation of hydrodynamic eigen-modes, the dissipation rates, i.e. shear, normal and entropic, were probed.
For this, standard tests were used as, e.g., given in~\cite{strässle2025a-fully-conservative, Ji24, Saadat2019, hosseini2020compressibility, ProbingDoubleDist2024}.
Note that all results reported here correspond to converged simulations in space and time. All setups were run in a fully periodic pseudo 1-D domain $x \in [0,1]$ with a resolution of $\delta x = L_x/512$.

\subsubsection{Shear viscosity}

\begin{figure*}
\centering
    \includegraphics[width=\linewidth]{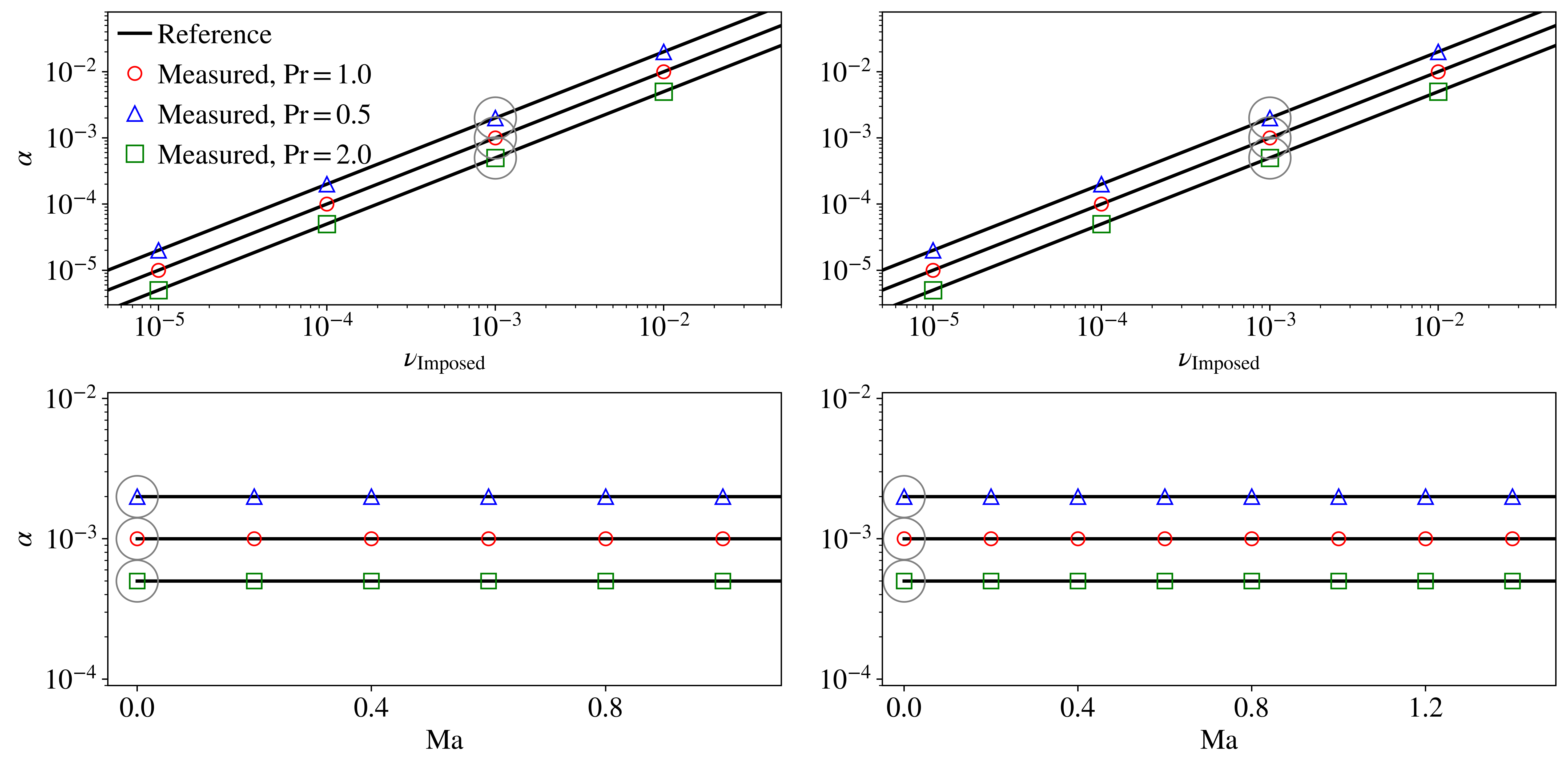}%
    \caption{
    Results of the dissipation tests for the 
entopic mode 
    with $\mathrm{Pr} = \{0.5,1,2\}$ for the total energy split in the left column and internal-non-translational in the right column. 
    The measured versus imposed values of
thermal diffusivity 
    is depicted in the top row at $\mathrm{Ma} = 0$, whereas the bottom row shows the measured values at various Mach numbers.
    The gray circles indicate corresponding measurement points between the two rows.
    }
    \label{fig:thermal}%
\end{figure*}

\begin{figure*}
    \centering
    \includegraphics[width=\linewidth]{Fig/ThermalCouette.png}%
    \caption{{
    Results of the thermal Couette problem with $\mathrm{Pr} = \{0.5,1,2\}$ for the total energy split in the left column and internal-non-translational in the right column. 
    The top row depicts a comparison with the reference solution at $\mathrm{Ma}=0.5$ and $\mathrm{Ec}=4$, whereas the bottom row shows $\mathrm{Ma}=1.2$ and $\mathrm{Ec}=40$.
    }}
    \label{fig:ThermalCouette}%
\end{figure*}

The kinematic shear viscosity $\nu$ was investigated by simulating a plane shear wave with a small sinusoidal perturbation superimposed to the initial velocity field.
The \acp{IC} read
\begin{equation}
    \rho = \rho_0, \hfill T = T_0, \hfill u_x = u_0, \hfill u_y = A \sin\left(2\pi x/L_x\right)
    ,
\end{equation}
where the initial density and temperature were set to $ \left( \rho _0,T_0 \right) = \left( 1, 1 \right)$. 
The perturbation amplitude was $A=\num{1e-6}$ and $u_0$, which is derived from the Mach number as $u_0 = \mathrm{Ma} \sqrt{\gamma R T_0}$, was varied in subsequent simulations. The evolution of the maximum velocity $u_y^{\rm max}$ in the domain was tracked over time and an exponential function was fitted to it.
The decay rate, i.e. the shear viscosity $\nu$, was then obtained via
\begin{equation}
    u_y^{\rm max}(t) \propto \exp{\left(-\frac{4\pi^2\nu}{L_x^2}t\right)}
    .
\end{equation}

The obtained results are depicted in Fig.~\ref{fig:Shear}. The measured viscosities are in excellent agreement with the imposed values and recover the expression from the Chapman-Enskog analysis, cf. Section~\ref{sec:CE}, for several $\mathrm{Ma}$ numbers.

\subsubsection{Bulk viscosity}

The kinematic bulk viscosity $\zeta$ was investigated via the decay rate of sound waves in the linear regime. For this purpose, a small perturbation with initial amplitude $A=\num{1e-6}$ was superimposed to the density field. The flow was initialized as
\begin{equation}
    \rho = \rho_0 + A \sin\left(2\pi x/L_x\right), T = T_0, u_x = u_0, u_y = 0,
\end{equation}
with ($\rho_0$, $T_0$) = ($1$, $1$) and $u_0$ derived from the imposed Mach number.  
The perturbation acoustic energy $E'(t) = u_x^2+u_y^2-u_0^2+c_s^2\rho '^2$ of the whole domain, with $\rho' = \rho - \rho_0$, was tracked over time and the exponential function, 
\begin{equation}
    E'(t) \propto \exp{\left(-\frac{4\pi^2\nu_e}{L_x^2}t\right)},
\end{equation}
as defined by~\cite{dellar2001bulk}, was fitted to it.
The recovered decay rate is the effective viscosity $\nu_e$, i.e. the combination of shear and bulk viscosities as
\begin{equation}
    \nu_e = \frac{4}{3}\nu + \zeta 
    .
\end{equation}

The obtained results are depicted in Fig.~\ref{fig:Bulk} for several imposed viscosities and specific heat ratios at various $\mathrm{Ma}$ numbers.
It can be seen that the measured bulk viscosities accurately recover the expression from the Chapman-Enskog analysis, cf. Section~\ref{sec:CE}, for all models.

\subsubsection{Thermal diffusivity}

A different type of perturbation was introduced in the ICs of the system to assess the thermal diffusivity $\alpha$. These are
\begin{equation}
    \rho = \rho_0 + A \sin\left(2\pi x/L_x\right), T = \rho_0 T_0/\rho, u_x = u_0, u_y = 0,
\end{equation}
with $ \left( \rho _0,T_0 \right) = \left( 1, 1 \right) $ and a perturbation amplitude of $A=\num{1e-6}$.
The thermal diffusivity was measured by fitting the exponential function,
\begin{equation}
    T'(t) \propto \exp{\left(-\frac{4\pi^2\alpha}{L_x^2}t\right)},
\end{equation}
to the temporal evolution of the maximum temperature difference $T' = T - T_0 $ in the domain.

From Fig.~\ref{fig:thermal} it becomes evident that the models also perform well in terms of thermal dissipation rates, importantly for all the different Prandtl numbers, as the exact expression, cf. Section~\ref{sec:CE}, is accurately recovered.

To summarize the insights from all dissipation tests:
All dissipation rates are correctly recovered for a wide range of imposed parameters.
The difference in the reported range of Mach numbers is tied to the stability limits associated to the employed velocity sets, and follows the same arguments as in~\ref{sec:speedSound}.
It is known that the D2Q25 which was employed with the internal non-translational energy split, possesses an increased Mach number range as compared to the D2Q16 velocity set, which was employed with the total energy split.
These results confirm that the models were correctly constructed for what concerns the Navier--Stokes--Fourier level solutions and dissipation of eigen-modes.

\subsection{{Thermal Couette flow}}

\begin{figure}[b]
    \centering
    \includegraphics[width=0.4\linewidth]{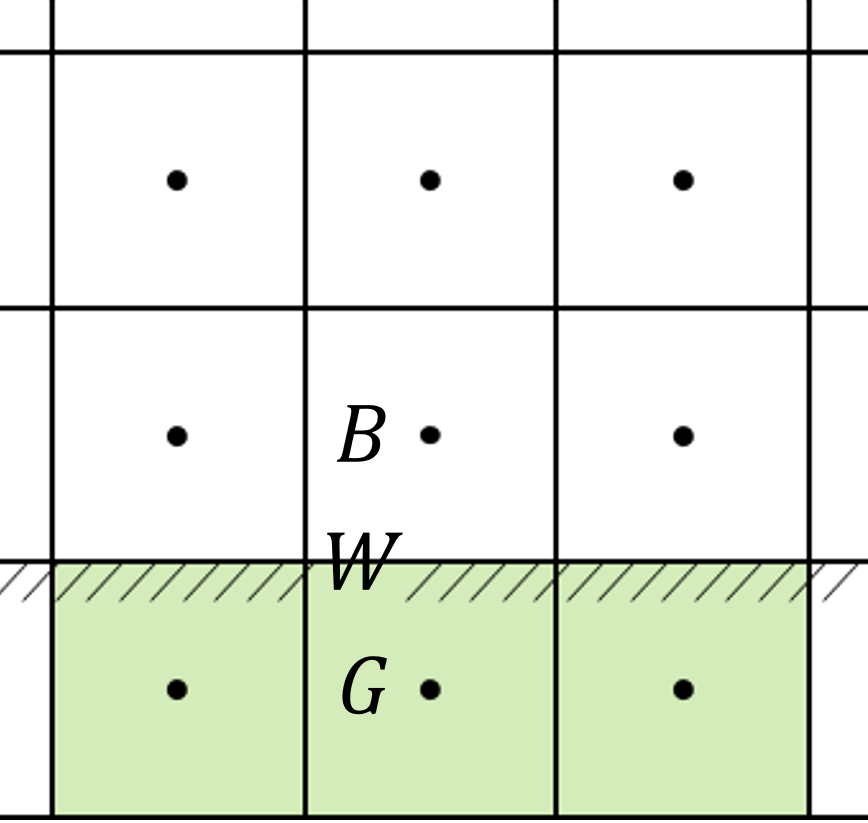}%
    \caption{{
    Schematic of the boundaries for implementation of the boundary conditions.
    }}
    \label{fig:BCinThermalCouette}%
\end{figure}

\begin{figure*}
    \centering
    \includegraphics[width=0.33\linewidth, height=0.33\linewidth]{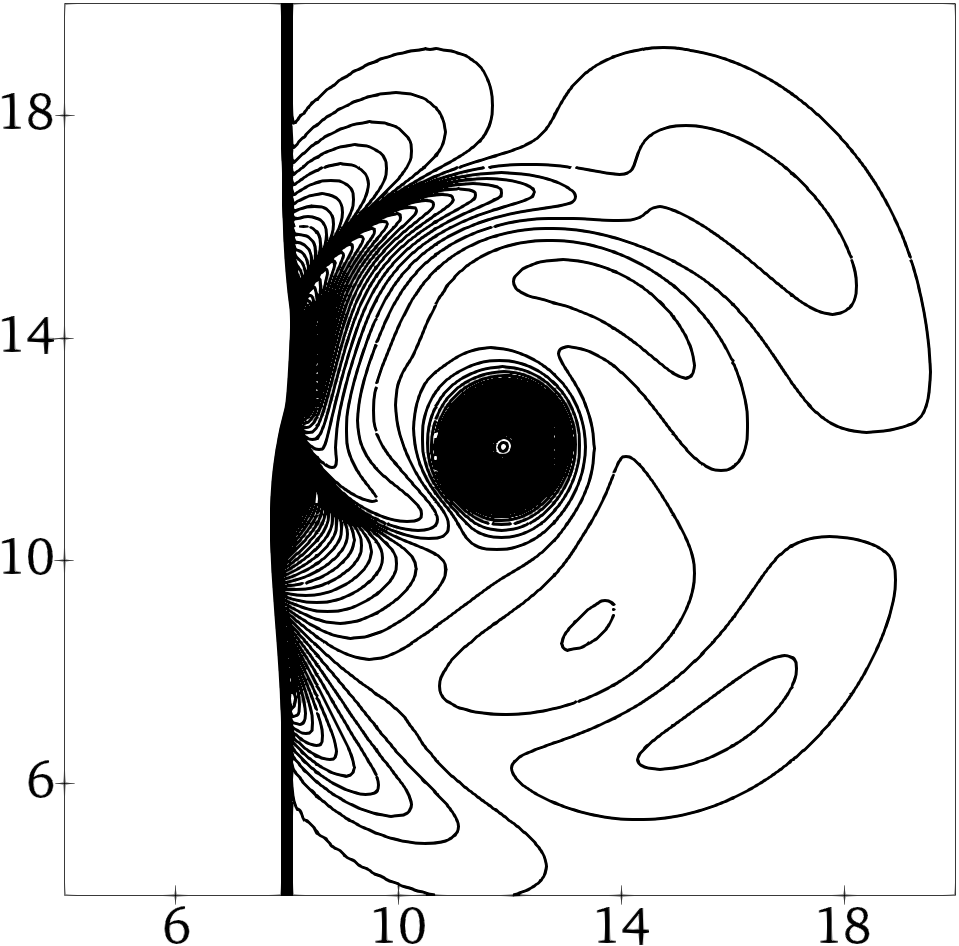}%
    \includegraphics[width=0.33\linewidth, height=0.33\linewidth]{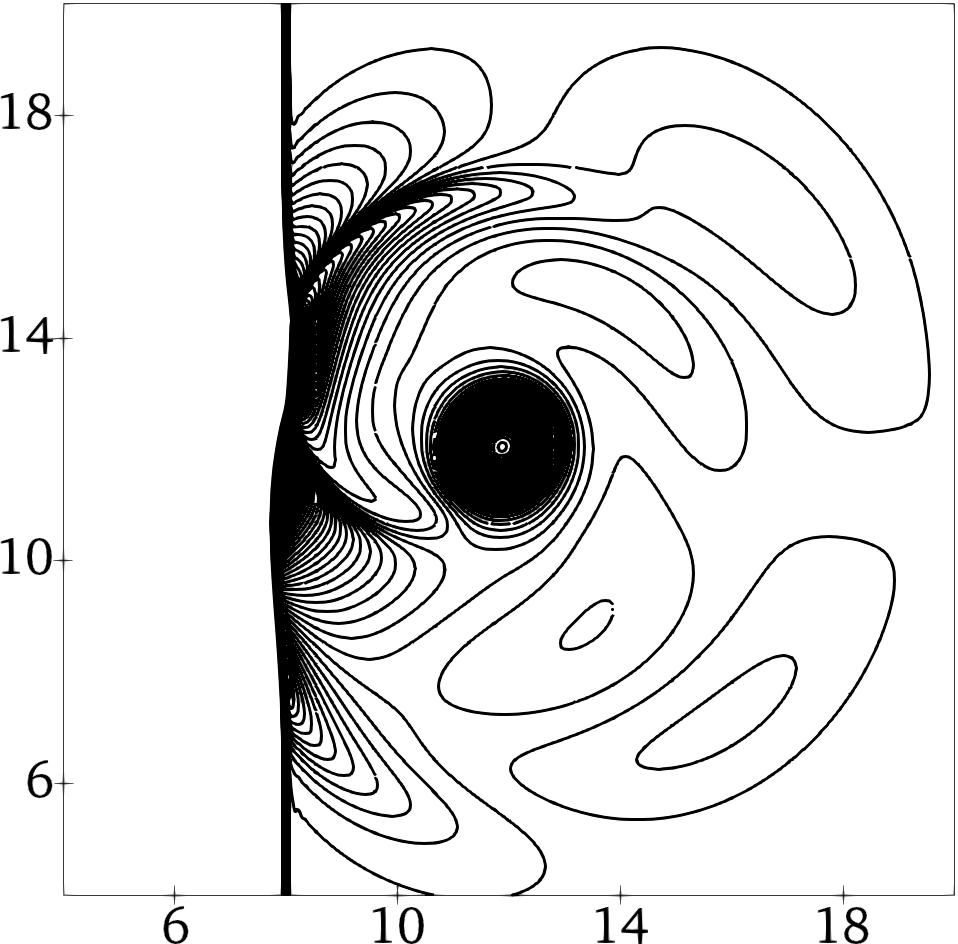}%
    \includegraphics[width=0.33\linewidth, height=0.33\linewidth]{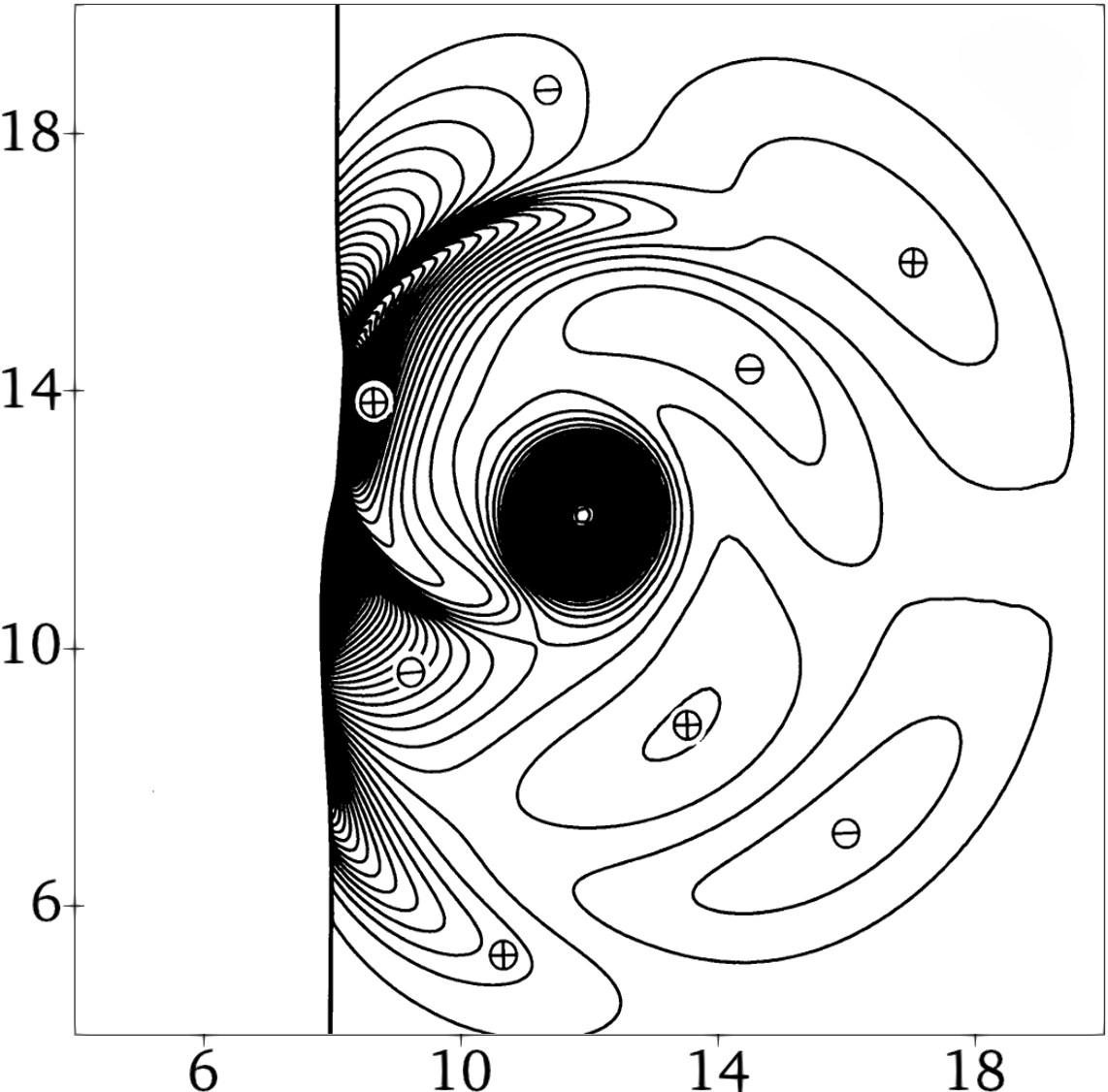}%
    \\
    \includegraphics[width=0.33\linewidth, height=0.33\linewidth]{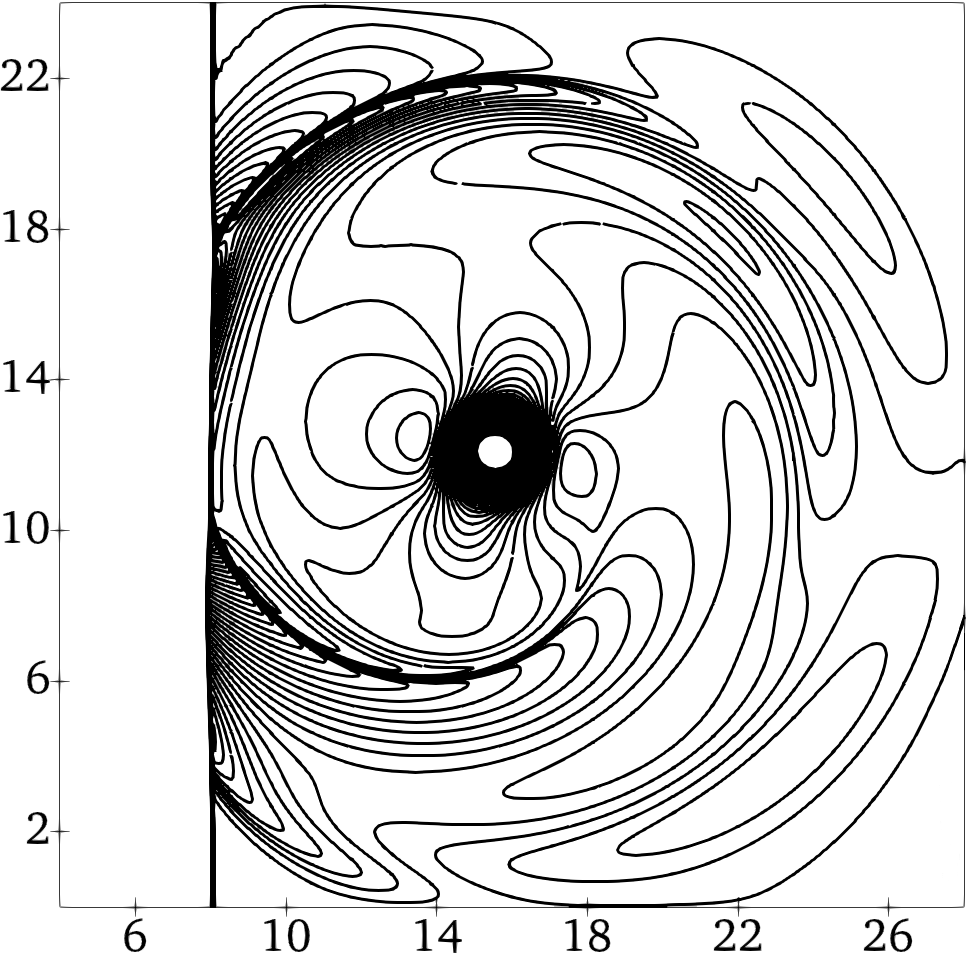}%
    \includegraphics[width=0.33\linewidth, height=0.33\linewidth]{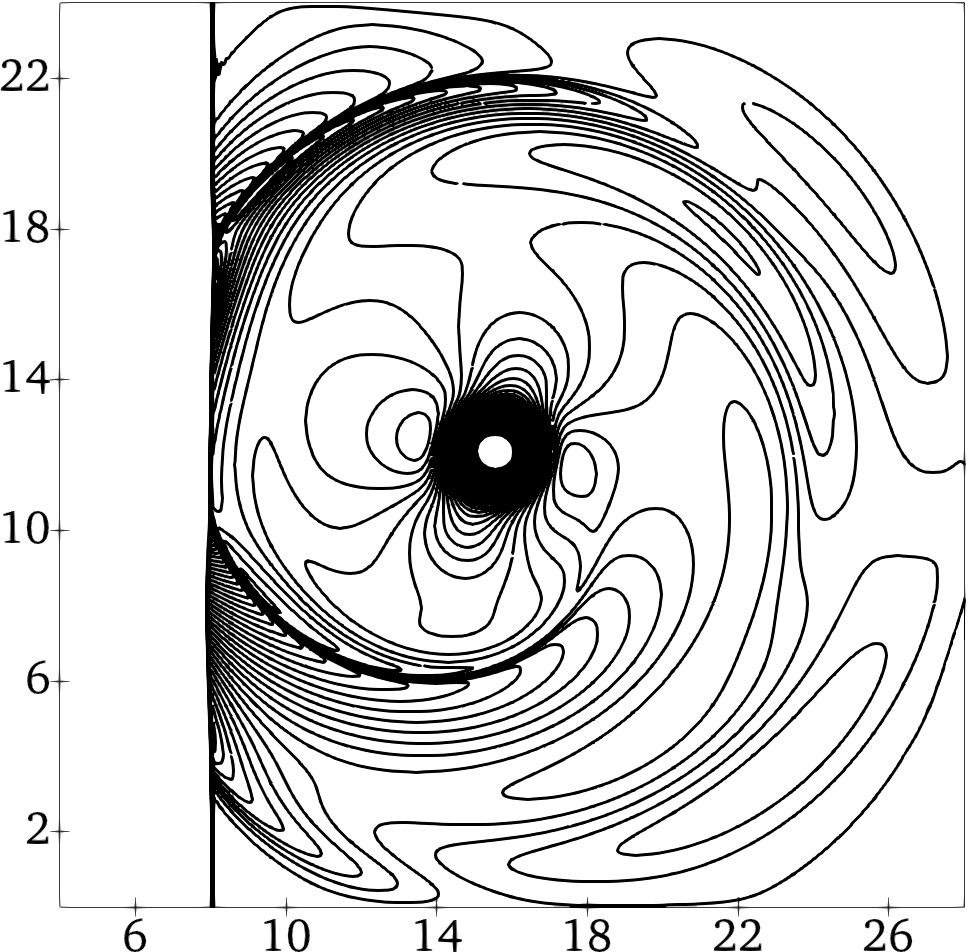}%
    \includegraphics[width=0.33\linewidth, height=0.33\linewidth]{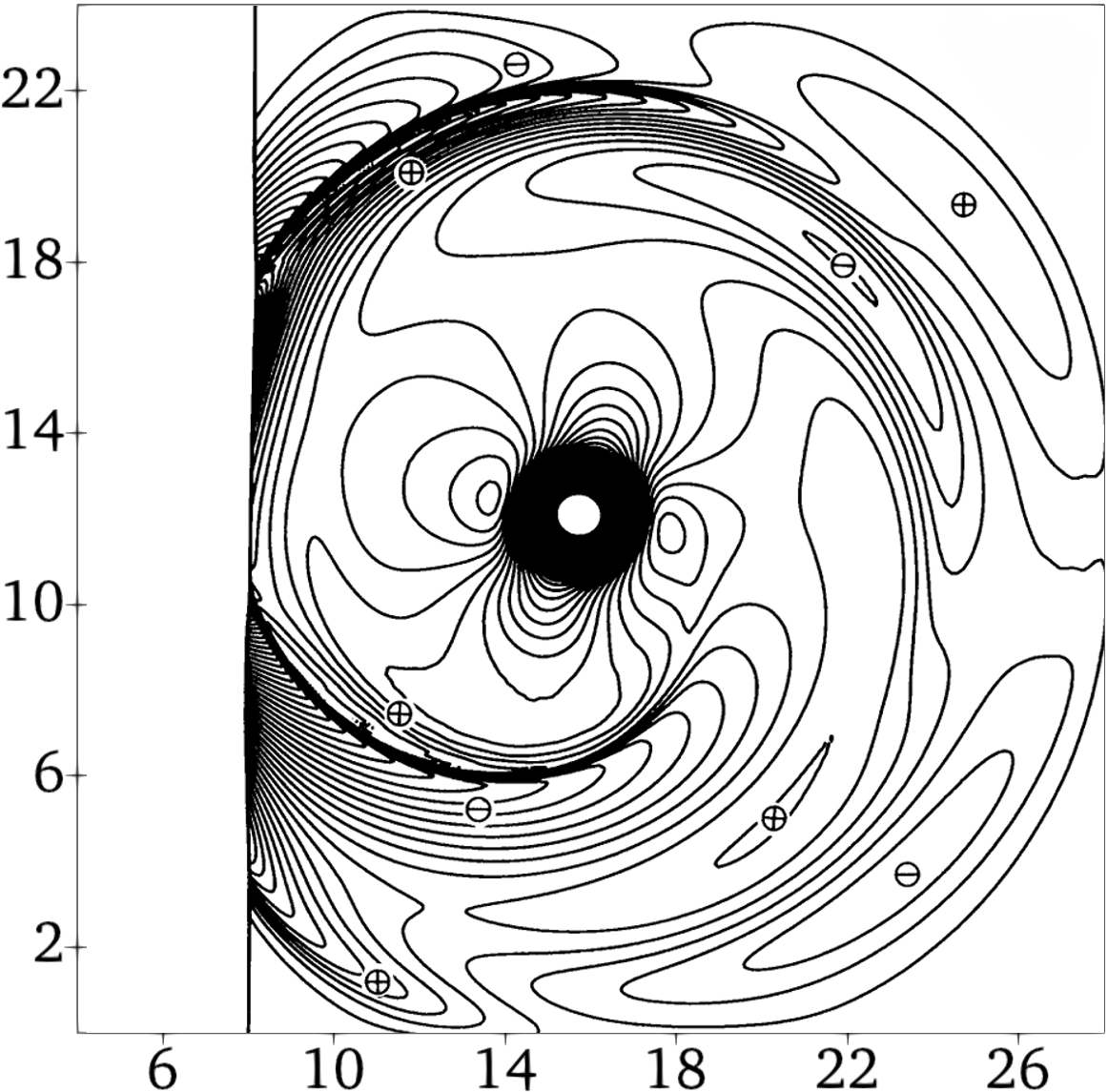}%
    \caption{
    Results of the shock-vortex interaction.
    The left and middle column show the model with total and internal-non-translational energy split, whereas
    Case C ($\mathrm{Ma}_s = 1.2$ and  $\mathrm{Ma}_v = 0.25$) at $t^*=6$ 
    and
    case G ($\mathrm{Ma}_s = 1.29$ and  $\mathrm{Ma}_v = 0.39$) at $t^*=10.3$ 
    are depicted with $140$ and $90$ equidistant contours of the sound pressure $p_{\rm Sound}$ in the top and bottom row, respectively.
    The reference solution for both cases C and G from Inoue \& Hattori~\cite{inoue1999sound} are displayed in the right column. 
    }
    \label{fig:SVcotours}%
\end{figure*}

{
The thermal Couette flow is the next benchmark considered here, to assess the proper recovery of the combined effect of viscous heating and thermal conductivity with the respective Prandtl number.
In this test case, an upper no slip wall with the higher temperature $T_{W} = T_H$ is in motion with a constant speed in $x$ direction $u_{x,W} = u_0$, while the lower no slip wall is at rest, $u_{x,W} = 0$, with a temperature $T_{W} = T_C$.
The parameters for the simulations were set to a Mach number of $\mathrm{Ma}  = u_0/ \sqrt{\gamma R T_C} = 0.5$ and $1.2$, as well as an 
Eckert number of $\mathrm{Ec} = u_0^2/(C_p (T_H - T_C))=4$ and $40$.
Furthermore, the Reynolds number was $\mathrm{Re} = \rho u_o L_y / \mu = 100$, 
the lower temperature $T_C = T_{ref}$ 
and the resolution $\delta y = L_y/128$ in $y \in [0,1]$.
The simulations were initialized with linear temperature profiles in temperature and velocity as   
\begin{equation}
  (u_x, T, \rho) = (u_0\frac{y}{L_y},  T_C + (T_H - T_C)\frac{y}{L_y}, \frac{1}{R T_C}), \quad 0 \leq y \leq L_y,
\end{equation}
and were run until the solution converged.

While periodic BCs were enforced in the horizontal direction, the BC at the no slip walls were implemented via the ghost node approach \cite{Kallikounis23, Tiwari_2012_GhostNodeBC} in combination with non-equilibrium extrapolation approach \cite{Bhaduria23, Guo_2002_BCextrapolation} by splitting the boundary populations into an equilibrium and non-equilibrium part (see also \cite{KarlinTwoPop, Saadat2019, ProbingDoubleDist2024} for other valuable approaches in combination with the thermal Couette case).
The approach consists of the following steps in accordance with the schematic given in Fig. \ref{fig:BCinThermalCouette} for the nomenclature.
First, the density, flow velocity and temperature are determined in the ghost cell 
With the fixed values at the wall and the known values at the boundary nodes, they read
\begin{align}
    \rho_G &= \rho_B
    \\
    \bm{u}_G &= 2 \bm{u}_W - \bm{u}_B,
    \\
    T_G &= 2 T_W - T_B.
\end{align}
Thereafter, the equilibrium is computed and the non-equilibrium part extrapolated from the neighboring nodes, yielding to the expressions for both $f_i$ and $g_i$ populations as
\begin{align}
    f_i^{\rm G} &= f_i^{\rm eq}(\rho_{G}, \bm{u}_{G}, T_{G}) + f_i^{\rm neq}(\rho_{B}, \bm{u}_{B}, T_{B})\Big|_{G},
    \\
    g_i^{\rm G} &= g_i^{\rm eq}(\rho_{G}, \bm{u}_{G}, T_{G}) + g_i^{\rm neq}(\rho_{B}, \bm{u}_{B}, T_{B})\Big|_{G},
\end{align}
}

{
Fig. \ref{fig:ThermalCouette} depicts the non-dimensional temperature profiles for three different Prandtl numbers at $\mathrm{Ma}=0.5$, $\mathrm{Ec}=4$ and the more pronounced case of $\mathrm{Ma}=1.2$, $\mathrm{Ec}=40$.
The analytical solution for the temperature can be written as \cite{Roshko1957ElementsOG}
\begin{equation}
    \frac{T-T_C}{T_H-T_C} = \frac{y}{L_y} + \frac{\mathrm{Pr} \ \mathrm{Ec}}{2} \frac{y}{L_y} \left(1-\frac{y}{L_y}\right).
\end{equation} 
It can clearly be seen that the results are in excellent agreement with the reference solutions. 
}

\subsection{Shock-Vortex interaction}

Lastly, a sensitive benchmark problem for viscous compressible flows, namely a shock–vortex interaction, was assessed. 
There, proper recovery of the dissipation rates for non-unity Prandtl numbers together with high resolution is crucial.
The setup of Inoue \& Hattori~\cite{inoue1999sound} as adopted in~\cite{saadat2021extended, strässle2025a-fully-conservative} was followed here, where the main field is separated by a stationary shock with Mach number  $\mathrm{Ma}_s$ and the left- and right-hand initial states satisfy the Rankine-–Hugoniot jump conditions.
For a pre-shock state of $(\rho, p, u_x=\mathrm{Ma}_s c_{s}, u_y=0)_l$ on the left-hand side, where the flow velocity is given from the imposed $\mathrm{Ma}$ number via $c_{s,l} = \sqrt{\gamma R T_l} = \sqrt{\gamma p_l/\rho_l}$, the post-shock state $(\rho, p, u_x, u_y=0)_r$ is found as
\begin{align}
    \rho_r  &= \rho_l \frac{(\gamma+1)\mathrm{Ma}_s^2}{(\gamma-1)\mathrm{Ma}_s^2+2},\\
    p_r     &= p_l \frac{2\gamma \mathrm{Ma}_s^2-(\gamma-1)}{(\gamma+1)},\\ 
    u_{x,r} &= u_{x,l} \frac{(\gamma-1)\mathrm{Ma}_s^2+2}{(\gamma+1)\mathrm{Ma}_s^2}. 
\end{align}
The resulting initial field $(\rho, p, u_x, u_y)_\infty$ is perturbed by an isentropic vortex which is advected through the shock. 
The maximum tangential velocity of the vortex defines the vortex Mach number as $\mathrm{Ma}_v = u_{\varphi}^{\rm max} / c_{s,l}$.
In Cartesian coordinates, the \acp{IC} for the vortex read
\begin{align}
    u_x &= u_{x,\infty} + c_{s,l} \mathrm{Ma}_v \frac{y-y_v}{r_v}  e^{(1-r^2)/2},\\
    u_y &= u_{y,\infty} - c_{s,l} \mathrm{Ma}_v \frac{x-x_v}{r_v}  e^{(1-r^2)/2},\\
    \rho &= \rho_{\infty} \left[ 1 - \frac{\gamma -1}{2} \mathrm{Ma}_v^2  e^{(1-r^2)} \right]^{1/(\gamma -1)}, \\
    p &= p_{\infty} \left[ 1 - \frac{\gamma -1}{2} \mathrm{Ma}_v^2  e^{(1-r^2)} \right]^{\gamma/(\gamma -1)}.
\end{align}
Note that the field is perturbed on both sides of the shock to match the reference solution in the DNS setup~\cite{inoue1999sound}, where the influenced region of the vortex in the IC overlaps the shock slightly.
The reduced radius $r$ is defined with the vortex center position ($x_v$, $y_v$) and the vortex radius $r_v$ as $r = \sqrt{(x-x_v)^2+(y-y_v)^2}/r_v$, where the vortex radius is connected to the dynamic viscosity of the fluid via the Reynolds number defined as $\mathrm{Re}_v = \rho_l c_{s,l} r_v / \mu$.

The shock position was initialized at $x_s$ = $8$ with $\rho_l = 1$ and $p_l=1$ in a domain $x \in [0, 28]$ and $y \in [0, 24]$.
The vortex with $r_v = 1$, rotating in clock-wise direction, was centered at \mbox{($x_v$, $y_v$)} \mbox{$= (6, 12)$}.
The Reynolds number was set to $\mathrm{Re}_v = 800$ and the Prandtl number to $\mathrm{Pr} = 0.75$.
As in the original setup, periodic BCs were used for the boundaries in y-direction.
Cases C and G from~\cite{inoue1999sound} were run.
For case C,  $\mathrm{Ma}_s = 1.2$ and  $\mathrm{Ma}_v = 0.25$ was imposed, 
whereas for case G, these numbers were set to  $\mathrm{Ma}_s = 1.29$ and  $\mathrm{Ma}_v = 0.39$.
A resolution of $\delta x = \delta y = L_x/1680 = L_y/1440$ was applied.

Fig.~\ref{fig:SVcotours} depicts the sound pressure contours at the non-dimensional time $t^* = 6$ and $t^* = 10.3$ for case C and G compared to the reference solution.
Thereby, the sound pressure is defined as $p_{\rm Sound} = p/p_r -1$, with $p_r$ being the initial pressure in the post-shock region, and the non-dimensional time is given by $t^* = t c_{s,l} / r_v$.
Note that the sound pressure usually amounts to a small perturbation in the order of $\lesssim 1\%$ of the hydrodynamic pressure on top of it and is therefore a rather sensitive quantity.
Excellent agreement of the pressure contours with the reference DNS solution of~\cite{inoue1999sound} can be observed. Importantly, the deformation of the shock, including the two shock reflections, is also well captured.

\begin{figure}[b]
    \centering
    \includegraphics[width=1.0\linewidth]{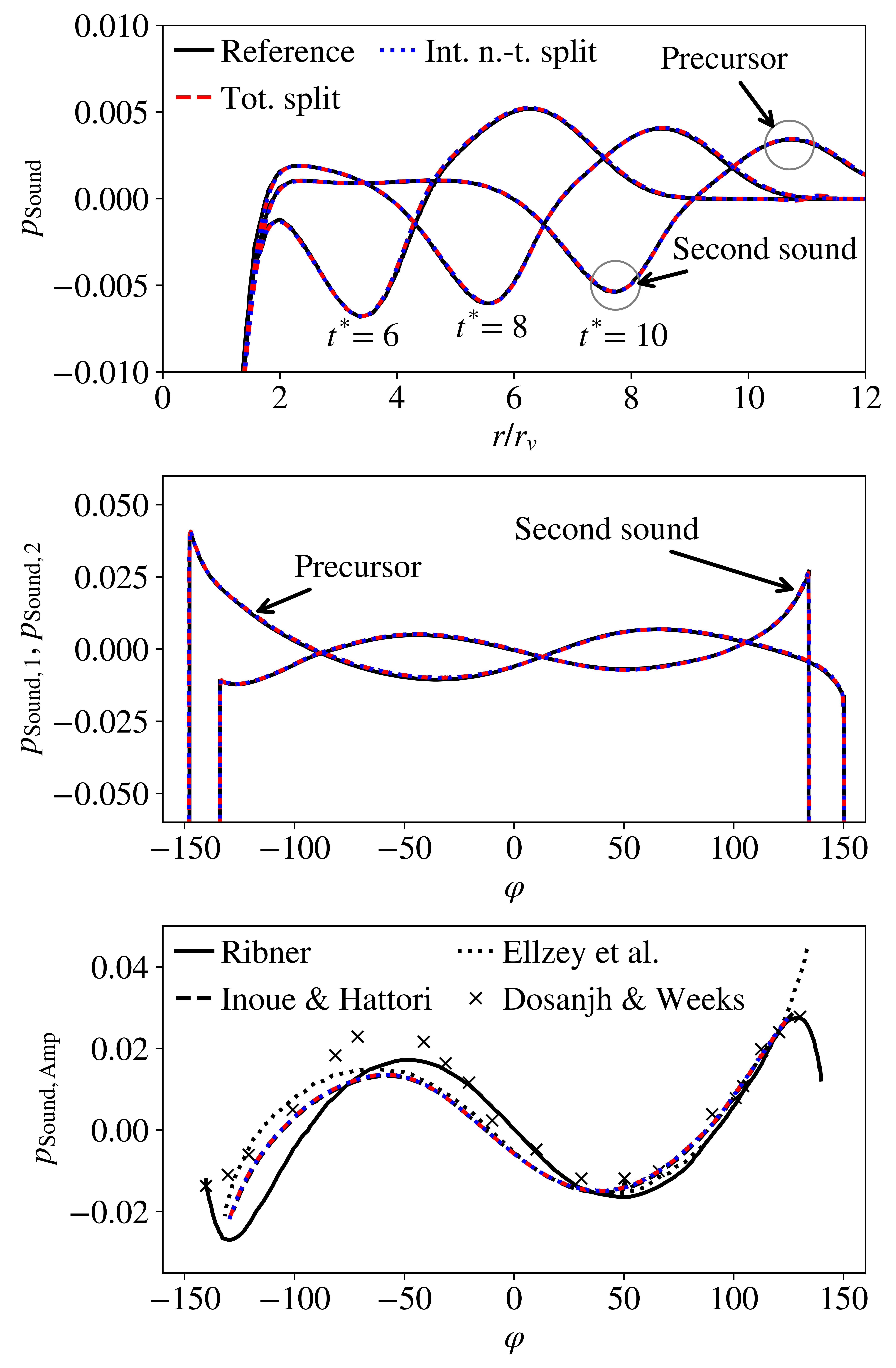}%
    \caption{
    Quantitative results of the shock-vortex interactions:
    (Top) Comparison of the radial sound pressure distribution of case C ($\mathrm{Ma}_s = 1.2$ and  $\mathrm{Ma}_v = 0.25$) at three different non-dimensional times $t^*=6, 8, 10$ with the reference solution from Inoue \& Hattori~\cite{inoue1999sound}. 
    The gray circles indicate the peak sound pressure of the precursor, $p_{\rm Sound,1}$, and the second sound, $p_{\rm Sound,2}$.
    (Middle) Comparison of the circumferential distribution of the peak sound pressure of the precursor, $p_{\rm Sound,1}$, and the second sound, $p_{\rm Sound,2}$, for case G ($\mathrm{Ma}_s = 1.29$ and  $\mathrm{Ma}_v = 0.39$) at non-dimensional time $t^*=10.3$ with the reference solution from Inoue \& Hattori~\cite{inoue1999sound}.
    (Bottom) Comparison of the circumferential distribution of the sound pressure amplitude, $p_{\rm Sound, amp}$, of case G ($\mathrm{Ma}_s = 1.29$ and  $\mathrm{Ma}_v = 0.39$) with
    theoretical results of Ribner~\cite{Ribner1985}, 
    viscous DNS simulations of Inoue \& Hattori~\cite{inoue1999sound}, 
    inviscid simulations by Ellzey et al.~\cite{Ellzey1995}, 
    and experimental results by Dosanjh \& Weeks~\cite{DOSANJH1965}.
    }
    \label{fig:SVquatitative}%
\end{figure}

For further quantitative comparison of case C, the radial distribution of the sound pressure was measured in the direction of $\varphi=45^\circ$ degrees with the origin at the vortex center.
The results are depicted in Fig.~\ref{fig:SVquatitative}~(top) for three different non-dimensional times $t^*$.
It can clearly be seen that the reference solutions are perfectly matched.
In particular, the temporal development of the position and magnitude of the peak sound pressure of the first sound (precursor) and the second sound emerging from the shock-vortex interaction are well captured.

For quantitative comparison of case G, the sound pressure amplitude $p_{\rm Sound, amp}$ is compared to inviscid and viscous DNS simulations, experimental results as well as theory.
Thereby, the sound pressure amplitude is computed as $p_{\rm Sound, Amp} = (p_2 - p_1)/p_r$, i.e. the difference between the peak sound pressure of the precursor, denoted as $p_{\rm Sound,1} = p_1/p_r - 1$, and the second sound, $p_{\rm Sound,2} = p_2/p_r - 1$, where $p_1$ and $p_2$ are measured equivalently to~\cite{inoue1999sound} at $r/r_v = 10.8$ and $r/r_v = 8.8$ from the vortex center, respectively. 
The results for the circumferential distribution of $p_{\rm Sound,1}$ and $p_{\rm Sound,2}$ are depicted against the reference solutions~\cite{inoue1999sound} as an intermediate result in Fig.~\ref{fig:SVquatitative}~(middle), and the results for the circumferential distribution of $p_{\rm Sound,Amp}$ are shown in Fig.~\ref{fig:SVquatitative}~(bottom). 
It can be seen that the reference solution of the viscous simulation~\cite{inoue1999sound} is perfectly matched for both models for non-unity Prandtl number with the total versus internal non-translational energy split, respectively. 
The results lie well within the knowledge band acquired by former viscous and inviscid simulations~\cite{inoue1999sound, Ellzey1995}, experimental measurements~\cite{DOSANJH1965} conducted at a much higher Reynolds number of Re $\approx \num{1.6e5}$, and theoretical results~\cite{Ribner1985}.

These insights further validate the presented models and demonstrate an excellent performance in obtaining sensitive quantities at the Navier--Stokes--Fourier level for complex setups in compressible and moderately supersonic flows with variable Prandtl numbers.

\section{Summary, conclusions and outlook\label{sec:summary}}

In this work, a consistent kinetic modeling and discretization approach for compressible flow simulation across arbitrary Prandtl numbers and specific heat ratios is discussed. Using the quasi-equilibrium method within two double-distribution function approaches, the models recover full Navier-–Stokes–-Fourier dynamics, including all correct macroscopic moments and dissipation rates. 
The construction of the quasi-equilibria is demonstrated through rigorous hydrodynamic analysis. Higher-order static velocity sets in the context of a discrete velocity Boltzmann method enable an accurate discrete representation without the necessity for correction terms. 

The models were validated against sensitive benchmarks, demonstrating physical fidelity, strict conservation, stability, and Galilean invariance across demanding flow conditions. 
Correct dispersion and dissipation behavior was recovered across a temperature range covering orders of magnitude and a broad range of Mach numbers.
Due to the application of the D2Q16 velocity set, the total energy split exhibited a superior temperature range over the internal non-translational split, which performed over a wider Mach number range due to the requirement of employing the D2Q25.
Lastly, a successful and very accurate reproduction of {the solutions of a thermal Couette flow and} a viscous shock–vortex interaction confirmed the models’ practical viability.
These insights showcase that the presented models offer an efficient and scalable framework for simulating compressible fluid dynamics with variable Prandtl numbers, while ensuring thermodynamic consistency, strict conservation of key physical quantities, and accurate reproduction of Navier-–Stokes-–Fourier-level behavior. 

Although restricted to higher-order lattices in this manuscript, future work shall also focus on the application of correction terms to accommodate standard lattices for usage with the lattice Boltzmann method, such as in~\cite{saadat2021extended}, as well as the extension with shifted, scaled, and adaptive reference frames, such as in the particles on demand method~\cite{PonD18}, in order to simulate high-Mach, strong discontinuities and hypersonic regimes.
Furthermore, application of space-time adaptive conservative refinement methods, such as in~\cite{strässle2025a-fully-conservative}, marks a promising avenue to increase computational efficiency.


\section*{Acknowledgments}
This work was supported by the European Research Council (ERC) Advanced Grant No. 834763-PonD and by the Swiss National Science Foundation (SNSF) Grant Nos. 200021-228065 and 200021-236715.
Computational resources at the Swiss National Super Computing Center (CSCS) were provided under Grant Nos. s1286, sm101 and s1327.
Open access funding was provided by the Swiss Federal Institute of Technology Zürich (ETH Zürich).

\section*{Author declarations}

\noindent\textbf{\textit{Conflict of interest}}\\
The authors have no conflicts to disclose.

\noindent\textbf{\textit{Ethical approval}}\\
The work presented here by the authors did not require ethics approval or consent to participate.

\noindent\textbf{\textit{Author contributions}}\\
R.M.S.:
Conceptualization of the study, 
formal analysis,
derivation and development of the methodology, 
implementation of the solver, 
evaluation and data analysis, 
writing- initial manuscript and revised versions. 
S.A.H.:
Conceptualization of the study, 
writing- initial manuscript and revised versions. 
I.V.K.:
Conceptualization of the study, 
writing- initial manuscript and revised versions,
funding acquisition, resources.
All authors approved the final manuscript.

\noindent\textbf{\textit{Data availability statement}}\\
The data that support the findings of this study are available within the article or from the corresponding author(s) upon reasonable request.

\noindent\textbf{\textit{Open access}}\\
This article is licensed under a Creative Commons Attribution 4.0 International License, which permits use, sharing, adaptation, distribution and reproduction in any medium or format, as long as you give appropriate credit to the original author(s) and the source, provide a link to the Creative Commons license, and indicate if changes were made. The images or other third party material in this article are included in the article’s Creative Commons license, unless indicated otherwise in a credit line to the material. If material is not included in the article’s Creative Commons license and your intended use is not permitted by statutory regulation or exceeds the permitted use, you will need to obtain permission directly from the copyright holder. To view a copy of this license, visit http://creativecommons.org/licenses/by/4.0/.

\appendix

\section{Relevant equilibrium and conserved moments\label{appendix:EqMoments}}

The moments of the Maxwell--Boltzmann distribution relevant to the models outlined in this manuscript are found as follows, denoted in index notation with summation convention.
The conserved moments read
\begin{align}
    & \phantom{(\mathrm{II})} \quad
    \int \{ f^{\rm eq}, f, f^{*} \} d\bm{v} 
    = \rho 
    ,
\\
    & \phantom{(\mathrm{II})} \quad
     \int v_{\alpha} \{ f^{\rm eq}, f, f^{*} \}  d\bm{v}
    = \rho u_\alpha
    ,
\\
    &\left.
    \begin{aligned}
        &\textcolor{black}{(\mathrm{I})}\phantom{\mathrm{I}} \quad
        \int  \{ g^{\rm eq}, g, g^{*} \} d\bm{v}
        \\
        &\textcolor{black}{(\mathrm{II})} \quad
        \int  \{ g^{\rm eq}, g, g^{*} \} +\frac{v^2}{2} \{ f^{\rm eq}, f^{*}, f \} d\bm{v} 
    \end{aligned} 
    \right\}
    \nonumber \\
    &\qquad \qquad \qquad \qquad \qquad \quad 
    = \rho \left(C_v T + \frac{u_\gamma u_\gamma}{2}\right)
    = E
    ,\label{eq:Relevantmoments_E}
\end{align}
the higher-order moments of the $f^{\rm eq}$-distribution read,
\begin{align}    \label{eq:f-equilibria-moments-MB}
    &\int v_{\alpha} v_{\beta} f^{\rm eq} d\bm{v}
     = \rho u_\alpha u_\beta + \rho R T \delta_{\alpha \beta} 
     ,
\\
    & \int v_{\alpha} v_{\beta} v_{\gamma} f^{\rm eq} d\bm{v}
    = \rho  u_\alpha u_\beta u_\gamma + \rho R T \left[u_\alpha \delta_{\beta \gamma}\right]_{cyc} 
    ,
\\
    &\begin{aligned}[b]
    \int v_{\alpha} v_{\beta} &v_{\gamma} v_{i\delta} f^{\rm eq} d\bm{v}
    = \ \rho  u_\alpha u_\beta u_\gamma u_\delta 
    \\
    &+ \rho R T \left[u_\alpha u_\beta \delta_{\gamma \delta}\right]_{cyc}
    + \rho (RT)^2 [\delta_{\alpha\beta}\delta_{\gamma\delta}]_{cyc}
    ,
    \end{aligned}
\end{align}
and the higher-order moments involving the $g^{\rm eq}$-distribution read,
\begin{align}
    &\left.
    \begin{aligned}[c]
        & \textcolor{black}{(\mathrm{I})}\phantom{\mathrm{I}} \quad
        \int v_{\alpha} g^{\rm eq} d\bm{v} 
    \\    
        & \textcolor{black}{(\mathrm{II})} \quad
        \int v_{\alpha} g^{\rm eq}
        + v_{\alpha}\frac{v^2}{2}  f^{\rm eq} d\bm{v} 
    \end{aligned}
    \right\}
        =   u_\alpha (E+\rho RT)
        , \label{eq:Relevantmoments_q}
\\
    &\left.
    \begin{aligned}
        &\textcolor{black}{(\mathrm{I})}\phantom{\mathrm{I}} \quad
        \int v_{\alpha}v_{\beta} g^{\rm eq} d\bm{v} 
    \\
        &\textcolor{black}{(\mathrm{II})} \quad
        \int v_{\alpha}v_{\beta} g^{\rm eq} 
        + v_{\alpha}v_{\beta}\frac{v^2}{2}f^{\rm eq} d\bm{v} 
    \end{aligned}  
    \right\}
    \nonumber \\ & \qquad \quad
    = \rho R T u_\alpha u_\beta + (E + \rho RT)(u_\alpha u_\beta + RT \delta_{\alpha \beta})   
    .\label{eq:Relevantmoments_R}
\end{align}
Furthermore, 
by substitution of the contracted higher-order moments of $f^{\rm eq}$ (divided by two), i.e.
\begin{align}
    &\int \frac{v^2}{2} f^{\rm eq} d\bm{v} = \rho \frac{u^2}{2} + \rho RT \frac{D}{2}
    ,
\\
    &\int v_{\alpha} \frac{v^2}{2} f^{\rm eq} d\bm{v} = \rho u_\alpha \frac{u^2}{2} + \rho R T \frac{u_\alpha (D+2)}{2}
    ,
\\
    &\begin{aligned}[b]
    \int v_{\alpha} v_{\beta} \frac{v^2}{2} f^{\rm eq} d\bm{v} = \rho u_\alpha u_\beta  \frac{u^2}{2} 
     & + \rho RT \frac{u_\alpha u_\beta(D+4)+u^2 \delta_{\alpha\beta}}{2} 
     \\ & + \rho (RT)^2 \frac{\delta_{\alpha\beta}(D+2)}{2}
     ,
    \end{aligned}
\end{align}
into Eqs.~\eqref{eq:Relevantmoments_E},~\eqref{eq:Relevantmoments_q} and~\eqref{eq:Relevantmoments_R},
one obtains the following equilibrium moments of $g$ for the internal non-translational energy split,
\begin{align}
    &
    \begin{aligned}[b]
    \textcolor{black}{(\mathrm{II})} \quad 
    \int g^{\rm eq} d\bm{v} &= E - \Bigl( \rho \frac{u^2}{2} + \rho RT \frac{D}{2} \Bigr) 
    = \rho \Bigl( C_v T -  RT \frac{D}{2} \Bigr)
    ,
    \end{aligned}
\\
    &\begin{aligned}[b]
    \textcolor{black}{(\mathrm{II})} \quad
    \int &v_{\alpha} g^{\rm eq} d\bm{v} 
    = u_\alpha (E+\rho RT) - \Bigl( \rho u_\alpha \frac{u^2}{2} 
    \Bigr. \\ \Bigl. & + \rho R T \frac{u_\alpha (D+2)}{2} \Bigr) = \rho \Bigl( C_v T -  RT \frac{D}{2} \Bigr)  u_{\alpha}
    ,
    \end{aligned}
\\
    &\begin{aligned}[b]
    \textcolor{black}{(\mathrm{II})} \quad
    \int & v_{\alpha} v_{\beta} g^{\rm eq} d\bm{v} 
     = \rho R T u_\alpha u_\beta + (E + \rho RT)(u_\alpha u_\beta + RT \delta_{\alpha \beta}) 
    \\& \quad - \Bigl( \rho u_\alpha u_\beta  \frac{u^2}{2} + \rho RT \frac{u_\alpha u_\beta(D+4)+u^2 \delta_{\alpha\beta}}{2} 
    \Bigr. \\ \Bigl. & \quad + \rho (RT)^2 \frac{\delta_{\alpha\beta}(D+2)}{2} \Bigr)
    \\
    & = \rho \Bigl( C_v T -  RT \frac{D}{2} \Bigr) \Bigl( u_{\alpha}u_{\beta} + RT \delta_{\alpha\beta}\Bigr)
    .
    \end{aligned}
\end{align}
Note that a moment of the Maxwell--Boltzmann distribution can also be computed from the next lower order moment by application of the operator
\begin{equation}
    \mathcal{O}_{\alpha} A = RT \partial_{u_\alpha} A + u_{\alpha} A
    .
\end{equation}
Hence, all Maxwell–Boltzmann energy moments can be written as the result of repeated application of operators on the generating function, i.e. Eq.~\eqref{eq:Total_energy_MB}.

\section{Hydrodynamic limit\label{appendix:CE}}
A multiscale analysis in the form of the Chapman-Enskog expansion~\cite{Chapman} is conducted hereafter.
The starting point is the following system of equations,
{\begin{multline}
	\partial_t \{f, g\} + \bm{v}\cdot\bm{\nabla} \{f, g\} 
    = 
    - \frac{1}{\tau_1} \left( \{f, g\} - \{f^{*}, g^{*}\}\right) 
    \\ 
    - \frac{1}{\tau_2} \left( \{f^{*}, g^{*}\} - \{f^{\rm eq}, g^{\rm eq}\}\right)
    ,
\end{multline}
}
where the following parameters are introduced: 
\begin{itemize}[itemsep=0cm,parsep=0.0cm]
    \item Characteristic flow velocity $\mathcal{U}$, 
    \item Characteristic flow scale $\mathcal{L}$, 
    \item Characteristic flow time $\mathcal{T}=\mathcal{L}/\mathcal{U}$,
    \item Characteristic density $\bar{\rho}$, 
    \item Speed of sound of an ideal gas $c_s=\sqrt{\gamma R T}$. 
\end{itemize}
With these, the variables are reduced as follows (primes denote non-dimensional variables): 
\begin{itemize}[itemsep=0cm,parsep=0.0cm]
    \item Time $t=\mathcal{T}t'$, 
    \item Space $\bm{r}=\mathcal{L}\bm{r}'$, 
    \item Flow velocity $\bm{u}=\mathcal{U}\bm{u}'$,  
    \item Particle velocity $\bm{v}=c_s\bm{v}'$,  
    \item Density $\rho$=$\bar{\rho}\rho'$,  
    \item Distribution function $f=\bar{\rho}c_s^{-3}f'$. 
\end{itemize}
Furthermore, the following non-dimensional groups are introduced:
\begin{itemize}[itemsep=0cm,parsep=0.0cm]
    \item Knudsen number, ${\rm Kn}={\tau c_s}/{\mathcal{L}}$, 
    \item Mach number, ${\rm Ma}={\mathcal{U}}/{c_s}$. 
\end{itemize}
With these, the equations are rescaled as
{\begin{multline}
	{\rm Ma}\,{\rm Kn}\left(\partial_t' \{f', g'\} + \bm{v'}\cdot\bm{\nabla}' \{f', g'\}\right) = 
    \\
    - \frac{1}{\tau_1'} \left( \{f', g'\} - \{f^{*'}, g^{*'}\}\right) 
    \\ 
    - \frac{1}{\tau_2'} \left( \{f^{*'}, g^{*'}\} - \{f^{\rm eq'}, g^{\rm eq'}\}\right)
    .
\end{multline}
}
Of interest here is ${\rm Ma}\sim1$, ${\rm Kn}\sim\epsilon$, i.e. the hydrodynamic limit.
After dropping the primes for the sake of readability,
{\begin{multline}
    \epsilon \left( \partial_t \{f, g\} + \bm{v}\cdot\bm{\nabla} \{f, g\} \right) =     - \frac{1}{\tau_1} \left( \{f, g\} - \{f^{*}, g^{*}\}\right) 
\\ 
    - \frac{1}{\tau_2} \left( \{f^{*}, g^{*}\} - \{f^{\rm eq}, g^{\rm eq}\}\right)
    .
\end{multline}
}
and introducing the multiscale expansions in the distribution functions,
\begin{multline}
    \{f, g, f^*, g^*\} = 
    \{f^{(0)}, g^{(0)}, f^{*(0)}, g^{*(0)}\} 
    \\+ \epsilon \{f^{(1)}, g^{(1)}, f^{*(1)}, g^{*(1)}\} 
    \\+ \epsilon^2 \{f^{(2)}, g^{(2)}, f^{*(2)}, g^{*(2)}\} 
    + O(\epsilon^3),
\end{multline}
as well as the time derivative operator,
\begin{equation}
	   \partial_t = \partial_t^{(1)} + \epsilon \partial_t^{(2)} + O(\epsilon^2)
       ,\label{eq:expansion_time}
\end{equation}
the following equations are recovered at scales $\epsilon^0$, $\epsilon^1$ and $\epsilon^2$:
\begin{align}
    \epsilon^0:\ \ &%
        0 = 
        {-\frac{1}{\tau_1} \left(  \{f^{(0)} , g^{(0)}\} -  \{f^{*(0)}, g^{*(0)}\} \right)
        }
        \nonumber\\ 
        &\phantom{0 = }%
        { -\frac{1}{\tau_2} \left(  \{f^{*(0)}, g^{*(0)}\} - \{f^{\rm eq},g^{\rm eq}\} \right)
        }
        ,\label{Eq:CE_Eq_orders_0}
\\
    \epsilon^1:\ \ &%
        \partial_t^{(1)} 
        \{f^{(0)}, g^{(0)}\}  + \bm{v}\cdot\bm{\nabla} \{f^{(0)}, g^{(0)}\} = 
        \nonumber\\ 
        &-\frac{1}{\tau_1} \{f^{(1)} , g^{(1)}\} + \left(\frac{1}{\tau_1}-\frac{1}{\tau_2}\right) \{f^{*(1)}, g^{*(1)}\}
        ,\label{Eq:CE_Eq_orders_1}
\\
    \epsilon^2:\ \ &%
        \partial_t^{(1)} 
        \{f^{(1)}, g^{(1)}\}  + \bm{v}\cdot\bm{\nabla} \{f^{(1)}, g^{(1)}\} + \partial_t^{(2)}\{f^{(0)}, g^{(0)}\} = 
        \nonumber\\ 
        &-\frac{1}{\tau_1} \{f^{(2)} , g^{(2)}\} + \left(\frac{1}{\tau_1}-\frac{1}{\tau_2}\right) \{f^{*(2)}, g^{*(2)}\}
        .\label{Eq:CE_Eq_orders_2}
\end{align}
Note that for the sake of readability, writing the factors {\{$\epsilon^0$, $\epsilon^1$, $\epsilon^2$\}} in front of every term was omitted for the remainder of the analysis. 
However it shall be kept in mind that, e.g. the whole order-$2$-in-$\epsilon$ equation would possess a prefactor of $\epsilon^2$.
An analysis on each order-in-$\epsilon$ equation will follow.

From order $\epsilon^0$ it directly follows that
\begin{equation}
    \{f^{(0)}, g^{(0)}\}=\{f^{*(0)}, g^{*(0)}\}=\{f^{\rm eq}, g^{\rm eq}\}
    ,
\end{equation}
{for any $\tau_1$, $\tau_2$.}
Note that, therefore, the solvability conditions for this system for \mbox{$\forall k>0$}, 
which can be inferred from the equations provided in Section~\ref{sec:model-description}, become
\begin{align}
    &\phantom{(\mathrm{II})} \quad
    \int \{ f^{(k)}, f^{*(k)} \} d\bm{v} = 0
    ,
\\
    &\phantom{(\mathrm{II})} \quad
    \int \bm{v} \{ f^{(k)}, f^{*(k)} \} d\bm{v}  = 0
    ,
\\  
    &\begin{aligned}[b]
        \textcolor{black}{(\mathrm{I})}\phantom{\mathrm{I}} \quad
        & \int \{ g^{(k)}, g^{*(k)} \} d\bm{v} = 0
        ,
    \\  
        \textcolor{black}{(\mathrm{II})} \quad
        & \int \{ g^{(k)}, g^{*(k)} \} + \frac{v^2}{2} \{ f^{(k)}, f^{*(k)} \} d\bm{v}  = 0
        ,
    \end{aligned}
\end{align}
for the conserved moments.
Further solvability conditions related to Prandtl numbers 
{$\{\rm{Pr}\leq1, \rm{Pr}\geq1\}$}
for \mbox{$\forall k>0$} read
\begin{align}
    &\phantom{(\mathrm{II})}\quad
    \int\bm{v}\otimes\bm{v} f^{*(k)}d\bm{v} = 0
    ,\label{eq:Solvability_QE_Pr<1_conserved}
\\
    &\begin{aligned}[b]
        &\begin{aligned}
            \textcolor{black}{(\mathrm{I})}\phantom{\mathrm{I}} \quad
            \int \bm{v} g^{*(k)} d\bm{v} 
            - &\bm{u}\cdot
        \overbrace{
            \int  \bm{v} \otimes \bm{v} f^{*(k)} d\bm{v} 
        }^{\hidewidth
            0, \rm{ \ Eq.~\eqref{eq:Solvability_QE_Pr<1_conserved}}
        \hidewidth}
        \\
            &= \int \bm{v} g^{(k)} d\bm{v} 
            - \bm{u}\cdot \int  \bm{v} \otimes \bm{v} f^{(k)} d\bm{v} 
            ,
        \end{aligned}
    \\  
        &\begin{aligned}
            \textcolor{black}{(\mathrm{II})} \quad
            \int &\bm{v} g^{*(k)} + \bm{v}\frac{v^2}{2}f^{*(k)}   d\bm{v} 
            - \bm{u}\cdot 
        \overbrace{
            \int  \bm{v} \otimes \bm{v} f^{*(k)} d\bm{v} 
        }^{\hidewidth
            0, \rm{ \ Eq.~\eqref{eq:Solvability_QE_Pr<1_conserved}}
        \hidewidth}
        \\
            &= \int \bm{v} g^{(k)} + \bm{v}\frac{v^2}{2}f^{*(k)}  d\bm{v} 
            - \bm{u}\cdot \int  \bm{v} \otimes \bm{v} f^{(k)} d\bm{v} 
            ,
        \end{aligned}
        \label{eq:Solvability_QE_Pr<1_quasiconserved}
    \end{aligned}
\end{align}
for 
\textcolor{black}{
{$\rm{Pr}\leq1$},
}
i.e. in case the quasi-equilibria are constructed using minimization under constraints of conserving 
mass, momentum, total energy, and the pressure tensor, while the heat flux vector is quasi-conserved.
For the opposite case where the quasi-equilibria are constructed under constrains of conserving 
mass, momentum, total energy, and the heat flux vector, while the pressure tensor is quasi-conserved, i.e. for 
\textcolor{black}{
{$\rm{Pr}\geq1$},
}
the solvability conditions read
\begin{align}
    &\phantom{(\mathrm{II})}\quad
\int\bm{v}\otimes\bm{v} f^{*(k)}d\bm{v} = \int\bm{v}\otimes\bm{v} f^{(k)} d\bm{v}    ,\label{eq:Solvability_QE_Pr>1_quasiconserved}
\\
    &\begin{aligned}
        &\begin{aligned}[b]
            \textcolor{black}{(\mathrm{I})}\phantom{\mathrm{I}} \quad
            \int \bm{v} g^{*(k)} d\bm{v} 
            - \bm{u}\cdot
        \overbrace{
            \int  \bm{v} \otimes \bm{v} f^{*(k)} d\bm{v} 
        }^{\hidewidth
            \int\bm{v}\otimes\bm{v} f^{(k)} d\bm{v}, \rm{ \ Eq.\eqref{eq:Solvability_QE_Pr>1_quasiconserved}}
        \hidewidth}
            = 0,
        \end{aligned}
    \\  
        &\begin{aligned}
            \textcolor{black}{(\mathrm{II})} \quad
            \int \bm{v} g^{*(k)} + \bm{v}\frac{v^2}{2}f^{*(k)} d\bm{v} 
            - \bm{u}\cdot
        \overbrace{
            \int  \bm{v} \otimes \bm{v} f^{*(k)} d\bm{v} 
        }^{\hidewidth
            \int\bm{v}\otimes\bm{v} f^{(k)} d\bm{v}, \rm{ \ Eq.\eqref{eq:Solvability_QE_Pr>1_quasiconserved}}
        \hidewidth }
            = 0.
        \end{aligned}
        \label{eq:Solvability_QE_Pr>1_conserved}
    \end{aligned}
\end{align}

Next, going up one order in $\epsilon$ and computing the moments $\int\{f, \bm{v} f\}d\bm{v}$ of the Chapman--Enskog-expanded equations at order $\epsilon$, 
{using the solvability conditions and the known equilibrium moments (listed in Appendix~\ref{appendix:EqMoments}),}
the continuity and momentum balance equations become
\begin{gather}
	    \partial_t^{(1)}\rho + \bm{\nabla}\cdot \rho\bm{u}  = 0
        ,
        \label{eq:Eulerlevel_continuity1}
\\
	    \partial_t^{(1)}\left( \rho \bm{u} \right) + \bm{\nabla}\cdot \left( \rho \bm{u}\otimes\bm{u} + p\bm{I} \right)= 0
        ,
        \label{eq:Eulerlevel_momentum1}
\end{gather}
written in vector notation with the identity tensor $\bm{I}$.
For the energy balance equation, computing the moment
\textcolor{black}{
\mbox{$\{ (\mathrm{I}) \int g d\bm{v}$}, 
\mbox{$ (\mathrm{II}) \int g+\frac{v^2}{2}fd\bm{v}\}$} 
}
results in
\begin{equation}
	\partial_t^{(1)}E + \bm{\nabla}\cdot \left( E + p \right) \bm{u} = 0
    ,
    \label{eq:Eulerlevel_totenergy1}
\end{equation}
for both energy splits.
Note that a truncation of the expansion at this level, i.e. $\epsilon$ times Eq.~\{\eqref{eq:Eulerlevel_continuity1},~\eqref{eq:Eulerlevel_momentum1},~\eqref{eq:Eulerlevel_totenergy1}\},
yields the compressible Euler equations as 
\begin{gather}
    \partial_t\rho + \bm{\nabla}\cdot \rho\bm{u}  
    = 0
    ,
\\
    \partial_t\left( \rho \bm{u} \right) + \bm{\nabla}\cdot \left( \rho \bm{u}\otimes\bm{u} + p\bm{I} \right)
    = 0
    ,
\\
    \partial_t E + \bm{\nabla}\cdot \left( E + p \right) \bm{u} 
    = 0
    ,
\end{gather}
after using Eq.~\eqref{eq:expansion_time} truncated at $\partial_t = \partial_t^{(1)} + \mathcal{O}(\epsilon)$ and transforming back to dimensional variables. 

Further transport equations for $p$, $p\bm{u}$ and $E\bm{u}$, which are relevant for the remainder of the multiscale analysis at order $\epsilon^2$, may also be derived at this stage.
A transport equation for $p$ is obtained via the balance equations for kinetic and internal energy.
A balance equation for kinetic energy can be derived by multiplying the Euler level momentum balance, i.e. Eq.~\eqref{eq:Eulerlevel_momentum1}, with $\bm{u}$, 
as
\begin{multline}
    \bm{u}\cdot \left[
    \partial_t\left( \rho \bm{u} \right) 
    + \bm{\nabla}\cdot \left( \rho \bm{u}\otimes\bm{u} + p\bm{I} \right)
    \right] =
\\
    \bm{u}\cdot\partial_t\left( \rho \bm{u} \right) 
    + \bm{u} \cdot \left( \bm{\nabla} \cdot \rho\bm{u}\otimes\bm{u} \right)
    + \bm{u}\cdot\bm{\nabla}p 
    = 0
    .
\end{multline}
Using the expansions
\begin{gather}
    \bm{u}\cdot\partial_t^{(1)} (\rho \bm{u}) 
    = \partial_t^{(1)}\Bigl(
\overbrace{
    \frac{1}{2}\rho u^2
}^{K}
    \Bigr) 
    + \frac{1}{2} u^2 
\overbrace{
    \partial_t^{(1)} \rho
}^{-\bm{\nabla}\cdot\rho\bm{u}}
    ,
\\
    \bm{u} \cdot \left( \bm{\nabla} \cdot \rho\bm{u}\otimes\bm{u} \right)
    =  \bm{\nabla} \cdot 
    \Bigl( 
\underbrace{
    \frac{1}{2} \rho u^2\bm{u} 
}_{K\bm{u}}
    \Bigr)
    + \frac{1}{2} u^2 \bm{\nabla} \cdot \rho\bm{u}
    ,
\end{gather}
where the time derivative is replaced in favor of a spatial derivative
by means of the Euler level continuity, i.e.~\eqref{eq:Eulerlevel_continuity1},
results in
\begin{equation}
    \partial_t^{(1)}K + 
    \bm{\nabla}\cdot K \bm{u} 
    + \bm{u}\cdot \bm{\nabla}p = 0
    .
\end{equation}
In turn, this can be used to derive a balance equation for internal energy by subtraction from the balance equation for total energy, i.e. Eq.\eqref{eq:Eulerlevel_totenergy1}, as
\begin{equation}
        \partial_t^{(1)} E - \partial_t^{(1)} K 
    + 
        \bm{\nabla}\cdot E\bm{u} - \bm{\nabla}\cdot K \bm{u} 
    + \bm{\nabla}\cdot p\bm{u}  - \bm{u}\cdot \bm{\nabla}p
    = 0
    .
\end{equation}
By applying the expansion 
\begin{equation}
    \bm{\nabla}\cdot p\bm{u} - \bm{u}\cdot \bm{\nabla}p = p\bm{\nabla} \cdot \bm{u}
    ,
\end{equation}
this results in
\begin{equation}
    \partial_t^{(1)} U + \bm{\nabla}\cdot U \bm{u} + p \bm{\nabla}\cdot\bm{u} = 0
    ,
\end{equation}
from which a balance equation for pressure
can be derived using $\partial_T U = \rho C_v$ together with the ideal gas equation of state as 
\begin{equation}
    \partial_p U = C_v / R,    
\end{equation}
resulting in
\begin{equation}
    \partial_t^{(1)} p + \bm{\nabla}\cdot p \bm{u} + \frac{R}{C_v} p \bm{\nabla}\cdot\bm{u} = 0
    .
    \label{eq:pressure_euler_CE}
\end{equation}
Furthermore, for the purpose of deriving a transport equation for $p\bm{u}$, Eq.~\eqref{eq:pressure_euler_CE} is multiplied by $\bm{u}$, as
\begin{multline}
    \bm{u} \Bigl[ \partial_t^{(1)} p + \bm{\nabla}\cdot p \bm{u} + \frac{R}{C_v} p \bm{\nabla}\cdot\bm{u} \Bigr] 
\\
    \bm{u} \partial_t^{(1)} p +  \bm{u} (\bm{\nabla}\cdot p \bm{u}) + \frac{R}{C_v} p  \bm{u} (\bm{\nabla}\cdot\bm{u}) = 0
    .
\end{multline}
Using Euler level momentum balance, i.e.~\eqref{eq:Eulerlevel_momentum1}, expanded as
\begin{multline}
    \partial_t^{(1)}(\bm{u})
    = - \frac{1}{\rho} \Bigl(
    \bm{u} 
\overbrace{
    \partial_t^{(1)}\rho
}^{-\bm{\nabla}\cdot\rho\bm{u}}
    +
\overbrace{
    \bm{\nabla}\cdot\rho\bm{u}\otimes\bm{u} 
}^{\rho\bm{u}\cdot\bm{\nabla}\bm{u}+(\bm{\nabla}\cdot\rho\bm{u})\bm{u}}
    +\bm{\nabla}p
    \Bigr)
\\
    = - \bm{u}\cdot\bm{\nabla}\bm{u} - \frac{1}{\rho}  \bm{\nabla}p
    ,\label{eq:CE_dtu-expanded}
\end{multline}
the first term can be expanded as  
\begin{align}    
    \bm{u}\partial_t^{(1)}p  
    &= \partial_t^{(1)}(p\bm{u}) 
    -p\partial_t^{(1)}\bm{u}
\nonumber \\
    &= \partial_t^{(1)}(p\bm{u}) + p \bm{u}\cdot\bm{\nabla}\bm{u} + \frac{p}{\rho}  \bm{\nabla}p
    ,
\end{align}
whereas the second term can be rewritten as
\begin{equation}
     \bm{u}(\bm{\nabla}\cdot p\bm{u}) = \bm{\nabla}\cdot p\bm{u}\otimes\bm{u} - p\bm{u}\cdot\bm{\nabla}\bm{u}
     ,
     \label{eq:unabladotpu_expansion_1}
\end{equation}
resulting in 
\begin{equation}
    \label{eq:eulerlevel_ptimesu}
    \partial_t^{(1)}p\bm{u} + \bm{\nabla}\cdot p \bm{u}\otimes\bm{u} + \frac{p}{\rho}\bm{\nabla}p + \frac{R}{C_v}p\bm{u}\bm{\nabla}\cdot\bm{u} = 0
    .
\end{equation}
Using the same approach to derive a balance equation for $E\bm{u}$, starting off with Eq.~\eqref{eq:Eulerlevel_totenergy1}, multiplying by $\bm{u}$ as
\begin{multline}
    \bm{u} \Bigl[ \partial_t^{(1)} E + \bm{\nabla}\cdot E \bm{u} + \bm{\nabla}\cdot p \bm{u} \Bigr] 
\\
    \bm{u} \partial_t^{(1)} E +  \bm{u}(\bm{\nabla}\cdot E \bm{u})  + \bm{u}(\bm{\nabla}\cdot p \bm{u}) = 0
    ,
\end{multline}
using the expansions
\begin{gather}
    \begin{aligned}[b]
        \bm{u}\partial_t^{(1)}E &=  \partial_t^{(1)}(E\bm{u}) -  E\partial_t^{(1)}(\bm{u})
    \\
        &= \partial_t^{(1)}(E\bm{u}) +E\bm{u}\cdot\bm{\nabla}\bm{u} + \frac{E}{\rho} \bm{\nabla}p
        ,
    \end{aligned}
\\
     \bm{u}(\bm{\nabla}\cdot E\bm{u}) = \bm{\nabla}\cdot E\bm{u}\otimes\bm{u} - E\bm{u}\cdot\bm{\nabla}\bm{u}
    ,
\end{gather}
and Eq.~\eqref{eq:unabladotpu_expansion_1},
the following additional balance equation is received as
\begin{equation}
    \partial_t^{(1)}E \bm{u} 
    + \bm{\nabla}\cdot\left(E + p\right)\bm{u}\otimes\bm{u} 
    + \frac{E}{\rho}\bm{\nabla}p 
    - p\bm{u}\cdot\bm{\nabla}\bm{u}
    = 0
    .
    \label{eq:eulerlevel_energytimesu}
\end{equation}

Next, going up one more order in $\epsilon$ and computing the moments $\int\{f, \bm{v} f\}d\bm{v}$ of the Chapman--Enskog-expanded equations at order $\epsilon^2$, the continuity equation becomes
	\begin{equation}
	    \partial_t^{(2)}\rho = 0
        ,
        \label{eq:NSFlevel_continuity1}
	\end{equation}
while for the momentum balance equation one has
    \begin{equation}
        \partial_t^{(2)}\left( \rho \bm{u} \right) + \bm{\nabla}\cdot \int\bm{v}\otimes\bm{v} f^{(1)} d\bm{v} = 0
        .
        \label{eq:approach2_momentum2_raw}
    \end{equation}
The second term can be further expanded using the moment $\int \bm{v}\otimes\bm{v} f d\bm{v}$ of the first-order-in-$\epsilon$ equation, rearranged as
\begin{multline}
    -\frac{1}{\tau_1} \int\bm{v}\otimes\bm{v} f^{(1)}d\bm{v} 
    +\left(\frac{1}{\tau_1}-\frac{1}{\tau_2}\right) \int\bm{v}\otimes\bm{v} f^{*(1)}d\bm{v} 
    =
\\ 
    \partial_t^{(1)} 
    \int\bm{v}\otimes\bm{v}f^{(0)}d\bm{v} 
    + \bm{\nabla}\cdot 
    \int\bm{v}\otimes\bm{v}\otimes\bm{v}f^{(0)}d\bm{v}
    .
\end{multline}
Here, the expansion depends on the construction of the quasi-equilibria attractors depending on the case of 
{\mbox{$\{\rm{Pr}\leq1, \rm{Pr}\geq1\}$}}.
After plugging the equilibrium moments into the terms on the RHS and applying the solvability conditions for the \ac{QE} term on the LHS, 
one gets
\begin{multline}
    \int\bm{v}\otimes\bm{v} f^{(1)}d\bm{v} = 
    -\{\tau_1, \tau_2\} \Bigl[ \partial_t^{(1)} \left( \rho\bm{u}\otimes\bm{u} + p\bm{I} \right) \Bigl. 
    \\ \Bigl.
    + \bm{\nabla}\cdot \rho\bm{u}\otimes\bm{u}\otimes\bm{u} + \bm{\nabla}p\bm{u} + (\bm{\nabla}p\bm{u})^{\dagger} + \bm{I}\bm{\nabla}\cdot p\bm{u} \Bigr]
    .
    \label{eq:pnoneq_tau12}
\end{multline}
Note that, in the prefactor \textcolor{black}{$\{\tau_1, \tau_2\}$}, the $\tau_1$ emerges after application of solvability condition~\eqref{eq:Solvability_QE_Pr<1_conserved}, whereas $\tau_2$ emerges after application of solvability condition~\eqref{eq:Solvability_QE_Pr>1_quasiconserved}.
Expanding the term
\begin{multline}
    \partial_t^{(1)}\left(\rho\bm{u}\otimes\bm{u} + p\bm{I}\right) = 
    \bm{u}\otimes\partial_t^{(1)}\rho \bm{u} 
    + (\bm{u}\otimes\partial_t^{(1)}\rho \bm{u})^{\dagger} 
    \\ - \bm{u}\otimes\bm{u}\partial_t^{(1)}\rho + \partial_t^{(1)}p\bm{I}
    ,
\end{multline}
and replacing the time derivatives in favor of a spatial derivatives
by means of the Euler level continuity, momentum and pressure balance equations, i.e. Eqs.~\eqref{eq:Eulerlevel_continuity1},~\eqref{eq:Eulerlevel_momentum1} and~\eqref{eq:pressure_euler_CE}, 
one arrives at
\begin{multline}
    \partial_t^{(1)}\left(\rho\bm{u}\otimes\bm{u} + p\bm{I}\right) = 
    -\bm{u}\otimes\left[\bm{\nabla}\cdot\rho\bm{u}\otimes\bm{u} + \bm{\nabla}p\right]
    \\-\left(\bm{u}\otimes\left[\bm{\nabla}\cdot\rho\bm{u}\otimes\bm{u} + \bm{\nabla}p\right]\right)^{\dagger}
     + \bm{u}\otimes\bm{u} \bm{\nabla}\cdot\rho \bm{u} 
    \\ - (\bm{\nabla} \cdot p \bm{u} )\bm{I} - \frac{R}{C_v} p (\bm{\nabla} \cdot \bm{u})\bm{I}
    .
\end{multline}
After applying the expansions
\begin{gather}
    \begin{aligned}[b]
        -\bm{u}\otimes\bm{\nabla}\cdot\rho\bm{u}\otimes&\bm{u}
        -\left(\bm{u}\otimes\bm{\nabla}\cdot\rho\bm{u}\otimes\bm{u}\right)^{\dagger}
        \\
        &+ \bm{u}\otimes\bm{u} \bm{\nabla}\cdot\rho \bm{u} 
        = -\bm{\nabla}\cdot\rho\bm{u}\otimes\bm{u}\otimes\bm{u}
        ,
    \end{aligned}
\\
    - \bm{u}\otimes\bm{\nabla}p = - (\bm{\nabla} p\bm{u})^{\dagger} + p (\bm{\nabla}\bm{u})^{\dagger}
    ,
\\
    -(\bm{u}\otimes\bm{\nabla}p)^{\dagger} = -\bm{\nabla} p\bm{u} + p \bm{\nabla}\bm{u}
    ,
\end{gather}
and plugging into Eq.~\eqref{eq:pnoneq_tau12},
most terms cancel and one obtains
\begin{multline}
    \int \bm{v}\otimes\bm{v} f^{(1)}d\bm{v} = - \{\tau_1, \tau_2\} p \left[ \bm{\nabla}\bm{u} + \bm{\nabla}\bm{u}^{\dagger}- \frac{R}{C_v}(\bm{\nabla}\cdot\bm{u})\bm{I}\right]
    .
\end{multline}
Plugging this final expression into the momentum balance equation at order $\epsilon^2$, i.e. Eq.~\eqref{eq:approach2_momentum2_raw}, results in
    \begin{equation}
        \partial_t^{(2)}(\rho \bm{u} )+ \bm{\nabla}\cdot \bm{\tau}_{\rm NS}  = 0
        ,\label{eq:NSFlevel_momentum1}
    \end{equation}
where the deviatoric Cauchy stress tensor in the NSF equations, i.e. the Navier--Stokes stress tensor, is correctly recovered as
    \begin{multline}
        \bm{\tau}_{\rm NS} = 
        {-} \{\tau_1, \tau_2\}p\left[ \bm{\nabla}\bm{u} + \bm{\nabla}\bm{u}^{\dagger}- \frac{R}{C_v}(\bm{\nabla}\cdot\bm{u})\bm{I} \right]
        \\ = 
        {-} \{\tau_1,\tau_2\} p\left[\bm{\nabla}\bm{u} + \bm{\nabla}\bm{u}^\dagger - \frac{2}{D}(\bm{\nabla}\cdot\bm{u})\bm{I} \right] 
        \\ 
        {-} \{\tau_1,\tau_2\} p\left(\frac{2}{D}-\frac{R}{C_v}\right)(\bm{\nabla}\cdot\bm{u})\bm{I}
        ,\label{eq:CE_Tns}
    \end{multline}
with the dynamic shear viscosity, second viscosity, and dynamic bulk viscosity given as
\begin{gather}
    \mu = \{\tau_1, \tau_2\}p
    ,
\\
    \lambda = - \{\tau_1, \tau_2\}p \frac{R}{C_v}
    , 
\\
     \eta =   \{\tau_1,\tau_2\}p\left(\frac{2}{D}-\frac{R}{C_v}\right)
     .
\end{gather}
Moving on to computing the moments
\textcolor{black}{
\mbox{$\{ (\mathrm{I}) \int g d\bm{v}$}, 
\mbox{$ (\mathrm{II}) \int g+\frac{v^2}{2}fd\bm{v}\}$} 
}
for the energy balance at order $\epsilon^2$,
\begin{align}
    \begin{aligned}[b]
        \textcolor{black}{(\mathrm{I})}\phantom{\mathrm{I}} \quad 
        &
        \partial_t^{(2)} E + \bm{\nabla}\cdot \int \bm{v} g^{(1)} d\bm{v} = 0,
    \\
        \textcolor{black}{(\mathrm{II})} \quad 
        &
        \partial_t^{(2)} E + \bm{\nabla}\cdot  \int\bm{v} g^{(1)} +\bm{v}\frac{v^2}{2} f^{(1)} d\bm{v} = 0,
        \label{eq:approach2_energy2_raw}
    \end{aligned}
\end{align} 
are obtained for the total energy split and the internal non-translational split, respectively.
The flux terms can be further expanded using the moments
\textcolor{black}{
\mbox{$\{ (\mathrm{I}) \int \bm{v} g d\bm{v}$}, 
\mbox{$ (\mathrm{II}) \int \bm{v} g + \bm{v} \frac{v^2}{2}fd\bm{v}\}$} 
}
of the first-order-in-$\epsilon$ equation, rearranged as
\begin{align}
    \begin{aligned}[b]
        \textcolor{black}{(\mathrm{I})}\phantom{\mathrm{I}} \quad 
        -\frac{1}{\tau_1}&\int\bm{v} g^{(1)}d\bm{v} 
        +\left( \frac{1}{\tau_1}-\frac{1}{\tau_2} \right) \int\bm{v}g^{*(1)} d\bm{v} 
        \\ &=
        \partial_t^{(1)}  \int\bm{v} g^{(0)}d\bm{v} 
        + \bm{\nabla}\cdot  \int\bm{v}\otimes\bm{v}g^{(0)}d\bm{v}  
        ,
    \\
        \textcolor{black}{(\mathrm{II})} \quad 
        -\frac{1}{\tau_1}&\int \bm{v} g^{(1)} +\bm{v}\frac{v^2}{2} f^{(1)} d\bm{v} 
        \\&+\left( \frac{1}{\tau_1}-\frac{1}{\tau_2} \right) \int \bm{v}g^{*(1)} + \bm{v}\frac{v^2}{2}f^{*(1)} d\bm{v}
        \\& =
        \partial_t^{(1)}  \int  \bm{v} g^{(0)} +\bm{v}\frac{v^2}{2} f^{(0)}d\bm{v} 
        \\& 
        + \bm{\nabla}\cdot \int \bm{v}\otimes\bm{v} g^{(0)} + \bm{v}\otimes\bm{v} \frac{v^2}{2} f^{(0)}d\bm{v}   
        ,
    \end{aligned}
\end{align}
respectively.
Plugging the equilibrium moments into the RHS results in
\begin{align}
    &\left.
    \begin{aligned}
        &\textcolor{black}{(\mathrm{I})}\phantom{\mathrm{I}} \quad 
        -\frac{1}{\tau_1}\int\bm{v} g^{(1)}d\bm{v} 
        +\left( \frac{1}{\tau_1}-\frac{1}{\tau_2} \right)  \int\bm{v}g^{*(1)} d\bm{v} 
    \\
        &\begin{aligned}
        \textcolor{black}{(\mathrm{II})} \quad 
        -\frac{1}{\tau_1}&\int \bm{v} g^{(1)} +\bm{v}\frac{v^2}{2} f^{(1)} d\bm{v} 
        \\&
        +\left( \frac{1}{\tau_1}-\frac{1}{\tau_2} \right) \int \bm{v}g^{*(1)} + \bm{v}\frac{v^2}{2}f^{*(1)} d\bm{v} 
        \end{aligned}
    \end{aligned}
    \right\} 
    \nonumber \\
    &\begin{aligned}[b]
    \qquad \qquad \qquad 
        = \partial_t^{(1)} \left(p+E\right)\bm{u}  + \bm{\nabla}\cdot \Bigl( \frac{p}{\rho}\left(E + p\right) \bm{I} 
        \\ 
        + p\bm{u}\otimes\bm{u} + \left(p + E\right)\bm{u}\otimes\bm{u} \Bigr)  
        ,
    \end{aligned}
\end{align}
which can be simplified by using the balance equation for $p\bm{u}$ and $E\bm{u}$, i.e. Eqs.~\eqref{eq:eulerlevel_ptimesu} and~\eqref{eq:eulerlevel_energytimesu}, to replace the time derivatives in favor of space derivatives. 
Applying $E/\rho = C_vT+1/2u^2$ and $p/\rho = RT$ after the expansions 
\begin{gather}
    \bm{\nabla}\left( \frac{pE}{\rho} \right) - \frac{E}{\rho}\bm{\nabla}p = p \bm{\nabla}\left( \frac{E}{\rho} \right) 
    = p C_v\bm{\nabla}T + p \bm{\nabla}\left(\frac{u^2}{2}\right) 
    ,
\\
    \bm{\nabla}\left( \frac{p^2}{\rho} \right) - \frac{p}{\rho}\bm{\nabla}p = p \bm{\nabla}\left( \frac{p}{\rho} \right) = p R\bm{\nabla}T
    ,
\end{gather}
together with the identity $\bm{\nabla}\left(\frac{u^2}{2}\right) = \bm{u}\cdot(\bm{\nabla}\bm{u})^{\dagger}$,
leads to most terms canceling and one obtains 
\begin{align}
    &\left.
    \begin{aligned}
        &\textcolor{black}{(\mathrm{I})}\phantom{\mathrm{I}} \quad 
        -\frac{1}{\tau_1}\int\bm{v} g^{(1)}d\bm{v} 
        +\left( \frac{1}{\tau_1}-\frac{1}{\tau_2} \right)  \int\bm{v}g^{*(1)} d\bm{v} 
    \\
        &\begin{aligned}
        \textcolor{black}{(\mathrm{II})} \quad 
        -\frac{1}{\tau_1}&\int \bm{v} g^{(1)} +\bm{v}\frac{v^2}{2} f^{(1)} d\bm{v} 
        \\&
        +\left( \frac{1}{\tau_1}-\frac{1}{\tau_2} \right)  \int \bm{v}g^{*(1)} + \bm{v}\frac{v^2}{2}f^{*(1)} d\bm{v} 
        \end{aligned}
    \end{aligned}
    \right\} 
    \nonumber \\
    &\begin{aligned}[b]
    \qquad \qquad \quad
        = p \left(C_v+R\right)\bm{\nabla}T + p \Bigl[ \bm{u}\cdot\bm{\nabla}\bm{u} + \bm{u}\cdot\bm{\nabla}\bm{u}^{\dagger} 
        \\ - \frac{R}{C_v}
        {\bm{u}(\bm{\nabla}\cdot\bm{u})} 
        \Bigr]
    \end{aligned}
\end{align}
Here again, the expansion depends on the construction of the quasi-equilibria attractors depending on the case of 
{\mbox{$\{\rm{Pr}\leq1, \rm{Pr}\geq1\}$}}.
After applying the solvability conditions for the \ac{QE} terms on the LHS, 
\begin{align}
    &\left. 
    \begin{aligned}
        \textcolor{black}{(\mathrm{I})}\phantom{\mathrm{I}} \quad 
        & \int\bm{v} g^{(1)} d\bm{v}
    \\    
        \textcolor{black}{(\mathrm{II})} \quad 
        & \int\bm{v} g^{(1)} +\bm{v}\frac{v^2}{2} f^{(1)} d\bm{v}
    \end{aligned}
    \right\}
    = - \{\tau_2,\tau_1\} p \left(C_v+R\right)\bm{\nabla}T  
    \nonumber \\ & \quad 
    - \{\tau_1, \tau_2\} p \Bigl[ \bm{u}\cdot\bm{\nabla}\bm{u} + \bm{u}\cdot\bm{\nabla}\bm{u}^{\dagger} - \frac{R}{C_v}
    {\bm{u}(\bm{\nabla}\cdot\bm{u})} 
    \Bigr]
\end{align} 
are received.
Note that the solvability conditions for \textcolor{black}{\mbox{$\rm{Pr}<1$}}, 
i.e.~\eqref{eq:Solvability_QE_Pr<1_quasiconserved}, 
result in $\tau_2$ as the prefactor in front of the first term and $\tau_1$ in front of the second term on the RHS, respectively, and vice versa for the case of \textcolor{black}{\mbox{$\rm{Pr}>1$}} with~\eqref{eq:Solvability_QE_Pr>1_conserved}.
This result leads to the balance equation for total energy at order $\epsilon^2$, i.e. after plugging into~\eqref{eq:approach2_energy2_raw}, as
\begin{equation}
    \partial_t^{(2)}E + \bm{\nabla}\cdot \bm{q}_{\rm NSF} = 0
    ,\label{eq:NSFlevel_energy1}
\end{equation}
where in
\begin{equation}
    \bm{q}_{\rm NSF} = \bm{q}_{\rm F} 
    {+}  \bm{q}_{\rm H} 
    ,
\end{equation}
the viscous heating vector 
\begin{equation}
    \bm{q}_{\rm H} = \bm{u}\cdot\bm{\tau}_{\rm NS},
\end{equation}
composed of the  Navier--Stokes stress tensor, cf.~\eqref{eq:CE_Tns}, can be identified. 
Also the Fourier heat flux can consistently be recovered as
 \begin{equation}
     \bm{q}_{\rm F} = - \kappa \bm{\nabla}T
     ,
\end{equation}
where the thermal conductivity is given by
 \begin{equation}
     \kappa = \{\tau_2, \tau_1\}p 
     \underbrace{\left(C_v+R\right)}_{C_p}
     .
 \end{equation}
The Prandtl number is readily shown to be,
\begin{equation}
    {\rm Pr} = \frac{\nu}{\alpha} = \frac{C_p \mu}{\kappa}
    = \frac{C_p \{\tau_1, \tau_2\}p}{\{\tau_2,\tau_1\}\left(C_v+R\right)p}
    = \frac{\{\tau_1, \tau_2\}}{\{\tau_2,\tau_1\}}
    ,
\end{equation}
with \textcolor{black}{${\rm Pr} = 1$} in the limit of a BGK collision operator, $\tau_1=\tau_2$.
Truncating at this level, 
i.e. $\epsilon$ times Eq.~\{\eqref{eq:Eulerlevel_continuity1},~\eqref{eq:Eulerlevel_momentum1},~\eqref{eq:Eulerlevel_totenergy1}\} plus $\epsilon^2$ times Eq.~\{\eqref{eq:NSFlevel_continuity1},~\eqref{eq:NSFlevel_momentum1},~\eqref{eq:NSFlevel_energy1}\} and using Eq.~\eqref{eq:expansion_time} truncated at $\partial_t = \partial_t^{(1)} + \epsilon \partial_t^{(2)} + \mathcal{O}(\epsilon^2)$,
results in the equations
\begin{gather}
    \partial_t\rho + \bm{\nabla} \cdot \rho \bm{u} 
    = 0
    ,
    \\
    \partial_t (\rho \bm{u}) + \bm{\nabla} \cdot \left( \rho \bm{u} \otimes \bm{u} + p\bm{I} 
    {+} \epsilon\bm{\tau}_{\rm NS} \right) 
    = 0
    ,
    \\
    \partial_t E + \bm{\nabla} \cdot \left[ (E + p)\bm{u} + \epsilon\bm{q}_{\rm F} 
    {+} \epsilon\bm{u} \cdot \bm{\tau}_{\rm NS} \right] 
    = 0
    ,
\end{gather}
after transforming back to dimensional variables.
The dissipative mechanisms, i.e. the Navier--Stokes stress tensor, the Fourier heat flux and the viscous heating vector, are of order $\mathcal{O}(\epsilon) \sim \mathcal{O}(\rm Kn)$.

This concludes the multiscale analysis recovering the \ac{NSF} equations.

\section{Background on Hermite polynomials and the Grad expansion\label{appendix:Gauss-Hermite}}

The Hermite polynomial of order $n$ is defined as
\begin{equation}
    \mathcal{H}_{n}(x) = \frac{(-1)^n}{w(x)} \partial_{x^n} w(x),
\end{equation}
where the normalized weight function is defined as
\begin{equation}
    w(x) = \frac{1}{\sqrt{2\pi}} e^{- \frac{x^2}{2}}
    .
\end{equation}
Hermite polynomials are mutually orthogonal with respect to the weight function $w(x)$, as
\begin{equation}
    \int_{-\infty}^{+\infty} \mathcal{H}_{m}(x) w(x) \mathcal{H}_{n}(x) dx = n! \delta_{mn}
    ,
\end{equation}
and form a complete orthonormal basis of Hilbert space functions $f(x)$ satisfying the weighted Lebesgue integral in $L_w^2(\mathbb{R})$, i.e.,
\begin{equation}
    \int_{-\infty}^{+\infty} |f(x)|^2 w(x) dx < \infty
    .
\end{equation}
Hence, any function $f(x)$ can be expanded in terms of Hermite polynomials as
\begin{equation}
    f(x) = \sum_{n=0}^{\infty} a_{n} \mathcal{H}_{n}(x)
    ,
\end{equation}
where $a_{n}$ is the Hermite coefficient of order $n$.
By multiplying both sides with $\mathcal{H}_{m}(x)w(x)$, integrating over $x$ and using the mutual orthogonality relation, the Hermite coefficients are found as
\begin{equation}
    a_{m} = \frac{1}{m!} \int_{-\infty}^{+\infty} \mathcal{H}_{m}(x) w(x) f(x) dx
    .
\end{equation}

Moving on to the multivariate case in $D$ dimensions, using $\bm{r}$ as the coordinate, 
the Hermite polynomial tensor of order $n$ is defined as
\begin{equation}
    \bm{\mathcal{H}}_{n}(\bm{r}) = \frac{(-1)^n}{w(\bm{r})} \bm{\nabla}^n w(\bm{r}),
\end{equation}
where $\bm{\mathcal{H}}_{n}$ and $\nabla^n$ are tensors of rank $n$ and $w(\bm{r})$ is the normalized multivariate Gaussian weight function defined as
\begin{equation}
    w(\bm{r}) = \frac{1}{(2\pi)^{D/2}} e^{-\frac{r^2}{2}}
    .
\end{equation}
The Hermite polynomial tensors are mutually orthogonal with respect to the weight function $w(\bm{r})$, as
\begin{equation}
    \int_{\mathbb{R}^D} w(\bm{r}) \bm{\mathcal{H}}_{\chi_1}(\bm{r}) : \bm{\mathcal{H}}_{\chi_2}(\bm{r}) \, d\bm{r} = n! \delta_{\bm{\chi_1},\bm{\chi_2}}
    ,
\end{equation}
where $\chi_1$ and $\chi_2$ denote the set of indices $\{\alpha, \beta, \gamma, \delta, \ldots\}$, and the Kronecker delta $\delta_{\chi_1,\chi_2} = 1$ if the indices are permutations of each other.
They form a complete orthonormal basis of Hilbert space functions $f(\bm{r})$ satisfying the weighted Lebesgue integral in $L_w^2(\mathbb{R}^D)$, i.e.,
\begin{equation}
    \int_{-\infty}^{+\infty} |f(\bm{r})|^2 w(\bm{r}) d\bm{r} < \infty
    .
\end{equation}
Hence, any function $f(\bm{r})$ can be expanded in terms of Hermite polynomials as
\begin{equation}
    f(\bm{r}) = \sum_{n=0}^{\infty} a_{n} : \bm{\mathcal{H}_{n}}(\bm{r})
    ,
\end{equation}
where $a_{n}$ is the Hermite coefficient tensor of order $n$.
By multiplying both sides with $\mathcal{H}_{m}(\bm{r})w(\bm{r})$, integrating over $\bm{r}$ and using the mutual orthogonality relation, the Hermite coefficients are found as
\begin{equation}
    a_{m} = \frac{1}{m!} \int_{-\infty}^{+\infty}  \bm{\mathcal{H}_{m}}(\bm{r}) w(\bm{r}) f(\bm{r}) dx
    .
\end{equation}

In principle, any orthonormal basis can be used to discretize the (Maxwell-Boltzmann) distribution function.
Note that the normalized weight function is of Gaussian nature, which marks an inherent advantage for Hermite polynomials as the orthonormal basis, besides the relation between the Hermite coefficients $\bm{a}_n$ and the macroscopic variables of interest.
Choosing the abscissae, i.e. the discrete velocity set, to be the roots of the orthogonal polynomial of the corresponding degree $n$ results in the maximum algebraic degree of precision of $2Q - 1$.
Hence the set of Hermite polynomials also guarantees best convergence of the approximation.
Harold Grad first introduced the Hermite based discretization of the distribution function in 1949~\cite{grad1949kinetic}.
Moments of the distribution function can be computed and reformulated using the Hermite basis as
\begin{equation}
    \bm{M}_{n} = \int_{\mathbb{R}^D} \overbrace{\bm{v} \otimes \dots \otimes \bm{v}}^{n} f d\bm{v} 
    =\int_{\mathbb{R}^D}   w(\bm{v})  P_{(\infty)}d\bm{v}
    ,
    \label{eq:Moment_Hermitexpanded_infty}
\end{equation}
where $P_{\infty}$ is the polynomial function of the variable $\bm{v}$ with order $\infty$, defined as
\begin{equation}
    P_{\infty}(\bm{v}, \rho, \bm{u}, T) = \frac{\overbrace{\bm{v} \otimes \dots \otimes \bm{v}}^{n} f(\bm{v}, \rho, \bm{u}, T)}{w(\bm{v})}
    .
\end{equation}
Since only a certain amount of moments are needed to correctly recover the the solution of the Boltzmann equation for a targeted regime, e.g. Navier--Stokes--Fourier equations,
the polynomial function $P_{(\infty)}$ can be truncated.
Note that higher-order polynomials have no effect on lower order terms as the Hermite basis marks weighted orthonormal functions.
Hence, the finite-order polynomial function,
\begin{equation}
    P_{M}(\bm{v}, \rho, \bm{u}, T) = \frac{\overbrace{\bm{v} \otimes \dots \otimes \bm{v}}^{n} f_{N}(\bm{v}, \rho, \bm{u}, T)}{w(\bm{v})}
    ,
\end{equation}
where $M = 2N$, and $N$ corresponds to the highest-order moment involved in the targeted hydrodynamics,
can be used to evaluate Eq.~\eqref{eq:Moment_Hermitexpanded_infty} as
\begin{equation}
    \bm{M}_{n} 
    = \int_{\mathbb{R}^D}  w(\bm{v})  P_{M} d\bm{v}
    = \sum_{i=0}^{Q-1} w_i P_{M}(\bm{v}_i, \rho, \bm{u}, T) 
    ,
\end{equation}
using the discrete sum via the Gauss--Hermite quadrature.
Note that $\bm{v}_i$ are the discrete abscissae used for the quadrature and $w_i$ the corresponding weights computed as
\begin{equation}
    w_i = n! \mathcal{H}_{n-1}(v_i)^{-2}
    ,
\end{equation}
which corresponds to the product of the $w_i$ in each direction in the multivariate case.
Hence, the Grad expansion truncated at order $n=N$ can be written as
\begin{equation}
    f_{i,N} = \sum_{n=0}^{N} a_{n} : \bm{\mathcal{H}_{n}}(\bm{v}_i)
    .
\end{equation}
Further, to correctly recover the highest order moment of the targeted hydrodynamics with maximum precision using the Gauss-Hermite quadrature, $M \leq 2Q-1$ must be fulfilled.

\section{Hermite polynomials and coefficients of the Grad expansion\label{appendix:Grad-Hermite}}

In notation mostly adopted in the lattice Boltzmann community, the Grad expansion~\cite{grad1949kinetic} of a discrete population reads
\begin{equation} \label{eq:Grad-Hermite-expansion-static}
     \{f_{i},g_{i}\} 
     = w_i \sum_{n=0}^{N} \frac{1}{n!(RT_{ref})^n} 
     \bm{a}_{n}(\{f,g\}) : \bm{\mathcal{H}}_{n}(\bm{v}_i)
     ,
\end{equation}
where $N$ is the order of the expansion. 
Further, $R$ is the specific gas constant, $w_i$, $\bm{v}_i$, and $T_{ref}$ are the weights, the discrete particle velocities and the reference temperature of the velocity set, and ":" denotes the Frobenius inner product, i.e. the inner product with full contraction between the Hermite coefficient tensors $\bm{a}_{n}$ and the Hermite polynomial tensors $\bm{\mathcal{H}}_{n}$.
The $\bm{\mathcal{H}}_{n}$ are parametrized by the discrete velocities, whereas the $\bm{a}_{n}$ take into account a set of moments up to order $N$, i.e. $\{ M_{n=0}, \dots, \bm{M}_{n=N} \}$ such that the populations $f_{i}$ and $g_{i}$ expanded up to order $N$ correctly reproduce the same set of moments, where 
\begin{equation}
    \bm{M}_{n}(\{f,g\}) = \sum_{i=0}^{Q-1} \overbrace{v_{i} \otimes\dots\otimes v_{i}}^{n} \{f_i,g_i\}
    .
\end{equation}
For the computation of equilibrium populations $f_i^{\rm eq}$ and $g_i^{\rm eq}$, the coefficients $\bm{a}^{\rm{\rm eq}}_{n}$ are found as a function of the set of equilibrium moments $\{ M_{n=0}^{\rm eq}, \dots, \bm{M}_{n=N}^{\rm eq} \}$.

In the following, the Hermite polynomial tensors $\bm{\mathcal{H}}_{n}$ are given up to $N=4$ using index notation, followed by the coefficient tensors $\bm{a}_{n}$ for the case with arbitrary moments and explicitly evaluated with equilibrium moments, respectively.
The Hermite polynomial tensors are found as
    \begin{align} \label{eq:Hermitepolynomials-0-4}
        \mathcal{H}_{0}
        = \ & 1 
        ,
\\
        \mathcal{H}_{\alpha}
        = \ & v_{i\alpha}
        ,
\\
        \mathcal{H}_{\alpha \beta}
        = \ & v_{i\alpha} v_{i\beta} - RT_{ref}\delta_{\alpha \beta} 
        ,
\\
        \mathcal{H}_{\alpha \beta \gamma}
        = \ & v_{i\alpha} v_{i\beta} v_{i\gamma} 
        - RT_{ref} [v_{i\alpha} \delta_{\beta \gamma}]_{cyc} 
        ,
\\  
        \mathcal{H}_{\alpha \beta \gamma \delta}
        = \ & v_{i\alpha} v_{i\beta} v_{i\gamma} v_{i\delta} 
        - RT_{ref} [v_{i\alpha} v_{i\beta} \delta_{\gamma\delta}]_{cyc} 
    \nonumber \\
        & + (RT_{ref})^2 [\delta_{\alpha\beta} \delta_{\gamma\delta}]_{cyc} 
        ,
    \end{align}
and the contracted Hermite polynomial tensors of order three and four as
\begin{align}
    \mathcal{H}_{\alpha\beta\beta}
    = \ & v_{i\alpha} \left(  v_i^2 - RT_{ref}\left( D + 2 \right) \right)
   , 
\\  
    \mathcal{H}_{\alpha \beta\gamma\gamma}
    = \ & v_{i\alpha} v_{i\beta}  \left(  v_i^2 - RT_{ref}\left( D + 4 \right) \right) 
    \nonumber \\
    &- RT_{ref} \delta_{\alpha \beta} \left(  v_i^2 - RT_{ref}\left( D + 2 \right) \right)
    ,
\end{align}
where $"[]_{cyc}"$ denotes cyclic non-repetitive permutation over the indices and $D$ is the physical dimension.
Depending on the model, the contracted tensors can be sufficient compared to the full tensors at the same expansion order.
The corresponding Hermite coefficient tensors for arbitrary moments, here denoted in semi-recursive form, are found as
\begin{align} \label{eq:Hermitecoefficients-static-regform}
        a_{0} 
        = \ & M_{0}
        ,
\\
        a_{\alpha}
        = \ & M_{\alpha}       
        ,
\\
        a_{\alpha\beta} 
        = \ & M_{\alpha \beta} - RT_{ref} a_{0} \delta_{\alpha \beta} 
        ,
\\
        a_{\alpha\beta\gamma} 
        = \ & M_{\alpha\beta\gamma} 
        -  R T_{ref} \left[ a_{\alpha} \delta_{\beta\gamma} \right]_{cyc} 
        ,
\\
        a_{\alpha\beta\gamma\delta} 
        = \ & 
        M_{\alpha\beta\gamma\delta}
        - R T_{ref} \left[ a_{\alpha\beta} \delta_{\gamma\delta} \right]_{cyc}  
    \nonumber \\ &
        - (R T_{ref})^2 a_{0}  \left[ \delta_{\alpha\beta} \delta_{\gamma\delta} \right]_{cyc} 
        ,
\end{align}
and the contracted tensors of order three and four as
\begin{align}
        a_{\alpha\beta\beta} 
        = \ & M_{\alpha\beta\beta} 
        -  R T_{ref}  a_{\alpha} (D+2) 
        ,
\\
        a_{\alpha\beta\gamma\gamma} 
        = \ & 
        M_{\alpha\beta\gamma\gamma}
        - R T_{ref} \left( a_{\alpha\beta} (D+4) + a_{\gamma\gamma}  \delta_{\alpha\beta} \right)
    \nonumber \\ &
        - (R T_{ref})^2 a_{0}  \delta_{\alpha\beta} (D+2)
        .
\end{align}
At equilibrium, the Hermite coefficient tensors written out explicitly read
\begin{align} 
\label{eq:Hermitecoefficients-feq-writtenout-static}
    a_{0}^{\rm{\rm eq}}
    = \ & \rho 
        ,
\\
    a^{\rm{\rm eq}}_{\alpha}
    = \ &   \rho u_{\alpha} 
        ,
\\
    a^{\rm{\rm eq}}_{\alpha\beta} 
    = \ & \rho u_\alpha u_\beta + RT_{ref} (\theta - 1) 
    \rho
    \delta_{\alpha\beta} 
        ,
\\
    a^{\rm{\rm eq}}_{\alpha\beta\gamma} 
    = \ & \rho u_\alpha u_\beta u_\gamma +  R T_{ref} (\theta -1) 
    [
    \rho  u_\alpha 
    \delta_{\beta\gamma}]_{cyc} 
        ,
\\
    a^{\rm{\rm eq}}_{\alpha\beta\gamma\delta} 
    = \ & \rho u_\alpha u_\beta u_\gamma u_\delta + R T_{ref} (\theta -1) [ \rho u_\alpha u_\beta \delta_{\gamma\delta}]_{cyc}  
    \nonumber \\ 
    &+ (R T_{ref})^2 (\theta -1)^2   [\rho \delta_{\alpha\beta} \delta_{\gamma\delta}]_{cyc} 
        ,
\\
        a^{\rm{\rm eq}}_{\alpha\beta\beta} 
        = \ & \rho u_{\alpha} \left( u^2 + R T_{ref}\left(\theta-1\right)\left(D+2\right)\right)
        , 
\\
        a^{\rm{\rm eq}}_{\alpha\beta\gamma\gamma} 
        = \ &    \rho u_{\alpha} u_{\beta} \left( u^2+RT_{ref}\left(\theta-1\right)\left(D+4\right)\right)
    \nonumber \\
        &+RT_{ref}\left(\theta-1\right) \delta_{\alpha\beta} \left(u^2
        +RT_{ref}\left(\theta-1\right)\left(D+2\right) \right)
        ,
\end{align}
where the normalized temperature is defined as $\theta = T / T_{ref}$.

\section{Requirements on the phase-space discretization\label{appendix:requirements}}

In order to capture the fundamental flow physical properties of the \ac{NSF} equations in the hydrodynamic limit,
all moments appearing in the phase-space continuous multiscale expansion have to be matched when computed with discrete quadratures.

In case of the total energy split, 
this means that the $f$-distribution needs to properly recover the continuity and momentum balance equations to the Navier--Stokes order in the Chapman--Enskog expansion, including the dissipative fluxes, i.e. the non-equilibrium moments, appearing in the Navier--Stokes stress tensor $\bm{\tau}_{\rm NS}$, which involve equilibrium moments up to order three of the $f$-distribution.
Hence,
\begin{equation}
    \textcolor{black}{(\mathrm{I})} \quad
    \int \bm{v} \otimes  \bm{v} \otimes  \bm{v} f^{\rm eq} d\bm{v} 
    = \sum_{i=0}^{Q-1}  \bm{v}_i \otimes  \bm{v}_i \otimes  \bm{v}_i f_i^{\rm eq} 
    ,
\end{equation}
and all equilibrium moments of lower order have to be satisfied, which requires at least an order three ($N=3$) expansion of the $f$-equilibria.
The same consideration for the energy balance equation leads to the requirement that the dissipative fluxes have to be correctly recovered to the NSF level, hence
\begin{equation}    
    \textcolor{black}{(\mathrm{I})} \quad
    \int\bm{v}\otimes\bm{v}g^{\rm eq} d\bm{v}  = \sum_{i=0}^{Q-1}  \bm{v}_i \otimes \bm{v}_i g_i^{\rm eq}  
    ,
\end{equation}
leading to a minimal expansion of order two ($N=2$) of the $g$-equilibria.
If the same velocity set is applied to both distributions, as the requirements on $f$ are higher, this means that the requirements on $g$ are automatically satisfied in case the requirements on $f$ satisfied. 
For the quadrature order, this means that a third-order Grad expansion (that of the $f$-equilibria) needs to be accommodated, which requires a fourth-order quadrature (e.g. D$D$Q$4^D$ lattice)
or a third-order quadrature (e.g. D$D$Q$3^D$ lattice) with correction terms~\cite{Prasianakis2007, saadat2021extended} for the incorrect third-order moments of $f$.

In case of the internal non-translational energy split, 
as some part of the $f$-distribution contributes to the balance equation for total energy,
the requirements for recovering the NSF equations are up to the moment
\begin{equation}
    \textcolor{black}{(\mathrm{II})} \quad
    \int \bm{v}\otimes\bm{v} g^{(0)} + \bm{v}\otimes\bm{v} \frac{v^2}{2} f^{(0)}d\bm{v}
    ,
\end{equation}
which means that the requirement for the $g$ distribution, 
\begin{equation}
\textcolor{black}{(\mathrm{II})} \quad
    \int\bm{v}\otimes\bm{v}g^{\rm eq} d\bm{v}  = \sum_{i=0}^{Q-1}  \bm{v}_i \otimes \bm{v}_i g_i^{\rm eq}  
    ,
\end{equation}
i.e. a second-order ($N=2$) expansion of the $g$-equilibria, still holds,
however, the fourth-order equilibrium moment of $f$ also has to be correctly recovered as
\begin{equation}
    \textcolor{black}{(\mathrm{II})} \quad
    \int \bm{v} \otimes \bm{v} \otimes  \bm{v} \otimes  \bm{v} f^{\rm eq} d\bm{v} 
    = \sum_{i=0}^{Q-1}  \bm{v}_i \otimes \bm{v}_i \otimes  \bm{v}_i \otimes  \bm{v}_i f_i^{\rm eq} 
    ,
\end{equation}
or at least the contracted fourth-order equilibrium moment of $f$, as
\begin{equation}
    \textcolor{black}{(\mathrm{II})} \quad
    \int \bm{v} \otimes \bm{v} v^2 f^{\rm eq} d\bm{v} 
    = \sum_{i=0}^{Q-1}  \bm{v}_i \otimes \bm{v}_i v_i^2 f_i^{\rm eq} 
    .
\end{equation}
This can be achieved with a fourth-order Grad expansion ($N=4$) for the $f$-equilibria 
and accommodated either through a fifth-order quadrature (e.g. D$D$Q$5^D$ lattice), 
a fourth-order quadrature (e.g. D$D$Q$4^D$ lattice) with correction terms for the incorrect (contracted) fourth-order moment of $f$,
or a third-order quadrature (e.g. D$D$Q$3^D$ lattice) with correction terms for the incorrect third-order and (contracted) fourth-order moments of $f$.

Note that, to obtain the compressible Euler equations, the dissipative fluxes, i.e. the non-equilibrium pressure and energy flux tensors, don't have to be matched,
hence only moments of up to one order less than for the NSF equations have to be correctly recovered.
This also results in requirements to the order of the Grad expansion and the quadrature of one order less as compared to \ac{NSF}.


\section{Direct parametrization of the second distribution\label{appendix:directParamEq}}

The discrete $g$-equilibria can also be directly parametrized by the discrete $f$-equilibria as in the phase-space continuous kinetic model, cf. Eq.~\eqref{energy_eq_1}.
This reads
\begin{align}
    \label{quad_energy_1_total}
    \textcolor{black}{(\mathrm{I})}\phantom{\mathrm{I}} \quad
    g_i^{\rm eq} &= 
    \left (C_vT-\frac{RDT}{2} + \frac{v_i^2}{2} \right) f_i^{\rm eq}
    ,
\\
    \label{quad_energy_1_intnontrans}
    \textcolor{black}{(\mathrm{II})} \quad
    g_i^{\rm eq} &= 
    \left (C_v T - \frac{RDT}{2} \right) f_i^{\rm eq}
    .
\end{align}
While this parametrization simplifies the computation of the $g_i^{\rm eq}$ for the internal non-translational split, for the total energy split in increases the requirements on the phase-space discretization.
This is due to the $v_i^2$ sitting in the link, increasing the requirements to a fourth-order ($N=4$) expansion of the $g_i$-equilibria and therefore a fifth-order quadrature ($Q=5^D$), i.e. the D$2$Q$25$ velocity set in $D=2$.
Note that an additional requirement for direct parametrization is the application of the same quadrature order, i.e. discrete velocity set, for both distributions, whereas this is not a necessity in case the $g_i$ are constructed independently of the $f_i$ using Grad--Hermite expansions.

\section{Specifications of the velocity sets}
\label{appendix:latticespecs}

The specifications, i.e. particle velocities, weights and reference temperatures, 
for the D$D$Q$4^D$ and D$D$Q$5^D$ velocity sets for $D=2$ built with the Gauss-Hermite quadrature are listed in Table~\ref{tab:theD2Q16Lattice} and~\ref{tab:theD2Q25Lattice}, respectively.

\begin{table}[h!]
\footnotesize
\centering
\begin{tabular}{ |l|l|l| }
 \hline
 \multicolumn{3}{|c|}{ D2Q16  (Hermite) for $T_{ref}=1$ } \\
 \hline
 $i$ & $ \bm{v}_i=(v_{ix},v_{iy}) $ & $w_i$ \\
 \hline
$0-3$  & $(\pm \sqrt{3-\sqrt{6}},\pm \sqrt{3-\sqrt{6}})$   &   $\frac{3+\sqrt{6}}{12} \ \frac{3+\sqrt{6}}{12}$  \\
$4-7$   &$(\pm \sqrt{3-\sqrt{6}} ,\pm \sqrt{3+\sqrt{6}})$  &   $\frac{3+\sqrt{6}}{12} \ \frac{3-\sqrt{6}}{12}$  \\
$8-11$ &   $(\pm \sqrt{3+\sqrt{6}},\pm \sqrt{3-\sqrt{6}})$   &   $\frac{3-\sqrt{6}}{12} \ \frac{3+\sqrt{6}}{12}$ \\
$12-15$  & $(\pm \sqrt{3+\sqrt{6}},\pm \sqrt{3+\sqrt{6}})$   &   $\frac{3-\sqrt{6}}{12} \ \frac{3-\sqrt{6}}{12}$  \\
 \hline
\end{tabular}
\caption{Fourth-order quadrature in two dimensions (D2Q16).}
\label{tab:theD2Q16Lattice}
\end{table}

\begin{table}[h!]
\footnotesize
\centering
\begin{tabular}{ |l|l|l| }
 \hline
 \multicolumn{3}{|c|}{ D2Q25 (Hermite) for $T_{ref}=1$ } \\
 \hline
 $i$ & $ \bm{v}_i=(v_{ix},v_{iy}) $ & $w_i$  \\
 \hline
$0$ & ($0$, $0$) &  $\frac{8}{15}$ \\
$1,3$ & ($\pm \sqrt{5-\sqrt{10}}$, $0$) &  $\frac{8}{15} \ (7+2\sqrt{10})$ \\
$2,4$ & ($0$, $\pm \sqrt{5-\sqrt{10}}$) &  $(7+2\sqrt{10}) \ \frac{8}{15} $ \\
 $5,6,7,8$ & ($\pm \sqrt{5-\sqrt{10}}$, $\pm \sqrt{5-\sqrt{10}}$) &  $(7+2\sqrt{10}) \ (7+2\sqrt{10})$ \\
 $9,11$ & ($\pm \sqrt{5+\sqrt{10}}$, $0$) &  $(7-2\sqrt{10}) \ \frac{8}{15}$ \\
 $10,12$ & ($0$, $\pm \sqrt{5+\sqrt{10}}$) &  $\frac{8}{15} \ (7-2\sqrt{10})$ \\
$13,16,17,20$ & ($\pm \sqrt{5+\sqrt{10}}, \pm \sqrt{5-\sqrt{10}})$ &  $(7-2\sqrt{10}) \ (7+2\sqrt{10})$ \\
$14,15,18,19$ & $(\pm \sqrt{5-\sqrt{10}}, \pm \sqrt{5+\sqrt{10}})$ &  $(7+2\sqrt{10}) \ (7-2\sqrt{10})$ \\
$21,22,23,24$ & $(\pm \sqrt{5+\sqrt{10}}, \pm \sqrt{5+\sqrt{10}})$ &  $(7-2\sqrt{10}) \ (7-2\sqrt{10})$ \\
 \hline
\end{tabular}
\caption{Fifth-order quadrature in two dimensions (D2Q25).}
\label{tab:theD2Q25Lattice}
\end{table}

\section*{References}
\bibliography{_references, _ownpubs}

\end{document}

%% file: _abbrev.tex
\usepackage{acro} 
\acsetup{patch/longtable=false}

\DeclareAcronym{NS}{
  short = NS ,
  long  = Navier--Stokes 
}
\DeclareAcronym{NSF}{
  short = NSF ,
  long  = Navier--Stokes--Fourier 
}
\DeclareAcronym{NSFE}{
  short = NSFE,
  long =  Navier--Stokes--Fourier equation,
  short-plural-form = NSFEs,
  long-plural-form = Navier--Stokes--Fourier equations
}
\DeclareAcronym{PDE}{
  short = PDE,
  long = partial differential equation,
  short-plural-form = PDEs,
  long-plural-form = partial differential equations
}
\DeclareAcronym{EoS}{
  short = EoS,
  long = equation of state,
  short-plural-form = EoS,
  long-plural-form = equations of states
}
\DeclareAcronym{w.r.t.}{
  short = w.r.t. ,
  long  = with respect to 
}
\DeclareAcronym{BE}{
  short = BE ,
  long  = Boltzmann transport equation 
}
\DeclareAcronym{BGK}{
  short = BGK ,
  long  = Bhatnagar--Gross--Krook
}
\DeclareAcronym{BGK-BE}{
  short = BGK-BE ,
  long  = Bhatnagar-Gross-Krook--Boltzmann equation 
}
\DeclareAcronym{MB}{
  short = MB ,
  long  = Maxwell--Boltzmann 
}
\DeclareAcronym{DF}{
  short = DF ,
  long  = distribution function
}
\DeclareAcronym{EDF}{
  short = EDF ,
  long  = equilibrium distribution function
}
\DeclareAcronym{QE}{
  short = QE ,
  long  = quasi-equilibrium
}
\DeclareAcronym{QEDF}{
  short = QEDF ,
  long  = quasi-equilibrium distribution function
}
\DeclareAcronym{QE-BGK}{
  short = QE-BGK ,
  long  = quasi-equilibrium Bhatnagar--Gross--Krook
}
\DeclareAcronym{DDF}{
  short = DDF ,
  long  = double-distribution function
}
\DeclareAcronym{TRT}{
  short = TRT ,
  long  = two-relaxation-time
}
\DeclareAcronym{CE}{
  short = CE ,
  long  = Chapman--Enskog
}
\DeclareAcronym{CFD}{
  short = CFD ,
  long  = computational fluid dynamics
}
\DeclareAcronym{FV}{
  short = FV ,
  long  = finite volume
}
\DeclareAcronym{FD}{
  short = FD ,
  long  = finite difference
}
\DeclareAcronym{AMR}{
  short = AMR ,
  long  = adaptive mesh refinement
}
\DeclareAcronym{AAR}{
  short = AAR ,
  long  = adaptive algorithm refinement
}
\DeclareAcronym{AMAR}{
  short = AMAR ,
  long  = adaptive mesh and algorithm refinement
}
\DeclareAcronym{SAMR}{
  short = SAMR ,
  long  = block-structured adaptive mesh refinement
}
\DeclareAcronym{LBM}{
  short = LBM ,
  long  = lattice Boltzmann method
}
\DeclareAcronym{DVBM}{
  short = DVBM ,
  long  = discrete velocity Boltzmann method
}
\DeclareAcronym{UGKS}{
  short = UGKS ,
  long  = unified gas kinetic scheme
}
\DeclareAcronym{DUGKS}{
  short = DUGKS ,
  long  = discrete unified gas kinetic scheme
}
\DeclareAcronym{ELBM}{
  short = ELBM ,
  long  = entropic lattice Boltzmann method
}
\DeclareAcronym{LBGK}{
  short = LBGK ,
  long  = lattice Boltzmann Bhatnagar--Gross--Krook scheme
}
\DeclareAcronym{IC}{
  short = IC ,
  long  = initial condition
}
\DeclareAcronym{BC}{
  short = BC ,
  long  = boundary condition
}